\documentclass[conference]{IEEEtran4}
\usepackage[hidelinks]{hyperref}
\usepackage{microtype}

\microtypecontext{spacing=nonfrench}

\usepackage{tikz}
\usepackage{amsmath}
\usepackage{enumitem}
\setitemize{noitemsep,topsep=0pt,parsep=0pt,partopsep=0pt}
\usepackage{xspace}
\usepackage{multirow}
\usepackage{booktabs}
\usepackage{makecell}
\usepackage{tabularx}
\usepackage{calc}
\usepackage{subcaption}
\usepackage{amssymb}
\usepackage{siunitx}
\DeclareSIUnit{\KB}{KB}
\DeclareSIUnit{\ops}{ops}
\usepackage{pifont}%
\usepackage{listings}
\usepackage[table, dvipsnames]{xcolor}
\definecolor{darkpastelgreen}{rgb}{0.01, 0.75, 0.24}
\definecolor{darkpastelred}{rgb}{0.76, 0.23, 0.13}
\newcommand{\cmark}{\textcolor{darkpastelgreen}{\ding{51}}}%
\newcommand{\xmark}{\textcolor{darkpastelred}{\ding{55}}}%
\newcommand{\nmark}{\textcolor{black!45}{--}}%
\newcommand{\lnote}[1]{\textsuperscript{\textit{#1}}}%
\newcommand{\cifm}[2]{\ifnum#1=1\textbf{#2}\else\textcolor{black!30}{#2}\fi}%

\usepackage{grumble}
\usepackage{circledsteps}

\definecolor{light-red}{HTML}{F0CFCC}
\definecolor{dark-red}{HTML}{B85450}
\definecolor{light-green}{HTML}{D5E8D4}
\definecolor{dark-green}{HTML}{82B366}
\definecolor{light-yellow}{HTML}{FFF2CC}
\definecolor{dark-yellow}{HTML}{D6B656}
\definecolor{light-purple}{HTML}{E1D5E7}
\definecolor{dark-purple}{HTML}{9673A6}
\definecolor{light-blue}{HTML}{DAE8FC}
\definecolor{dark-blue}{HTML}{6C8EBF}

\def\UrlBreaks{\do\/\do\a\do\b\do\c\do\d\do\e\do\f\do\g\do\h\do\i\do\j\do\k\do\l\do\m\do\n\do\o\do\p\do\q\do\r\do\s\do\t\do\u\do\v\do\w\do\x\do\y\do\z\do\A\do\B\do\C\do\D\do\E\do\F\do\G\do\H\do\I\do\J\do\K\do\L\do\M\do\N\do\O\do\P\do\Q\do\R\do\S\do\T\do\U\do\V\do\W\do\X\do\Y\do\Z\do0\do1\do2\do3\do4\do5\do6\do7\do8\do9\do=\do/\do.\do:}

\usepackage[most]{tcolorbox}
\usepackage[ruled,vlined,linesnumbered]{algorithm2e}

\makeatletter
\newcommand{\removelatexerror}{\let\@latex@error\@gobble}
\makeatother

\newcommand{\myfontsizeapi}{\fontsize{8}{9}\selectfont}

\newcommand{\coloropacity}{32}

\newcommand{\kvscolor}{CornflowerBlue!\coloropacity!white}
\newcommand{\logcolor}{Apricot!\coloropacity!white}
\newcommand{\fscolor}{SpringGreen!\coloropacity!white}
\newcommand{\provcolor}{Lavender!\coloropacity!white}

\newlength{\bandpad}
\newcommand{\bandtop}{\rule{0pt}{\dimexpr\ht\strutbox+\bandpad\relax}}
\newcommand{\bandbot}{\rule[\dimexpr-\dp\strutbox-\bandpad\relax]{0pt}{0pt}}
\newcommand{\headpad}{\bandtop\bandbot}
\newcommand{\crowlast}[5]{\cellcolor{#1}\multirow{-#2}{*}{#3} & \cellcolor{#1}\bandbot\emph{#4} & \cellcolor{#1}#5 \\}
\newcommand{\crowreg}[4]{\cellcolor{#1} & \cellcolor{#1}\emph{#2} & \cellcolor{#1}#3 #4}
\newcommand{\crowfirst}[4]{\cellcolor{#1}\bandtop & \cellcolor{#1}\emph{#2} & \cellcolor{#1}#3 #4}

\definecolor{codegreen}{rgb}{0,0.6,0}
\definecolor{codegray}{rgb}{0.5,0.5,0.5}
\definecolor{codepurple}{rgb}{0.58,0,0.82}
\definecolor{backcolour}{rgb}{0.95,0.95,0.92}
\lstdefinestyle{customc}{
    backgroundcolor=\color{backcolour},   
    commentstyle=\color{codegreen},
    keywordstyle=\color{magenta},
    numberstyle=\tiny\color{codegray},
    stringstyle=\color{codepurple},
    basicstyle=\ttfamily\footnotesize,
    breaklines,
    tabsize=2,
    numbers=left,
    columns=fullflexible,
    keepspaces=true,
    frame=lines,
    numbersep=2pt,
    escapechar=|,
    captionpos=b,
    language=c,
    keywords={PMEMoid, PMEMobjpool, size_t, void, for, int, struct, return, TX_BEGIN, TX_END, TX_ONABORT, long}
}
\renewcommand{\subsectionautorefname}{\S\!}%

\usepackage[subtle,margins=normal,lists=normal]{savetrees}

\usepackage{setspace}

\newcommand{\myparagraph}[1]{\noindent\textit{\textbf{{#1}.}~}}

\newcommand{\out}[1] {}

\newcounter{codeLineCntr}

\reversemarginpar
\newif\ifnotes
\notestrue

\newcommand{\punt}[1]{}

\renewcommand{\eqref}[1]{Equation~(\ref{eq:#1})}

\newcommand{\proc}[1]{\ifmmode\mbox{\textsc{#1}}\else\textsc{#1}\fi}

  \newcommand{\func}[1]{\ifmmode\mathrm{#1}\else\textrm{#1}fi}

\newcounter{remark}[section]

\newcounter{rqcounter}
\newcounter{takeawaycounter}
\newtcolorbox{takeaway}{
  enhanced,
  unbreakable,
  sharp corners,
  frame hidden,
  colback=dark-green!15!white,
  left=3pt,
  right=3pt,
  top=0pt,
  bottom=0pt,
  before skip=0pt,
  after skip=0pt,
  overlay unbroken and first={
    \draw[line width=2.5pt, color=dark-green, line cap=rect]
      ([xshift=0pt]interior.north west) -- ([xshift=0pt]interior.south west);
  },
  overlay middle and last={
    \draw[line width=2.5pt, color=dark-green, line cap=rect]
      ([yshift=0pt]interior.north west) -- ([yshift=0pt]interior.south west);
  },
  before upper={\stepcounter{takeawaycounter}\noindent\textbf{Takeaway~\thetakeawaycounter:}$\;$},
  fontupper=\normalfont
}

\RequirePackage{xcolor}
\RequirePackage{ifthen}
\RequirePackage[normalem]{ulem}

\definecolor{rev1}{HTML}{0000FF}
\definecolor{rev2}{HTML}{0000FF}
\definecolor{rev3}{HTML}{0000FF}
\definecolor{rev4}{HTML}{0000FF}
\definecolor{rev5}{HTML}{0000FF}
\definecolor{revdelcolor}{HTML}{9B1C1C}

\makeatletter

\@ifundefined{ifpublicversion}%
  {\newboolean{publicversion}\setboolean{publicversion}{false}}{}
\newif\ifrevmarkup
\ifthenelse{\boolean{publicversion}}{\revmarkupfalse}{\revmarkuptrue}

\newcommand{\revtagraise}{1.7ex}
\newcommand{\revtagfont}{\fontsize{4.8}{5}\selectfont\bfseries}
\newcommand{\revtagsep}{1pt}

\newcommand{\revtag}[2][rev1]{%
  \rlap{\smash{\raisebox{\revtagraise}{%
    \setlength{\fboxsep}{\revtagsep}%
    {\color{#1}\revtagfont #2}}}}}

\newcommand{\rev}[3][rev1]{%
  \ifrevmarkup\textcolor{#1}{\revtag[#1]{#2}#3}\else#3\fi}

\newcommand{\revnoref}[2][rev1]{%
  \ifrevmarkup\textcolor{#1}{#2}\else#2\fi}

\NewDocumentEnvironment{revblock}{O{rev1}}%
  {\@ifnextchar\bgroup{\rev@block{#1}}{\rev@block{#1}{}}}%
  {\ignorespacesafterend}
\newcommand{\rev@block}[2]{%
  \ifrevmarkup
    \if\relax\detokenize{#2}\relax\else\textcolor{#1}{\revtag[#1]{#2}}\fi
    \color{#1}%
  \fi\ignorespaces}

\NewDocumentCommand{\revbox}{O{rev1}O{0.7em}mm}{%
  \begingroup
    \ifrevmarkup
      \setlength{\fboxrule}{1pt}%
      \setbox0=\hbox{#4}%
      \setbox1=\hbox{\colorbox{#1}{\color{white}\scriptsize #3}}%
      \leavevmode
      \hbox{%
        \rlap{\copy0}%
        \rlap{\smash{\raise\dimexpr\ht0-\fboxrule\relax
            \hbox{\color{#1}\rule{\wd0}{\fboxrule}}}}%
        \rlap{\smash{\raise\dimexpr#2-\dp0\relax
            \hbox{\color{#1}\rule{\wd0}{\fboxrule}}}}%
        \rlap{\smash{\raise\dimexpr#2-\dp0\relax
            \hbox{\color{#1}%
              \rule{\fboxrule}{\dimexpr\ht0+\dp0-#2\relax}}}}%
        \rlap{\smash{\hspace{\dimexpr\wd0-\fboxrule\relax}%
            \raise\dimexpr#2-\dp0\relax
            \hbox{\color{#1}%
              \rule{\fboxrule}{\dimexpr\ht0+\dp0-#2\relax}}}}%
        \rlap{\smash{\raise\dimexpr\ht0-\ht1\relax\hbox{\copy1}}}%
        \phantom{\copy0}%
      }%
    \else
      #4%
    \fi
  \endgroup
}

\newbox\rev@float@body
\newbox\rev@float@tag
\newdimen\rev@float@ht
\NewDocumentEnvironment{revfloat}{O{rev1}O{0pt}m}{%
  \ifrevmarkup
    \def\rev@float@color{#1}%
    \def\rev@float@pad{#2}%
    \setbox\rev@float@tag=\hbox{\colorbox{#1}{\color{white}\scriptsize #3}}%
    \setbox\rev@float@body=\vbox\bgroup
  \fi
}{%
  \ifrevmarkup
    \par\egroup
    \setlength{\fboxrule}{1pt}%
    \rev@float@ht=\dimexpr\ht\rev@float@body+\dp\rev@float@body+\rev@float@pad+\rev@float@pad\relax
    \hbox{%
      \rlap{\copy\rev@float@body}%
      \rlap{\smash{\raise\dimexpr\ht\rev@float@body+\rev@float@pad-\fboxrule\relax
          \hbox{\color{\rev@float@color}\rule{\wd\rev@float@body}{\fboxrule}}}}%
      \rlap{\smash{\lower\dimexpr\dp\rev@float@body+\rev@float@pad\relax
          \hbox{\color{\rev@float@color}\rule{\wd\rev@float@body}{\fboxrule}}}}%
      \rlap{\smash{\lower\dimexpr\dp\rev@float@body+\rev@float@pad\relax
          \hbox{\color{\rev@float@color}\rule{\fboxrule}{\rev@float@ht}}}}%
      \rlap{\smash{\hskip\dimexpr\wd\rev@float@body-\fboxrule\relax
          \lower\dimexpr\dp\rev@float@body+\rev@float@pad\relax
          \hbox{\color{\rev@float@color}\rule{\fboxrule}{\rev@float@ht}}}}%
      \rlap{\smash{\raise\dimexpr\ht\rev@float@body+\rev@float@pad-\ht\rev@float@tag\relax
          \hbox{\copy\rev@float@tag}}}%
      \hskip\wd\rev@float@body
    }%
  \fi
}

\makeatother

\usepackage{graphicx}

\newcommand{\projectname}{\textsc {CVM-Store}\xspace}

\newif\ifextended
\ifdefined\cameraready\extendedfalse\else\extendedtrue\fi

\ifextended
  \newcommand{\extendedonly}[1]{\autoref{#1}}
  \newcommand{\extversion}{the extended version of this paper}
\else
  \newcommand{\extendedonly}[1]{\extversion}
  \newcommand{\extversion}{the extended version of this paper~\cite{cvmstoreExtended}}
\fi

\ifextended\else
  \usepackage[available,functional,reproduced]{ndssbadges}
\fi

\begin{document}

\date{}

\title{\projectname: A Secure Storage System \\ for Confidential Virtual Machines}

\author{\IEEEauthorblockN{Thore Sommer, Masanori Misono, Dimitrios Stavrakakis, Pramod Bhatotia}
\IEEEauthorblockA{Technical University of Munich, Germany%
\ifextended\else\\
\{thore.sommer, masanori.misono, dimitrios.stavrakakis, pramod.bhatotia\}@tum.de\fi}
}

\IEEEoverridecommandlockouts
\ifextended\else
\makeatletter\def\@IEEEpubidpullup{6.5\baselineskip}\makeatother
\IEEEpubid{\parbox{\columnwidth}{
		Network and Distributed System Security (NDSS) Symposium 2027\\
		22--26 March 2027, Seoul, Republic of Korea\\
		ISBN 978-1-970672-09-1\\
		https://dx.doi.org/10.14722/ndss.2027.230799\\
		www.ndss-symposium.org
}\hspace{\columnsep}\makebox[\columnwidth]{}}
\fi

\maketitle

\begin{abstract}

\rev[rev1]{G.02, G.15}{Modern stateful cloud workloads require high-performance persistent storage when running on untrusted cloud infrastructure.
Examples are AI inference and training pipelines that load multi-gigabyte model weights, checkpoints, and large datasets that must not be silently rolled back,
and other workloads such as container runtimes that provision immutable image layers and assets.}
Confidential Virtual Machines (CVMs) promise to protect these workloads through hardware-enforced memory encryption and isolation without requiring application changes.
However, they protect only data in use; persistent storage lies outside the CVM trust boundary, leaving all stateful data exposed to the cloud provider.

Extending the trust boundary to persistent storage presents inherent challenges.
High-bandwidth workloads require near-raw NVMe throughput, yet the default kernel CVM I/O stacks are inadequate, and userspace I/O libraries are not designed with CVM use cases in mind.
Access patterns and storage semantics vary between workloads, necessitating access-pattern-aware security abstractions rather than uniform block-level encryption.
In order to mitigate against rollback, freshness is needed, but CVMs lack hardware primitives for this purpose, requiring alternative solutions.

We present \projectname{}, a storage system that co-designs a high-throughput userspace I/O engine with CIF (confidentiality, integrity, freshness) data protection.
The key insight is that neither authenticated encryption nor I/O is a bottleneck by itself, but their naive combination is. 
\projectname{} exploits this by interleaving them, eliminating redundant copies and OS/hypervisor mediation.
CIF guarantees are provided as composable primitives that adapt protection to workload access patterns, from which we derive three concrete authenticated abstractions: a key-value store, a shared log, and a filesystem.
An asynchronous trusted counter service provides freshness guarantees with epoch-based batching, which in turn enables rollback protection without dedicated hardware support.
\revnoref{\projectname{} delivers full CIF guarantees within $\approx$1\% (write) and 1--19\% (read) of raw NVMe bandwidth for large objects.}

\end{abstract}

\section{Introduction}

A broad range of sensitive workloads are increasingly deployed in public clouds and require performant persistent storage.
These include \rev[rev1]{G.02}{container runtimes that provision large, immutable image layers and datasets}~\cite{linLargescaleDataSet2020,zhaoLargeScaleAnalysisDocker2019,confidentialcontainersConfidentialContainersSecure}, \rev[rev1]{G.02, G.15}{checkpointing and audit-logging services}~\cite{michaelForensicValidityApproximated2020,angelNimbleRollbackProtection2023}, and HPC, AI training and inference pipelines~\cite{wangFastPersistAcceleratingModel2024,mohanVelaVirtualizedLLM2025, quocSecureTFSecureTensorFlow2020, zhangCitadelProtectingData2021}, spanning sensitive domains such as healthcare record management and financial transaction processing.
In all cases, the cloud provider controls the entire workload lifecycle and may try to undermine the security guarantees such workloads depend on~\cite{popaEnablingSecurityCloud2011,mahajanDepotCloudStorage2011}.

Recently, Confidential Virtual Machines (CVMs) have become the dominant model for confidential computing~\cite{confidential_computing} in the cloud.
CVMs protect entire VMs with hardware-enforced memory encryption and isolation, enabling unmodified applications to run securely on untrusted infrastructure.
Major cloud providers~\cite{gcp-confidential-computing,azure-confidential-computing, ibm-confidential-computing} have deployed CVMs, offered by various vendors (e.g., AMD SEV-SNP~\cite{amdAMDSEVSNPStrengthening2020}, Intel TDX~\cite{tdx}, Arm CCA~\cite{cca}), at scale, making them the de facto foundation for running security-sensitive workloads.

However, CVMs protect only data \emph{in use}, leaving persistent storage outside their trust boundary.
NVMe devices, storage controllers, and the I/O stack remain under full control of the untrusted cloud provider.
Thus, a malicious operator can read, corrupt, or replay persistent data, exposing every workload that requires storage outside CVM memory and relies on local storage such as NVMe SSDs.
Importantly, fully closing this threat requires not only cryptographic protection of stored data, but also a mechanism to bind that data to a verified, trusted identity. 
Therefore, designing such secure storage systems for CVMs is necessary but non-trivial.

\rev[rev1]{G.04}{Simply layering existing confidential-storage techniques~\cite{liBifrostAnalysisOptimization2023,misonoConfidentialVMsExplained2024} onto the CVM I/O path compounds the problem:
CVMs encrypt their guest memory by default, preventing devices from performing direct memory access (DMA) to it.
All I/O must instead pass through \emph{bounce buffers}~\cite{bounce_buffer}, unencrypted shared memory regions where data is copied to and copied back, along with multiple context switches, further degrading I/O performance compared to conventional VMs~\cite{liBifrostAnalysisOptimization2023,qiu-cvmdb-2024}. 
Block-level encryption then adds another layer on top.
Neither cost is inherent. Userspace I/O frameworks~\cite{yang2017spdk, dpdk} bypass the kernel to optimize I/O operations, and modern AEAD ciphers can saturate NVMe bandwidth on a single core.
However, userspace I/O frameworks assume a trusted device model, which does not hold for CVM deployments: a malicious NVMe controller can read from DMA buffers, corrupt control structures, or selectively drop and reorder I/O operations, rendering these frameworks incompatible with the CVM threat model in their current form.
This mismatch between a fast I/O path built for trusted devices and the CVMs' untrusted-device reality makes securing CVM storage fundamentally challenging (\autoref{sec:motivation}).}

Further, protecting persistent data requires ensuring three properties: confidentiality, integrity, and freshness (CIF).
Existing solutions fall short, as generic block-level encryption~\cite{CryptsetupCryptsetupGitLab2026,chakrabortiDmxProtectingVolumelevel2017} protects only data uniformly across all access patterns, performing redundant cryptographic operations, and application-level schemes~\cite{bailleuSPEICHERSecuringLSMbased2019,anchor} bind their guarantees to a single workload.
Both remain vulnerable to rollback attacks where adversaries replay cryptographically valid but stale data after crashes or migrations.
Although counter-based freshness mechanisms exist~\cite{chuRollbaccineHerdImmunity2025, angelNimbleRollbackProtection2023}, they rely on hardware primitives (e.g., Intel SGX monotonic counters) or external distributed systems~\cite{mateticROTERollbackProtection2017} that suffer from limited write endurance~\cite{bailleuSPEICHERSecuringLSMbased2019, anchor} or extreme latency, rendering them unsuitable for high-throughput CVM storage workloads.

\revnoref{As a solution,} we present \projectname{}, a storage system tailored for CVM deployments that delivers end-to-end CIF guarantees.
Our key insight is that CVM storage security and I/O performance are \revnoref{\textit{not}} fundamentally at odds. Neither encryption nor userspace I/O is the bottleneck in isolation, only their naive composition is.
\rev[rev1]{G.04, G.08}{\projectname{}'s novelty lies in four principles that make this combination work under the CVM threat model, rather than in any single building block: \textit{(i)} hardening the userspace I/O trust boundary instead of assuming a trusted device, \textit{(ii)} interleaving cryptographic operations with in-flight DMA instead of serializing them, \textit{(iii)} aligning integrity structures with each abstraction's access pattern instead of applying uniform block-level protection, and \textit{(iv)} decoupling freshness into an external, epoch-batched service behind a minimal interface.}
\revnoref{Precisely, \projectname{} combines a hardened, kernel-bypass I/O engine, access-pattern-aware CIF primitives instantiated across a key-value store, shared log, and filesystem, an asynchronous trusted counter service that externalizes monotonic state via epoch-based batching, and an attestation and sealing pipeline that binds provisioned keys and counter state to the attested CVM identity.}

We implement \projectname{} for AMD SEV-SNP, contributing upstream fixes to SPDK DMA handling and the SEV-SNP attestation pipeline.
\revnoref{Our evaluation demonstrates that \projectname{} with full CIF protections stays within $\approx$1\% (write) and 1--19\% (read) of raw NVMe bandwidth on large objects from a single core (up to \qty{6.6}{\gibi\byte\per\second} and \qty{13.3}{\gibi\byte\per\second}), outperforming unencrypted CVM kernel I/O.}
End-to-end attestation completes in $\approx$\qty{188}{\milli\second}, and the trusted counter interval (\qtyrange{1}{281}{\milli\second}) allows applications to tune the performance-freshness trade-off according to their needs.

Overall, this paper makes four key contributions:
\begin{itemize}[leftmargin=*]
\item \rev[rev1]{G.08}{A \textit{hardened kernel-bypass I/O engine} that minimizes exposed state through CVM-aware DMA allocations and hardens the NVMe driver against untrusted infrastructure.}
\item \rev[rev1]{G.08}{\textit{Composable CIF primitives} that interleave AEAD encryption with in-flight DMA, hiding the CIF cost behind I/O.}
\item \rev[rev1]{G.08}{Per use-case \textit{integrity structures} matched to each access pattern, demonstrated through a KVS, shared log, and filesystem.}
\item \rev[rev1]{G.08}{A \textit{pluggable trusted counter service} behind a minimal interface with configurable epoch granularity to bound the rollback window, anchored by an attestation and sealing pipeline binding keys and counter state to the attested CVM identity.}
\end{itemize}

\section{Background}
\label{sec:background}

\subsection{Confidential Virtual Machines (CVMs)}
CVMs~\cite{amdAMDSEVSNPStrengthening2020, tdx, cca} enable confidential computing in untrusted clouds through hardware-enforced isolation and memory encryption for entire VMs~\cite{misonoConfidentialVMsExplained2024}.
During launch, the hypervisor
loads guest code, data, and state, but thereafter cannot access or modify guest memory or state\revnoref{, and the guest firmware configures the private memory and performs necessary measurements.}
Further, CVMs support generating attestation evidence to verify their initial state and platform integrity.

\myparagraph{Sealing and provisioning}
To enable secure data persistence and provisioning, sealing mechanisms bind encrypted data to specific evidence.
Sealing is implemented by generating an ephemeral asymmetric key and including its public-key hash in the user-controllable portion of the attestation evidence.
After verifying the evidence and public-key hash, the sealing party encrypts sensitive data (e.g., encryption keys) so only the CVM with the sealing key can decrypt it.

\begin{figure}
    \centering
    \includegraphics[width=0.9\linewidth,trim=0 0 0 0.7,clip]{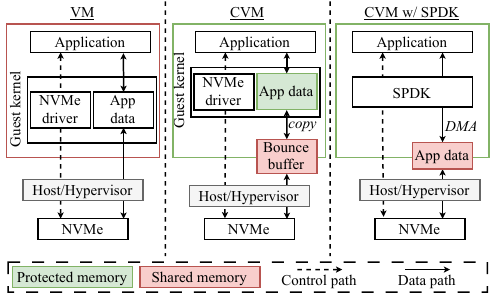}
    \vspace{-0.2em}
    \caption{I/O mechanism comparison in normal VMs, CVMs with kernel I/O, and a strawman design of SPDK in CVMs.}
    \label{fig:storage-stack}
    \vspace{-0.2em}
\end{figure}

\subsection{CVM I/O Architecture}
In CVMs, the hypervisor cannot directly access encrypted memory; thus, CVMs rely on an \textit{unencrypted} shared buffer (\emph{bounce buffer}~\cite{bounce_buffer}) for I/O communication.
The typical I/O process involves~\cite{liBifrostAnalysisOptimization2023}: \emph{(i)}~the guest kernel encrypts and copies data to a bounce buffer, \emph{(ii)}~the hypervisor or device (i.e., DMA) accesses data from this shared region, and \emph{(iii)}~for received data, the guest copies and decrypts it from the bounce buffer.

\myparagraph{Userspace I/O frameworks}
To reduce I/O overhead, kernel-bypass frameworks (e.g., SPDK~\cite{yang2017spdk}, DPDK~\cite{dpdk}) \revnoref{map device registers directly into userspace, so applications interact with devices without syscalls and poll for completion instead of receiving interrupts.}
In virtualized environments, userspace I/O frameworks can be combined with device passthrough mechanisms (VFIO~\cite{vfio}, iommufd~\cite{iommufd}).
In CVMs, userspace I/O still requires bounce buffers for DMA, and direct device access exposes the CVM to potentially malicious device behavior (\autoref{fig:storage-stack}).
While recent work has explored userspace I/O in CVMs~\cite{lamBoostingNetworkPerformance2024, yangFastSecureDPDK2023, liBifrostAnalysisOptimization2023, liBridgeFutureHighPerformance2024}, 
these efforts focus on the network I/O path and do not explore combining the I/O layer with CIF properties\revnoref{, and there is no publicly available userspace I/O framework for modern CVMs.}

\myparagraph{Trusted I/O devices}
The recent TEE Device Interface Security Protocol (TDISP)~\cite{pci-sigPCIExpressBase2024, pcie-ide-tdisp} extends the CVM trust boundary to include I/O devices.
It provides device attestation and encrypts PCIe transactions,
enabling devices to directly access encrypted memory.
TDISP \revnoref{requires hardware support} and is not yet widely commercially available.
Furthermore, it does not ensure security properties for data at rest; it still depends on software or support from the storage medium (e.g., self-encrypting drives~\cite{trustedcomputinggroupTCGStorageSecurity2025} or in-storage computing~\cite{kangIceClaveTrustedExecution2021}).

\begin{figure}[t]
\centering
\includegraphics[width=\linewidth,trim=0 0 0 6.8,clip]{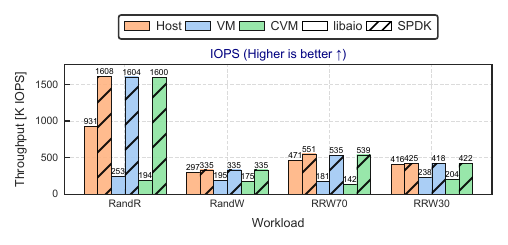}
\vspace{-2.4em}
\caption{Difference in throughput between host, VM, and CVM using the kernel libaio and the userspace SPDK driver.}
\label{fig:dma-overhead}
\vspace{-0.7em}
\end{figure}

\subsection{Modern Cloud Storage Devices}
Modern, high-performance NVMe SSDs communicate with the host via a queue-based interface comprising submission queues (SQs) and completion queues (CQs), enabling parallel processing and reduced latency compared to legacy storage protocols~\cite{nvme-spec}.
The host submits I/O commands to SQs. The controller fetches them, performs the requested DMA transfers, and posts completion entries to CQs.

\section{Motivation and Key Ideas}
\label{sec:motivation}

Protecting persistent storage under the CVM threat model is essential for modern stateful cloud workloads.
\rev[rev1]{G.04}{However, although AEAD ciphers, Merkle trees, and userspace I/O are each well-established individually, naively composing them under the CVM trust model leaves significant performance on the table.
Below, we identify three storage requirements arising from this intersection (experimental setup in \autoref{subsec:experimental_setup}).}

\myparagraph{Requirement \#1: Secure, performant storage I/O} Container runtimes load multi-GB image layers before pod start, AI inference loads tens-of-GB model weights on every cold start, and ledger services require continuous append-only logging.
However, standard CVM I/O stacks fail to meet such demands.
To this end, we measure I/O throughput using fio~\cite{FioFlexibleTester} across virtualization architectures with varying access patterns (\autoref{fig:dma-overhead}). We observe that CVMs achieve only up to 90\% of VM write throughput and 77\% of read throughput via the kernel I/O stack.
This is mainly because devices cannot perform DMA into protected CVM memory, making bounce buffers mandatory for I/O operations.
As a result, each operation incurs at least two extra memory copies.
Additional context switches across the trust boundary further increase the incurred overhead~\cite{misonoConfidentialVMsExplained2024,liBifrostAnalysisOptimization2023}.

Userspace I/O~\cite{yang2017spdk}
\revnoref{eliminates kernel overhead but does not resolve the bounce buffer under the CVM threat model:
devices cannot DMA into protected CVM memory, and an attacker can snoop data in DMA buffers or manipulate NVMe control data structures, requiring their drivers to be hardened.}

\projectname{} integrates protection into a hardened kernel-bypass I/O engine.
CVM-aware DMA allocations keep sensitive driver state out of shared memory, while in-path encryption/decryption subsumes unnecessary bounce-buffer copies.
On top of that, completion validation and bounds checks harden the device interface against untrusted infrastructure.

\begin{figure}[t]
\includegraphics[trim=0 0 0 6.8,clip]{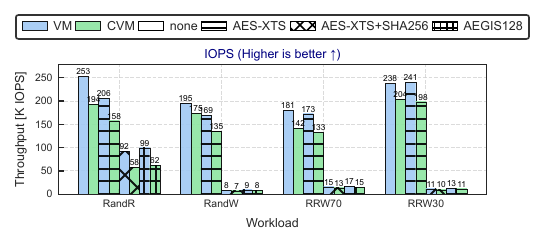}
\vspace{-2.4em}
\caption{Disk encryption performance: LUKS overhead with different cipher suites vs. baseline kernel driver inside a VM.}
\label{fig:luks-overhead}
\vspace{-0.7em}
\end{figure}

\begin{table}[t]
    \centering
    \footnotesize
    \revnoref[rev1]{%
    \setlength{\tabcolsep}{3pt}%
    \begin{tabular*}{\linewidth}{@{\extracolsep{\fill}}llll@{}}
        \toprule
        Application & Stored data & Pattern & Volume \\
        \midrule
        Confidential AI~\cite{wangFastPersistAcceleratingModel2024} & Model weights & Write-once & GB--TB \\
        Cloud apps~\cite{linLargescaleDataSet2020,zhaoLargeScaleAnalysisDocker2019} & Image layers & Write-once & MB--GB \\
        Forensic media~\cite{xuVStoreDataStore2019,rodriguez-silvaVideoSurveillanceBased2012,silvaPACSArchiveArchitecture2012} & Video, imaging & Write-once & GB--PB \\
        Object caches~\cite{luWiscKeySeparatingKeys2016,nguyenWhyFilesIf2024} & Cached objects & Keyed R/W & KB--GB \\
        Event logs~\cite{michaelForensicValidityApproximated2020} & Audit trails & Append-only & KB--MB \\
        \bottomrule
    \end{tabular*}}
    \vspace{-0.6em}
    \caption{\rev[rev1]{G.02}{Motivating large-object \projectname{} workloads.}}
    \label{tbl:application-patterns}
    \vspace{-0.8em}
\end{table}

\myparagraph{Requirement \#2: Workload-tailored abstractions}
CVMs provide no built-in data-at-rest protection, and application-level encryption is error-prone and often incomplete.
\rev[rev1]{G.02}{These large-object workloads exhibit fundamentally different access patterns that collapse into three classes, as summarized in \autoref{tbl:application-patterns}.
For instance, container image layers, model weights, and forensic media are large and write-once, with sequential writes followed by sequential or random reads~\cite{linLargescaleDataSet2020,zhaoLargeScaleAnalysisDocker2019,wangFastPersistAcceleratingModel2024,xuVStoreDataStore2019,rodriguez-silvaVideoSurveillanceBased2012,silvaPACSArchiveArchitecture2012}.
Audit trails and event logs are append-only and rarely read~\cite{michaelForensicValidityApproximated2020,angelNimbleRollbackProtection2023}, whereas object caches store variable-size blobs retrieved by key~\cite{luWiscKeySeparatingKeys2016,welchOptimizingHybridSSD2013}.}

\begin{table}[t]
    \centering
    \footnotesize
    \begin{tabular*}{\linewidth}{@{\hspace{1pt}\extracolsep{\fill}}l@{\hspace{4pt}}c@{\hspace{8pt}}c@{\hspace{1pt}}}
        \toprule
        Scheme  & Throughput (GB/s) [VM/CVM] &  CPU util. (\%) \\
        \midrule
        No encryption & 13.41 / 6.56 & 26.30 / 43.78 \\
        LUKS~(AES-XTS) & 10.33 / 4.10 & 38.43 / 46.15 \\
        LUKS~(AES-XTS+sha256) & 5.94 / 4.11 & 73.51 / 68.24 \\
        LUKS~(AEGIS-128) & 8.14 / 5.07 & 46.38 / 52.95 \\
        dm-verity & 3.42 / 2.53 & 54.89 / 50.23 \\
        \bottomrule
    \end{tabular*}
    \vspace{-0.6em}
    \caption{Block-level read encryption/integrity performance.}
    \label{tbl:luks-performance}
    \vspace{-1.1em}
\end{table}

Generic block-level encryption~\cite{dm-crypt} ignores access-pattern structure, incurring unnecessary overheads.
For write-once files, dynamic Merkle tree updates on every write are redundant; a static tree computed at write time suffices.
For append-only logs, Merkle tree integrity \revnoref{poorly matches} the access pattern; instead, counter-chained MACs provide integrity and ordering more efficiently.
Sometimes, a simple KV store is sufficient to \revnoref{securely persist application data, making the full} block-level storage overhead unnecessary.
\autoref{fig:luks-overhead} and \autoref{tbl:luks-performance} quantify this cost: authenticated block encryption with AES-XTS+sha256 reduces throughput by 37--56\%, while AEGIS-128 achieves up to \qty{8.14}{\giga\byte\per\second}.

\begin{table}[!tb]
\centering
\footnotesize
\begin{revfloat}[rev1]{G.04}
\begin{tabular*}{\linewidth}{@{\hspace{4pt}\extracolsep{\fill}}lccc@{\hspace{4pt}}}
\toprule
\multirow{2}{*}[-.5ex]{Algorithm}& \multicolumn{3}{c}{Throughput (GB/s) [Kernel/OpenSSL]} \\
\cmidrule(lr){2-4}
        & Host & VM & CVM \\
\midrule
AES-256-GCM & 4.52 / 4.38 & 4.47 / 4.31 & 4.43 / 4.22 \\
AES-256-XTS & 6.15 / 7.97 & 6.19 / 7.83 & 5.98 / 7.77 \\
ChaCha20 & 1.83 / 3.20 & 1.79 / 3.14 & 1.73 / 3.13 \\
ChaCha20-Poly1305 & 1.18 / 2.24 & 1.12 / 2.21 & 1.09 / 2.20 \\
SHA3-256 & 0.14 / 0.30 & 0.13 / 0.29 & 0.13 / 0.29 \\
SHA-256 & 1.03 / 1.04 & 1.02 / 1.03 & 0.98 / 1.01 \\
AEGIS-128 & 4.19 / -- & 4.38 / -- & 3.90 / -- \\
AEGIS-128L & -- / 9.47 & -- / 9.33 & -- / 9.25 \\
AEGIS-128X2 & -- / 20.07 & -- / 19.76 & -- / 19.59 \\
AEGIS-128X4 & -- / 27.06 & -- / 26.70 & -- / 26.38 \\
\bottomrule
\end{tabular*}
\vspace{-0.5em}
\caption{Encryption throughput on 4k block size.}
\label{tab:crypto-throughput}
\end{revfloat}
\vspace{-1em}
\end{table}

\begin{revblock}[rev1]{G.04}
\autoref{tab:crypto-throughput} shows the encryption throughput of several algorithms at a 4K block size.
We measure each algorithm using both the Linux kernel's crypto implementation and OpenSSL in three environments: on the host, in a VM, and in a CVM.
AEGIS-128 is available only in the kernel, while AEGIS-128\{L, X2, and X4\} are available only in OpenSSL.
Modern AEAD ciphers (i.e., AEGIS-128X4) are $\sim6 \times$ faster than traditional AES-256-GCM, and therefore, \projectname{} builds on AEGIS-128X4.
\end{revblock}

Beyond this, CIF enforcement must be lifted to the abstraction level to exploit per-pattern structure. Co-designing each storage API with the I/O engine reduces cryptographic overhead while preserving full CIF guarantees.
\projectname{} provides composable CIF primitives that enable access-pattern-aware protection, with pattern-specific abstractions layered on top of a hardened I/O engine.
We demonstrate this through three instantiations: a KV store, a shared log, and a filesystem, each composing these primitives into workload-specific CIF enforcement.

\myparagraph{\revnoref{Requirement \#3: Fresh, attested stateful data}}
Confidentiality and integrity alone do not protect against rollback attacks. 
A stale snapshot retains valid authentication tags: reverting an audit log violates non-repudiation~\cite{angelNimbleRollbackProtection2023,chuRollbaccineHerdImmunity2025}, and rolling back model weights can reintroduce known vulnerabilities (e.g., model backdoors).
\revnoref{However, generic} rollback defenses often require replication and synchronization, while host-level freshness mechanisms contradict the CVM threat model.

\begin{table}[t]
\centering
\footnotesize
\setlength{\tabcolsep}{3pt}
\revnoref[rev1]{%
\begin{tabular*}{\columnwidth}{@{\hspace{2pt}\extracolsep{\fill}}lll@{\hspace{6pt}\extracolsep{0pt}}c@{\hspace{2pt}}c@{\hspace{2pt}}c@{\hspace{6pt}}l@{\hspace{2pt}}}
\toprule
Solution & Hardware & Scope & C & I & F & Perf. \\
\midrule
\multicolumn{7}{@{}l}{\emph{Unprotected CVM I/O paths}} \\
Bounce buffer~\cite{bounce_buffer} & CVM & Kernel & \xmark & \xmark & \xmark & Medium \\
SPDK~\cite{yang2017spdk}\lnote{f} & CVM & Userspace & \xmark & \xmark & \xmark & High\textsuperscript{$\dagger$} \\
\addlinespace[0.15em]
\multicolumn{7}{@{}l}{\emph{Block-level protection}} \\
Kernel FDE~\cite{fde,dm-crypt} & CVM & Kernel & \cmark & \cmark & \xmark & Medium \\
MlsDisk~\cite{xuMlsDiskTrustedBlock2026}\lnote{c} & SGX, CVM & Kernel & \cmark & \cmark & \cmark & Low \\
Rollbaccine~\cite{chuRollbaccineHerdImmunity2025}\lnote{c,g} & CVM & Kernel & \cmark & \cmark & \cmark & Medium \\
\addlinespace[0.15em]
\multicolumn{7}{@{}l}{\emph{Shielded runtimes}} \\
Haven~\cite{baumannShieldingApplicationsUntrusted2015} & SGX & Userspace & \cmark & \cmark & \xmark & Low \\
SCONE~\cite{arnautovSCONESecureLinux2016a} & SGX & Userspace & \cmark\lnote{b} & \cmark\lnote{b} & \xmark & Medium \\
rkt-io~\cite{thalheimRktioDirectStack2021}\lnote{f} & SGX & Userspace & \cmark & \xmark & \xmark & High\textsuperscript{$\dagger$} \\
\addlinespace[0.15em]
\multicolumn{7}{@{}l}{\emph{Application-specific secure storage}} \\
ShieldStore~\cite{kimShieldStoreShieldedInmemory2019a}\lnote{a} & SGX & KV store & \cmark & \cmark & \xmark & Medium \\
Speicher~\cite{bailleuSPEICHERSecuringLSMbased2019} & SGX & KV store & \cmark & \cmark & \cmark & Low \\
Pesos~\cite{krahnPesosPolicyEnhanced2018}\lnote{d} & SGX & Objects & \cmark & \cmark & \cmark & Low \\
EnclaveDB~\cite{priebeEnclaveDBSecureDatabase2018}\lnote{a} & SGX & SQL DB & \cmark & \cmark & \cmark & Medium \\
NeXUS~\cite{djokoNeXUSPracticalSecure2019}\lnote{e} & SGX & Files & \cmark & \cmark & \xmark & Low \\
\addlinespace[0.15em]
\multicolumn{7}{@{}l}{\emph{Freshness and state-continuity services}} \\
Memoir~\cite{rollback_parno} & TPM & Module state & \nmark & \nmark & \cmark & Low\\
ROTE~\cite{mateticROTERollbackProtection2017}\lnote{c} & SGX & Counters (repl.) & \nmark & \nmark & \cmark & Medium\\
ADAM-CS~\cite{martinADAMCSAdvancedAsynchronous2021}\lnote{c} & SGX, TPM & Counters (async) & \nmark & \nmark & \cmark & Medium\\
Nimble~\cite{angelNimbleRollbackProtection2023}\lnote{c} & SGX, CVM & Ledgers & \nmark & \nmark & \cmark & Medium\\
\midrule
\projectname{}\lnote{c} & CVM & KV/Log/FS & \cmark & \cmark & \cmark & High\textsuperscript{$\dagger,\ddagger$} \\
\bottomrule
\end{tabular*}}

\vspace{0.25em}
\revnoref[rev1]{\parbox{\columnwidth}{\scriptsize
\lnote{a}\,whole dataset must fit in memory;
\lnote{b}\,read-only files, writes are ephemeral;
\lnote{c}\,freshness needs an external trusted store or replicas;
\lnote{d}\,freshness needs trusted storage hardware;
\lnote{e}\,per-file versions only, hierarchy open to forking attacks;
\lnote{f}\,driver not hardened against a malicious device;
\lnote{g}\,rollback prevention via replication.
\textsuperscript{$\dagger$}\,kernel-bypass userspace I/O path;
\textsuperscript{$\ddagger$}\,on MB-scale objects and larger;
\nmark{}\,out of scope.}}
\vspace{-0.3em}
\caption{\rev[rev1]{G.01}{TEE secure storage approaches, grouped by type.
}}
\label{tab:cvm-storage-comparison}
\vspace{-1.1em}
\end{table}

Precisely, CVMs lack a local hardware freshness primitive, such as SGX monotonic counters.
A prior work \revnoref{augments} CVMs with SGX monotonic counters~\cite{mahhoukMoeMahhoukGraminesealingkeyprovider2025}, but this incurs \qtyrange{60}{250}{\milli\second} latencies and wears out within days~\cite{bailleuSPEICHERSecuringLSMbased2019}.
Counter state must therefore be externalized to a trusted party, incurring a round trip on each freshness update.

Furthermore, external counters alone are insufficient: the trusted party must also verify \emph{which} CVM instance may receive storage keys, initialize counters, and resume persistent state.
Without this binding, an attacker can replay sealed state into a modified instance or bind a valid counter to the wrong identity.
Attestation is therefore not an auxiliary feature but a foundational design requirement: 
\revnoref{storage keys are released only after the requester's identity is verified, and provisioning seals those keys and the counter state to that attested identity.}

\projectname{} externalizes monotonic counter state and binds initialization, recovery, and key provisioning to its attestation flow.
Per-abstraction counter chains embed values in persistent structures for read-time rollback detection.
Epoch-based batching sets the performance/freshness trade-off: 
per-write synchronization \revnoref{bounds the rollback window to one write}.
In contrast, batching lets a crash revert the store only to the last committed epoch.

\myparagraph{Summary and problem statement}
The three requirements highlighted above are interdependent.
Secure high-throughput I/O requires integrating CIF protection directly into the I/O engine.
Workload-tailored abstractions require lifting CIF enforcement to the abstraction level, where per-pattern properties can be exploited.
Fresh, attested state requires embedding counter values in the persistent structures provided by those abstractions.
Addressing any one requirement in isolation leaves the others unsolved.
Unfortunately, existing approaches fail to simultaneously satisfy all three (\autoref{tab:cvm-storage-comparison}): 
\rev[rev1]{G.01}{Existing block-level systems protect every block uniformly and ignore access patterns; shielded runtimes mainly focus on their volatile states and do not provide integrity or freshness to persistent data; application-specific designs bind their guarantees to a single workload and to enclave-only primitives (e.g., SGX monotonic counters); and freshness services add rollback protection at \qtyrange{1}{76}{\milli\second} per state advance but provide no data path.}
This leads to our core research question: \textit{how can we co-design an attested CVM I/O engine and CIF storage abstractions to simultaneously achieve high-throughput secure I/O, access-pattern-aware protection, and freshness?}

\section{Overview}

\begin{figure}[t]
    \centering
    \includegraphics[width=0.9\linewidth,trim=0 0 0 0.35,clip]{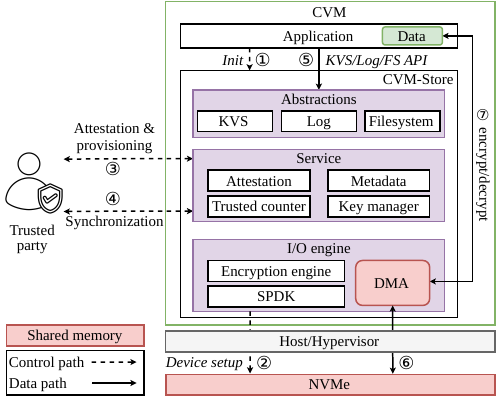}
    \caption{Overview of \projectname{}'s system architecture.}
    \label{fig:overview}
    \vspace{-0.5em}
\end{figure}

\subsection{System Architecture}
\label{subsec:overview:system_arch}

\projectname{} targets CVM deployments where applications demand high-performance storage access, and the underlying devices and storage stack are exposed to an untrusted cloud provider.
\autoref{fig:overview} depicts \projectname{}'s architecture, highlighting its key components, trusted entities, and workflow. 

\projectname{}'s I/O engine provides general CIF building \rev[rev1]{G.09}{blocks: authenticated encryption, trusted counters, and integrity structures.
From these, \projectname{} exposes three access-pattern-aware storage abstractions to support common cloud workloads:
First, the \textit{key-value store} manages large immutable values, read back in full and replaced as a whole, with minimal overhead for CIF protections.
Second, the \textit{shared log} serves append-only workloads, such as audit logs, where entries are continuously appended, periodically truncated, and rarely read except during recovery.
Third, the \textit{filesystem} covers write-once bulk data, spanning input provisioning (i.e., container images, ML model weights, system libraries, signed configuration) and output archives (i.e., model checkpoints, VM disk image snapshots, container layer tarballs).
It supports random reads, trading flexibility for more CIF overhead via a Merkle tree.
}
Unlike generic block-level protections, each abstraction exploits its specific access pattern to minimize cryptographic overhead while preserving full CIF guarantees.

At the core, \projectname{} implements an \textit{I/O engine} built on SPDK Blobstore~\cite{spdk_blobstore} for high-performance userspace storage access.
It integrates an \textit{encryption engine} that applies modern AEAD ciphers directly in the I/O path, optimizing bounce buffer copying with encryption and integrity checks.
Further, the I/O engine hardens SPDK's device-facing boundary against untrusted NVMe controllers (\autoref{fig:storage-stack}).

The \textit{\projectname{} Service} manages all control-path operations.
It handles metadata for storage abstractions, ensures crash consistency via copy-on-write updates, and enforces CIF properties by integrating the I/O engine's cryptographic primitives with an asynchronous \emph{Trusted counter} mechanism.
The trusted counter maintains monotonic values synchronized with an external trusted party to prevent rollback attacks.
Beyond this, the \emph{Service} provides remote attestation and sealing/unsealing primitives for secure provisioning.

\subsection{System Deployment and Workflow}
\label{subsec:overview:deployment}

\projectname{}'s deployment involves the \textit{application user}, and the \textit{trusted parties} responsible for attestation, provisioning, and counter synchronization.
\projectname{}'s cryptographic material includes the user-provided symmetric keys for blob encryption and the CVM-generated sealing key used to bind the provisioned configuration to the attested system identity.

\autoref{fig:overview} illustrates the system workflow.
During \textit{initialization} (\Circled{1}), the CVM boots, and the application selects \projectname{} as its storage backend.
\projectname{}'s I/O engine then performs the \textit{device setup} (\Circled{2}), configuring the NVMe via SPDK and allocating untrusted DMA buffers for data transfers.

Next, the user uses the \projectname{} Service to run the \textit{attestation and provisioning} protocol (\Circled{3}) (\autoref{des:subsec:attestation}).
The Service collects CVM attestation evidence, which the user verifies through an external verifier, and, upon success, the user provisions symmetric encryption keys and \projectname{} configuration sealed to the attested CVM.
The Service unseals and applies this configuration. 
Then, it establishes \textit{trusted counter synchronization} (\Circled{4}) with an external trusted party, retrieving or initializing monotonic counter values and periodically synchronizing them according to \projectname{}'s configuration.

At runtime, applications interact with \projectname{} via its storage abstractions (\Circled{5}, \autoref{tab:design-application-apis}).
For each operation, \projectname{} transparently enforces CIF guarantees.
The \projectname{} Service manages the blobs and metadata, while the I/O engine issues NVMe commands and performs DMA transfers using its untrusted DMA buffers (\Circled{6}).
On writes, \projectname{} encrypts data and adds integrity metadata and trusted counter values before writing it to the DMA buffer, ensuring that only CIF-protected data is written to storage (\Circled{7}).
On reads, \projectname{} decrypts the fetched data and performs CIF verification (\Circled{7}) by validating trusted counter values and cryptographic authentication tags before returning it to the application.

\begin{table}[t]
    \centering
    \footnotesize
    \setlength{\aboverulesep}{0pt}\setlength{\belowrulesep}{0pt}%
    \begin{tabularx}{\linewidth}{p{0.07\textwidth}p{0.13\textwidth}X}
    \toprule
        \headpad\textbf{Use case} & \textbf{API} & \textbf{Description}\\
    \midrule
        \crowfirst{\kvscolor}{get(key)}{Retrieve value of KV pair}{\\}
        \crowreg{\kvscolor}{put(key, value)}{Store/update value for KV pair}{\\}
        \crowreg{\kvscolor}{exists(key)}{Check the existence of a key}{\\}
        \crowlast{\kvscolor}{4}{KVS}{delete(key)}{Delete KV pair}
    \midrule
        \crowfirst{\logcolor}{create(name)}{Create named shared log}{\\}
        \crowreg{\logcolor}{append(data)}{Append data to log end}{\\}
        \crowreg{\logcolor}{truncate(pos)}{Truncate log to position}{\\}
        \crowlast{\logcolor}{4}{Shared log}{read(pos)}{Read data at position}
    \midrule
        \crowfirst{\fscolor}{create(name)}{Create named file}{\\}
        \crowreg{\fscolor}{open(name, op)}{Open file for read/write}{\\}
        \crowreg{\fscolor}{close(fd)}{Close file descriptor}{\\}
        \crowreg{\fscolor}{read(fd, pos, length)}{Read data at position/length}{\\}
        \crowlast{\fscolor}{5}{Filesystem}{write(fd, data)}{Write data to file}
    \midrule
        \crowfirst{\provcolor}{load(blob)}{Load keys \& counter from blob}{\\}
        \crowreg{\provcolor}{store()}{Export keys \& counter as blob}{\\}
        \crowlast{\provcolor}{3}{Provision}{evidence(nonce)}{Get evidence of CVM}
    \bottomrule
    \end{tabularx}
    \vspace{-0.5em}
    \caption{\projectname{} application and provisioning APIs.}
    \label{tab:design-application-apis}
    \vspace{-1em}
\end{table}

\subsection{Programming Model and APIs}
\label{subsec:overview:programming}

\autoref{tab:design-application-apis} lists \projectname{}'s APIs organized by storage abstraction.
\projectname{} transparently provides CIF guarantees to the application \revnoref{through} each abstraction: confidentiality and integrity are guaranteed \revnoref{by} its authenticated encryption scheme, and freshness is ensured \revnoref{by} trusted counter validation.

\revnoref{
The \emph{KVS API} stores and retrieves immutable values by key, with key management backed by persistent storage.
The \emph{Shared Log API} supports append-only workloads with efficient truncation and sequential access.
The \emph{Filesystem API} provides write-once files with sequential and random reads.
The \emph{Provisioning API} generates attestation evidence and seals cryptographic material and system state to the attested CVM identity.}
\rev[rev1]{G.15}{All APIs treat objects as a whole, without in-place mutations.
Application developers should pick the API that best matches their access pattern and adopt it instead of POSIX, embedding the \projectname{} library into the application.}

\begin{figure*}[t]
    \centering
    \includegraphics[width=0.9\linewidth,trim=0 0 0 0.35,clip]{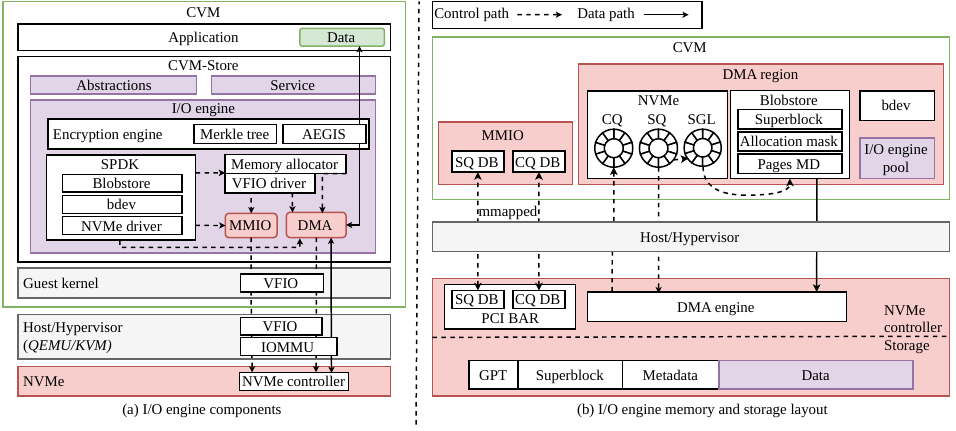}
    \vspace{-0.5em}
    \caption{Inner components of the I/O engine and the view of storage and shared memory region between CVM and host.}
    \label{fig:design:io-engine}
    \vspace{-1.5em}
\end{figure*}

\subsection{\texorpdfstring{\revnoref{Threat Model and Security Goal}}{Threat Model and Security Goal}}
\label{subsec:overview:threat_model}

\myparagraph{\revnoref{Threat model}}
\projectname{} adopts the standard confidential computing threat model~\cite{cheng_tdxsurvey_2023, linux_coco_threat_model, tdx_google_security_review}.
\revnoref{The} TCB comprises the CVM firmware and hardware, and the software running in the CVM. 
The rest of the system software and infrastructure, including the hypervisor, host OS, and storage devices, is untrusted. 
We consider attackers who have full control over the host environment and can inspect, modify, or replay any data outside the CVM boundaries, including storage, DMA buffers, and I/O traffic. 
We assume that the confidential computing hardware functions correctly and that cryptographic primitives are secure.
We \revnoref{further} assume all CVM components can be attested and that \projectname{} exclusively uses attestation primitives to prevent multi-protocol attacks~\cite{segallUsingTPMMachine2012}.
\revnoref{External services used for attestation and counter synchronization are likewise assumed to be trusted and correctly implemented.}
Side-channel~\cite{alam_cc-side-channel_2023,intel_secure_coding_guideline,wichelmann_cipherfix_2023,sev-cipher-sidechannel-mitigation,deng_cipherh_2023,firmware-sidechannel-mitigation,wang_pwrleak_2023}, denial-of-service, and physical attacks, including physical RAM attacks~\cite{tee-fail,batteringramsp26}, are considered out of scope.

\rev[rev1]{G.11}{%
\myparagraph{Security goal}
Under this threat model, \projectname{} aims to preserve the confidentiality and integrity of all data written through its interfaces, even across system restarts and crashes.
Additionally, \projectname{} offers two guarantees: \textit{(i)} \emph{freshness}: \projectname{} never exposes stale or rolled-back state, and \textit{(ii)} \emph{durability}: unsynchronized operations may be lost after an adversarial restart, but never reappear as stale data.
\projectname{} therefore admits \emph{data loss} only up to the last synchronization point, whose frequency is a configurable trade-off between consistency and performance.
\projectname{} does not provide metadata (e.g., file size, I/O pattern) confidentiality, but ensures its integrity and freshness.
}

\section{\projectname{} Design}
\label{sec:design}

\projectname{} consists of five building blocks: the I/O engine (\autoref{subsec:io-engine}), 
the trusted counter (\autoref{subsec:trusted_counter}), 
the \projectname{} service (\autoref{subsec:service}), 
and the attestation (\autoref{des:subsec:attestation}) and sealing (\autoref{des:subsec:sealing}) mechanisms, 
which form the basis of the CIF-protected abstractions (\autoref{sec:abstractions}).
\revnoref{The implementation details are presented in \autoref{sec:implementation}.}

\subsection{I/O Engine}
\label{subsec:io-engine}
\projectname{}'s I/O engine (\autoref{fig:design:io-engine}~(a)) builds on SPDK~\cite{yang2017spdk} and  provides high-performance userspace NVMe access from within the CVM.
It ensures confidentiality and integrity through authenticated data encryption.
On writes, \projectname transparently handles the encryption from application memory to an unprotected shared memory ring buffer before NVMe DMA, and follows the same process in reverse on reads.

\myparagraph{SPDK Blobstore}
SPDK Blobstore~\cite{spdk_blobstore} is a userspace, power-fail-safe data allocator designed for storage services without implementing a POSIX filesystem.
It organizes NVMe device space into a four-level hierarchy (logical blocks, pages, clusters of \qty{1}{\mebi\byte} by default, and blobs), where a blob is an ordered list of clusters forming a variable-sized, contiguous device allocation.
Blobstore manages all logical-to-physical block mappings and provides allocation, read, and write primitives via the SPDK NVMe driver.
\projectname{} adopts blobs as its fundamental storage unit, delegating physical block management to Blobstore and focusing its own logic on CIF-protected blob layout.

On the control path, the I/O engine uses Blobstore to manage storage operations, and combines it with confidentiality and integrity checks.
It organizes each blob on disk into three sections: unprotected metadata (nonces, MACs, file size), optional encrypted metadata, and encrypted data chunks.

\myparagraph{Encryption and integrity protection}
\begin{revblock}[rev1]{G.04}
\projectname{}'s I/O engine interleaves encryption and integrity protection in the data path.
Combining encryption with data movement between DMA-shared memory and application memory eliminates the need for additional copies.
On writes, the I/O engine submits a ring-buffer slot only after it is filled with ciphertext, then encrypts later chunks while earlier DMA transfers are in flight.
On reads, it posts multiple DMA reads into ring-buffer slots and decrypts completed slots in order, while further reads are in flight.

Its crypto subsystem provides \textit{authenticated encryption} for confidentiality and integrity, and \textit{Merkle tree} verification for tamper detection.
It is built to be cipher-agnostic and by default employs the modern AEGIS-128X4 cipher~\cite{wuAEGISFastAuthenticated2013, denisAddingMoreParallelism2023, AegisaeadOpenssl2026}, which is nearly 6$\times$ faster than the common AES-256-GCM (further details in \autoref{tab:crypto-throughput}).
For the Merkle tree, \projectname{} uses MACs as leaves rather than separate hashing, removing an additional cryptographic operation~\cite{burkeScalableIntegrityChecking2025}.
Caching validated Merkle tree nodes further reduces verification overhead.
By interleaving CIF protections, using modern ciphers, and eliminating copies in the I/O engine, \projectname{} saturates a modern NVMe device on a single core for large data-intensive workloads.
\end{revblock}

\myparagraph{Hardening}
\rev[rev1]{G.04}{In CVMs, the NVMe controller is untrusted and may read DMA buffers, corrupt control structures, or forge completion entries.
Therefore, \projectname{} must harden SPDK itself, which by default assumes a trusted controller and co-locates driver state with device-accessible memory.
We reviewed the SPDK codebase for its interactions with external devices and identify the following issues. 
}
Each \texttt{nvme\_tracker} embeds the PRP list, SGL descriptors, and metadata SGL in DMA memory, exposing request pointers, callback pointers and arguments, and queue links to a malicious host or NVMe controller.
\projectname{} hardens this by splitting the \texttt{nvme\_trackers} into private driver state and a separate per-tracker DMA buffer containing only the NVMe-required PRP/SGL descriptors.
The device receives the physical address of the descriptor buffer, while the tracker array remains outside DMA-accessible memory.
\projectname{} \revnoref{further verifies} that the controller-supplied completion identifier (CID) falls within the allocated tracker range before using it to index the array.
Invalid CIDs fail the queue, preventing a forged completion from turning an untrusted DMA field into an out-of-bounds access to SPDK's internal tracker state.

\myparagraph{Device passthrough}
\rev[rev1]{G.13}{
\projectname{} uses a pass-through NVMe device to achieve near-raw bandwidth (this assigns a device to a single guest and SR-IOV can relax this restriction~\cite{pci-sigPCIExpressBase2024}).
For this, the I/O engine employs VFIO~\cite{vfio} to bind the NVMe to \projectname{}'s CVM.}
\revnoref{CVM memory protection alone ensures that devices cannot write to arbitrary guest memory without IOMMU.}
Therefore, \projectname{} \revnoref{securely uses} VFIO without an IOMMU in the CVM.
The VFIO PCI BAR space is mapped directly into guest memory (\autoref{fig:design:io-engine}~(b)), which the guest then maps to userspace.
This direct mapping eliminates the need for expensive MMIO hypercalls for device access.

\subsection{Trusted Counter}
\label{subsec:trusted_counter}
\begin{revblock}[rev1]{G.05, G.11}
Persistent storage remains vulnerable to rollback attacks~\cite{rollback_parno,chuRollbaccineHerdImmunity2025,angelNimbleRollbackProtection2023,mateticROTERollbackProtection2017}.
As CVMs lack monotonic counters, they require external synchronization with a trusted party for freshness (e.g., a local SGX sidecar~\cite{mahhoukMoeMahhoukGraminesealingkeyprovider2025} or a remote service).
\projectname{} treats this trusted party as a pluggable service, implemented as a library exposing a trusted counter interface.
It distinguishes \emph{durability}, persisting metadata to disk, from freshness, synchronizing the covering epoch with the trusted party.

\projectname{} groups operations into \emph{epochs}, the granularity at which it maintains durability and freshness.
An uncommitted epoch defines the vulnerability window, configured by the \emph{epoch length} and \emph{synchronization mode}: the host can revert to the last \textit{committed} one.
In \emph{blocking} mode, further mutations are held until the in-flight epoch is durable and committed, so a crash loses at most one epoch; the \emph{non-blocking} mode lets them proceed while it synchronizes, trading the risk of losing up to two epochs for lower latency.
\ifextended Appendix~\ref{sec:appendix:counter}\else\expandafter\MakeUppercase\extversion\fi{} details the counter interface, the required service properties, and the durability and freshness modes.
\end{revblock}

\subsection{\projectname{} Services}
\label{subsec:service}
The \projectname{} Service manages all control-path operations (i.e., metadata, crash consistency, CIF enforcement), handles the data lifecycle, and reclaims unused resources.

\myparagraph{Metadata management}
The \projectname{} Service maintains Blobstore metadata for low-level storage management and application-specific metadata for the storage abstractions.
SPDK's Blobstore handles blob allocation and its metadata on the physical disk.  
Application-specific metadata is persisted via the I/O engine with explicit synchronization control to balance consistency with performance.
To ensure crash consistency, the Service employs \revnoref{a copy-on-write,}
writing modified metadata to new locations before swapping between them, ensuring atomic updates.
Cleanup of stale blobs is deferred till a successful metadata synchronization to prevent inconsistencies after crashes.
\projectname{} omits journaling-based recovery to maintain simplicity and minimize overhead.

\begin{figure}[t]
    \centering
    \begin{revfloat}[rev1]{G.12}
    \includegraphics[trim=0 0 0 1.45,clip]{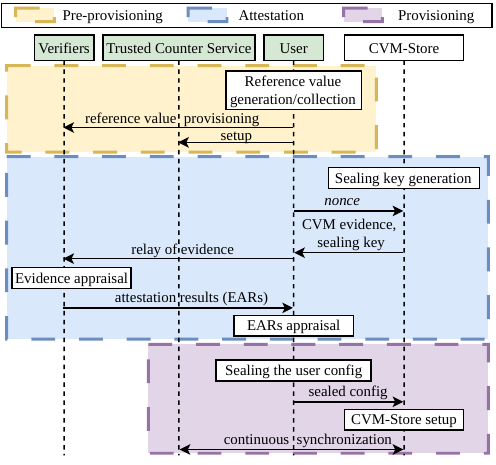}
    \vspace{-0.6em}
    \caption{\revnoref{Attestation and provisioning flow in \projectname{}.}}
    \label{fig:design:attestation}
    \end{revfloat}
    \vspace{-1em}
\end{figure}

\begin{figure*}[t]
    \centering
    \includegraphics[width=0.9\linewidth,trim=0 0 0 0.35,clip]{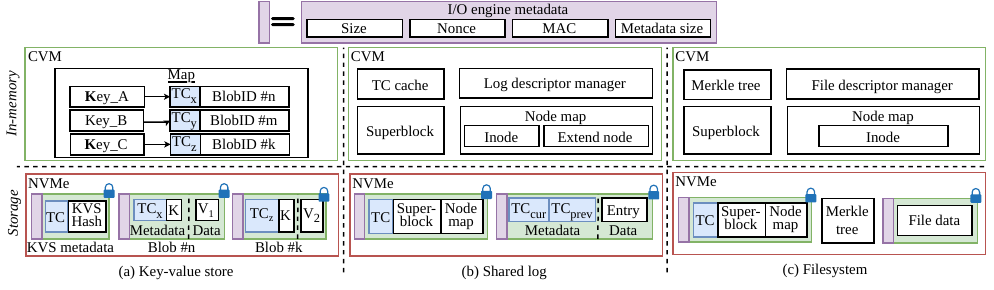}
    \vspace{-0.5em}
    \caption{\projectname{}'s storage interface in-memory metadata and persistent data structure organization.}
    \label{fig:accelstore-interfaces}
    \vspace{-0.5em}
\end{figure*}

\myparagraph{CIF enforcement}
The \projectname{} Service ensures the CIF properties through two mechanisms: the I/O engine's cryptographic primitives enforce confidentiality and integrity, while the trusted counter mechanism ensures freshness.
Trusted counters are embedded in persistent data, binding each storage operation to a counter state, ensuring no stale data is replayed after crashes or restarts.
Counter increments are batched with storage operations to minimize synchronization overhead.
At startup, the Service validates all CIF properties, verifying counter values against the external trusted party and authenticating all metadata structures, before granting application access.

\myparagraph{Attestation and provisioning service}
The \projectname{} Service serves as the central authority for attestation evidence collection and provisioning, coordinating with external verifiers and trusted counter services.
It implements sealing/unsealing primitives that enable the secure provisioning of secrets and configuration, allowing \projectname{} to bind persistent state to CVM instances via attestation-derived sealing keys.

\subsection{Attestation and Provisioning}
\label{des:subsec:attestation}
\projectname{} demonstrates a complete attestation and provisioning pipeline \revnoref{(\autoref{fig:design:attestation})}.
Before deployment, the user collects platform endorsements and reference values~\cite{amdVersionedChipEndorsement2025}, provides them to a verifier, and sets up the trusted counter service.

\myparagraph{Attestation} 
\revnoref{On CVM startup, \projectname{} generates a sealing key pair, and the Service embeds a user-supplied nonce and the public key in the CVM attestation report.
The user has the evidence appraised by the verifier (an EAT Attestation Result~\cite{fossatiEATAttestationResults2025}) and checks that it names the expected public sealing key.}

\myparagraph{Provisioning}
\rev[rev1]{G.12}{The user seals the \projectname{} configuration to the sealing key; the Service unseals it, initializes \projectname{}, and connects to the trusted counter service.
The configuration carries the symmetric on-disk key and the freshness-service parameters.
The sealing key is per boot, so a rebooted or migrated instance is re-attested and re-provisioned by the user, and the counter service rejects cloned or restored copies\ifextended{} (\autoref{fig:appendix:lifecycle}, Appendix~\ref{app:lifecycle})\fi.}

\subsection{Data Sealing and Unsealing} 
\label{des:subsec:sealing}
\projectname{} provides four operations for secure data management.
User-to-\projectname{} sealing encrypts data with CVM-bound keys, ensuring only the attested CVM instance can access provisioned material.
\projectname{}-to-user sealing encrypts internal state using the configuration-provisioned symmetric key, enabling the user to export and later restore system state.
The corresponding unseal operations perform decryption and system setup or allow data processing as opaque blobs.
For data provisioning, the trusted party runs a local \projectname{} with the target symmetric key and simulated counters. This allows the data to be written without any trusted counter round trips. 
The resulting counter value is then provisioned into the freshness service so that the receiving CVM can validate the data freshness and consistency.

\section{\projectname{} Abstractions}
\label{sec:abstractions}

To support diverse cloud workloads with high-performance I/O while remaining easy for applications to adopt, \projectname{} provides three storage abstractions: KVS, shared log, and filesystem.
\rev[rev1]{G.04}{
All three build on \projectname{}'s I/O engine (\autoref{subsec:io-engine}) and Service (\autoref{subsec:service}), and share copy-on-write and deferred cleanup policies for crash consistency and counter-based rollback protection.
However, rather than applying uniform block-level protection, each abstraction co-designs its integrity metadata with its access pattern, enabling workload-specific CIF enforcement and overlap of encryption with I/O.
}

\subsection{Key-Value Store (KVS)}
\label{subsec:abstractions:kvs}
\projectname{}'s KVS (\autoref{fig:accelstore-interfaces}~(a)) stores one KV pair per blob and models updates as object replacement, targeting large values that are read sequentially.

\myparagraph{Data organization}
\projectname{}'s KVS maintains a key map in memory along with CIF-protected on-disk blobs.
The map stores keys, trusted counter values, and blob identifiers, enabling existence checks and freshness validation without keeping values in memory.
This design trades memory proportional to the number of keys for fast lookup and rollback checks, targeting workloads with MB-to-GB objects rather than high-cardinality small records.
\rev[rev1]{G.11}{The metadata blob stores a multiset hash~\cite{clarkeIncrementalMultisetHash2003} over all key-counter pairs and the latest trusted counter value.
}
Each KV blob has an encrypted metadata section (trusted counter value, key), and a data section for the value.

\myparagraph{Crash consistency}
\rev[rev1]{G.11}{
\projectname{} stores the KVS hash with the trusted counter value in the encrypted KVS metadata blob, updated atomically. 
Blobs are removed only after synchronization, preventing dangling references.
}
On startup, \projectname{} reads the metadata blob to retrieve the trusted counter value and hash, then reconstructs the key mapping by walking all on-disk blobs, discarding those with counters exceeding the metadata counter and selecting the higher valid counter for duplicate keys.
Integrity is verified by recomputing the hash over all reconstructed key-counter pairs and comparing it against the stored value.

\myparagraph{Operations}
The \emph{Put} operation inserts or updates a KV pair.
\projectname{} allocates a blob, obtains and increments the trusted counter, encrypts the metadata (trusted counter, key) and data section (value), writes them to the blob, and places the metadata in the in-memory map.
Then, if requested, it atomically synchronizes the metadata blob.

The \emph{Get} operation retrieves a value given a key.
\projectname{} looks up the key in the in-memory map to obtain the blob identifier and expected trusted counter, reads and decrypts the blob, \revnoref{and validates the counter and the key in} the encrypted metadata against the looked-up key \revnoref{before returning the value.}

\revnoref{For the \emph{Delete} operation,} \projectname{} looks up the key in the map, marks the blob as deleted, removes the key from the in-memory map, and defers blob deallocation to the next metadata synchronization.

\subsection{Shared Log}
\label{subsec:abstractions:log}

\projectname{}'s shared log (\autoref{fig:accelstore-interfaces}~(b)) is suitable for append-heavy workloads, 
\rev[rev1]{G.09}{such as audit logs, where entries are appended, truncated, and rarely read outside recovery.}

\myparagraph{Data organization}
\revnoref{A \emph{super-block} contains a map of \textit{inodes} and \textit{extend nodes}.}
The \textit{inode} stores the log name, entry count, flags, first/last-entry counters
and direct references to the first and last extend nodes to avoid full list traversal on append and retrieval.
The \textit{extend nodes} form a doubly linked list of fixed-size \revnoref{segment-range arrays}.
Each \textit{log entry blob} stores \revnoref{encrypted metadata} (the current and previous entry's counters) and \revnoref{encrypted log data}; the MAC covers both, preventing tampering with the counter chain or the entry data.
\projectname{} caches expected counter values in memory to validate entries without full log traversal.

\myparagraph{Crash consistency}
\projectname{} updates the super metadata block atomically\rev[rev1]{G.11}{ by rewriting only dirty nodes into the cold blob before swapping it with the hot one.
Embedding the trusted counter value into the metadata guarantees freshness.}
\revnoref{On first read/open of a log}, it traverses all entries and builds the counter cache, verifying for each entry that the stored previous counter matches the predecessor's counter value, with the authenticated encryption MAC protecting the binding between counters and entry data.
Finally, \projectname{} confirms that the inode's recorded first and last counter values match those encountered during traversal.

\myparagraph{Operations}
The \emph{Append} operation adds new data at the end of the log.
\projectname{} \revnoref{first} marks the inode with an \emph{append-in-progress} flag to prevent concurrent modifications\revnoref{, then allocates a new blob and} obtains the last log entry's counter and the current counter from the \projectname{} service.
\revnoref{If the blob extends the last node's last range or fits a free slot, it is recorded there; otherwise, \projectname{} allocates a new extend node, links it from the inode and the previous node, and records the blob as its first range.}
\revnoref{It then} encrypts the metadata and data sections, and updates the trusted counter cache before writing to the blob.
The inode metadata is updated to reflect the new entry, including incrementing the entry count and updating the last entry's trusted counter value.
Finally, the append-in-progress flag is cleared, and the trusted counter is incremented.

The \emph{Retrieve} operation reads a log entry by an entry number.
\projectname{} uses the log descriptor to identify the inode, returns an error
if an append is in progress, and \revnoref{checks the requested number against the inode's entry count.}
\revnoref{It then traverses extend nodes backward} from the inode's last extend node reference to locate the \revnoref{target,} retrieves the blob identifier, reads and decrypts the blob, and \revnoref{validates the counter in the blob's metadata against the trusted counter cache.}
For the first and last entries, \projectname{} additionally validates counters
against the inode's boundary values. 
Only after CIF verification is the decrypted entry returned to the application.

The \emph{Truncate} operation removes old log entries.
An application specifies how many entries to retain\revnoref{; the oldest entries are removed first} to maintain ordering.
\projectname{} traverses the extend node linked list from the beginning, marking the blobs for deletion and \revnoref{any node whose reference count is zero} for deallocation.
\revnoref{To keep} subsequent consistency and freshness checks, \revnoref{it} updates the entry count and \revnoref{the first entry's counter} in the inode to reflect the truncation.
\revnoref{It also} updates the cached trusted counter mapping to remove \revnoref{metadata for} truncated entries.
Under \projectname{}'s deferred cleanup policy, marked blobs are \revnoref{freed only after metadata synchronization.}

\subsection{Filesystem}
\label{subsec:abstractions:fs}
\projectname{}'s filesystem \revnoref{abstraction} (\autoref{fig:accelstore-interfaces}~(c)) exposes file-oriented APIs (\autoref{tab:design-application-apis}) for write-once, read-many workloads \revnoref{such as input provisioning (e.g., container images) and output archives (e.g., snapshotting)}.
\projectname{} therefore \revnoref{adopts the design choice of treating} files as immutable, name-addressed objects in a flat namespace\revnoref{, computing a \textit{static} Merkle tree once at write time instead of the dynamic authenticated structures required by mutable data}.

\myparagraph{Data organization}
\projectname{}'s filesystem reuses the shared log's inode and node map, \revnoref{without extend nodes}, but adapts the integrity mechanism to support Merkle tree verification for random access to file chunks.
Each file consists of an encrypted blob for chunked file data and an \textit{unencrypted} blob for its Merkle tree, computed over all \revnoref{chunks in} the file's data blob.
The inode contains the Merkle tree root hash and direct references to the encrypted file data and the Merkle tree blob, enabling efficient integrity checks without exposing file contents. 
For newly created files, the inode is initialized with a file size of zero and no blob references.

\myparagraph{Crash consistency}
\rev[rev1]{G.11}{The superblock is updated with the same mechanism as the shared log. If a blob is no longer referenced in the metadata on disk, then it gets reclaimed (\autoref{subsec:service}).}

\myparagraph{Operations}
The \emph{Write} operation stores data to an \textit{empty} file.
\projectname{} marks the inode as \emph{write-in-progress} to prevent concurrent accesses\revnoref{, and chunks} the data according to a configurable policy.
\revnoref{It} encrypts each chunk using \revnoref{a nonce} derived from a global file nonce and the chunk number.
Because files are write-once, the global file nonce is generated once at creation and never reused, guaranteeing nonce uniqueness across \revnoref{all file chunks.}
As chunks are written, \projectname{} incrementally builds the Merkle tree using MAC-based leaf hashes and stores the tree in a separate blob.
The root hash and Merkle tree blob identifier are stored in the inode.
Upon completion, the file becomes read-only, and the inode is marked as available.

The \emph{Read} operation retrieves data from a file at a specified position and length.
\projectname{} uses the file descriptor to retrieve the inode, checks that no write is in progress, and verifies the requested range against the inode's stored size.
On the first read, \revnoref{it} loads and caches the Merkle tree from disk.
It then identifies the chunks for the requested range, reads and decrypts them directly into an application-provided destination buffer, and verifies them against the cached Merkle tree, skipping already-verified nodes.
Only after successful CIF verification does \projectname{} return control to the application.

\section{Security Analysis}
\label{sec:security}
We analyze \projectname{} against attack vectors from an untrusted host under our considered threat model (\autoref{subsec:overview:threat_model}).
\autoref{tab:security:attack-vectors} summarizes these attacks and \projectname{}'s mitigations.

\begin{table}[t]
    \centering
    \footnotesize
    \setlength{\aboverulesep}{0pt}\setlength{\belowrulesep}{0pt}%
    \begin{tabular}{p{0.11\linewidth}p{0.40\linewidth}p{0.36\linewidth}}
    \toprule
    \headpad\textbf{Category} & \textbf{Attack Vector} & \textbf{Mitigation} \\
    \midrule
        \cellcolor{\provcolor}\bandtop & \cellcolor{\provcolor}Loading compromised instance & \cellcolor{\provcolor}Attestation \\
        \cellcolor{\provcolor} & \cellcolor{\provcolor}CVM memory access & \cellcolor{\provcolor}CVM protection \\
        \cellcolor{\provcolor} & \cellcolor{\provcolor}Storage data snooping & \cellcolor{\provcolor}AEAD encryption \\
        \cellcolor{\provcolor} & \cellcolor{\provcolor}Storage data tampering & \cellcolor{\provcolor}AEAD authentication \\
        \cellcolor{\provcolor} & \cellcolor{\provcolor}Device descriptor manipulation & \cellcolor{\provcolor}Bounds checking on DMA \\
        \cellcolor{\provcolor} & \cellcolor{\provcolor}Rollback of storage state & \cellcolor{\provcolor}Trusted counter \\
        \cellcolor{\provcolor} & \cellcolor{\provcolor}\multirow{2}{*}{Cross-instance key reuse} & \cellcolor{\provcolor}\revnoref{Attestation-bound key provisioning} \\
        \cellcolor{\provcolor}\multirow{-9}{*}{General} & \cellcolor{\provcolor}\bandbot\multirow{2}{*}{State inconsistency exploitation} & \cellcolor{\provcolor}Copy-on-write \& CIF revalidation on startup \\
    \midrule
        \cellcolor{\kvscolor}\bandtop & \cellcolor{\kvscolor}\multirow{1}{*}{KV value rollback} & \cellcolor{\kvscolor}\revnoref{Per-blob counter checks} \\
        \cellcolor{\kvscolor}\multirow{-2}{*}{KVS} & \cellcolor{\kvscolor}\bandbot\multirow{1}{*}{KV unauthorized deletion} & \cellcolor{\kvscolor}\revnoref{Hash over key--counter pairs} \\
    \midrule
        \cellcolor{\logcolor}\bandtop & \cellcolor{\logcolor}Log entry splicing & \cellcolor{\logcolor}Counter-chained MACs \\
        \cellcolor{\logcolor} & \cellcolor{\logcolor}Log truncation (front/back) & \cellcolor{\logcolor}Boundary counter validation \\
        \cellcolor{\logcolor}\multirow{-3}{*}{Log} & \cellcolor{\logcolor}\bandbot Log entry reordering & \cellcolor{\logcolor}Monotone counter chain \\
    \midrule
        \cellcolor{\fscolor}\bandtop & \cellcolor{\fscolor}Chunk/block replacement & \cellcolor{\fscolor}Merkle tree leaf MAC \\
        \cellcolor{\fscolor}\multirow{-1}{*}{FS} & \cellcolor{\fscolor}\bandbot\multirow{2}{*}{File replacement} & \cellcolor{\fscolor}Root hash in trusted-counter-protected metadata \\
    \bottomrule
    \end{tabular}
    \vspace{-0.5em}
    \caption{Attack vectors and \projectname{}'s mitigations.}
    \label{tab:security:attack-vectors}
    \vspace{-1em}
\end{table}

\subsection{General Attack Vectors}
\label{subsec:security:general}

\myparagraph{Loading compromised instance}
An attacker may try to bypass \projectname{} by loading a compromised instance; attestation detects this through invalid measurements.

\myparagraph{CVM memory access}
CVM protection blocks unauthorized reads or writes to private memory.

\myparagraph{Storage data snooping}
An attacker may observe storage or DMA-shared memory.
The I/O engine encrypts data inside the CVM before it enters DMA-shared memory. Thus, \revnoref{DMA buffers and the storage device see only ciphertext or control structures} (\autoref{subsec:io-engine}).
Keys are provisioned through attestation and never leave the protected CVM memory (\autoref{des:subsec:attestation}).

\myparagraph{Storage data tampering}
An attacker may alter an encrypted \projectname{} blob while preserving the storage layout.
Each blob written through the I/O engine carries an AEAD authentication tag (\autoref{subsec:io-engine}), allowing \projectname{} to detect tampering  on a subsequent read operation.
For KVS and Shared Log, ciphertext, metadata, or tag changes lead to a failed decryption; filesystem cross-chunk replacement is handled by Merkle-tree checks, as explained in \autoref{subsec:security:fs}.

\myparagraph{Manipulating device descriptors}
An attacker may manipulate device descriptors in the shared memory, injecting adversarial metadata (e.g., invalid descriptor offset)~\cite{hetzeltAnalyzingDeviceInterfaces2021}.
\projectname{} mitigates this by limiting DMA-shared state and performing bounds-checking when reading DMA-received fields (\autoref{subsec:io-engine}).

\myparagraph{Rollback of storage state}
An attacker may replay an older disk state with valid authentication tags.
\projectname{} mitigates this by embedding trusted counters in persisted data and synchronizing them with an external trusted party (\autoref{subsec:trusted_counter}).
These values are validated before returning data (\autoref{subsec:service}).
The asynchronous synchronization interval bounds freshness, as discussed in \autoref{sec:security:limitations}.

\myparagraph{Cross-instance key reuse}
An attacker may replay sealed configuration or storage keys into another CVM.
\projectname{} binds provisioned keys and configuration to the public sealing key in attested evidence in such a way that a different or compromised CVM cannot unseal them (\autoref{des:subsec:attestation}).

\myparagraph{State inconsistency exploitation}
An attacker may crash a CVM to induce \revnoref{an} inconsistent \projectname{} state.
\projectname{}'s copy-on-write metadata updates, deferred cleanup, and restart revalidation ensure each synchronized state is a valid snapshot, effectively eliminating inconsistent state (\autoref{subsec:service}).

\subsection{Attacks on KVS}
\label{subsec:security:kvs}

\projectname{} \revnoref{persists all key--counter mappings on disk via a multiset hash} in the CIF-protected metadata blob (\autoref{subsec:abstractions:kvs}), \revnoref{enabling the detection of the following attacks.}

\myparagraph{\revnoref{Value rollback}}
An attacker may replace a value with an older, valid blob for the same key.
On every \textsc{Get}, \projectname{} compares the blob's key and counter with the in-memory key map \revnoref{to prevent} stale valid blobs from replacing current values (\autoref{subsec:abstractions:kvs}).

\myparagraph{KV unauthorized deletion}
An attacker may remove a valid blob while leaving others intact.
\revnoref{Startup reconstruction is then a different set of key--counter pairs, so the hash set recomputed over them no longer matches the stored one, catching the deletion before exposing the key space.}
Legitimate deletions atomically update the metadata hash and defer physical deallocation until the next synchronization.

\subsection{Attacks on Shared Log}
\label{subsec:security:log}

The Shared Log's potential attack surface consists of rearrangements that leave individual entries valid.
\revnoref{Its security invariant is an AEAD-protected counter chain, along with inode boundary counters for the first and last entries. Any rearrangement becomes visible during reads or at log-open validation} (\autoref{subsec:abstractions:log}).

\myparagraph{Log entry splicing}
An attacker may replace an entry in the middle of the log with another valid entry.
However, \projectname{} maintains the trusted counter so that its values are unique across all log entries, including those from different logs.
Successfully replacing an entry in the middle of the log would require mutually consistent counter values and MACs; specifically, the replacement must match its own counter, its predecessor counter, and the successor's stored predecessor counter.
Without access to the encryption key, the adversary cannot forge such an entry or modify neighboring entries accordingly.

\myparagraph{Log truncation (front/back)}
An attacker may remove valid entries from either log end.
\revnoref{The log inode stores boundary counters and entry count, validated at log open and boundary-entry retrieval. Therefore, a truncation that changes boundary counters or entry count would fail validation (\autoref{subsec:abstractions:log}).}

\myparagraph{Log entry reordering}
An attacker may swap valid entries.
The monotonic counter chain commits each entry to its predecessor's counter, so any swap breaks at least one predecessor relationship and exposes tampering to \projectname{}.

\subsection{Attacks on Filesystem}
\label{subsec:security:fs}

The Filesystem attack surface consists of block-level and whole-file replacement.
Our filesystem invariant is that immutable file chunks are verified against a static Merkle tree whose root is stored in CIF-protected filesystem metadata (\autoref{subsec:abstractions:fs}).

\myparagraph{Chunk/block replacement}
An attacker may replace, reorder, or replay individual chunks within a file.
At write time, \projectname{} commits the encrypted chunks to the file's Merkle root; on read, it releases plaintext only after the requested chunks get verified against this root (\autoref{subsec:abstractions:fs}).
These operations ensure that replacing, reordering, or replaying a chunk invalidates the corresponding leaf and propagates to a root mismatch before plaintext is returned.

\myparagraph{File replacement}
Although an old data blob and a matching Merkle-tree blob can verify internally when replayed together, \projectname{} detects the file replacement as read operations check chunks against the current file root hash recorded in trusted-counter-protected filesystem metadata.
Replaying the metadata, including the superblock state, is rejected during startup because its trusted counter value is detected as stale.

\begin{figure*}[t]
    
    \begin{revfloat}[rev1]{G.06}
    \begin{center}
    \includegraphics[]{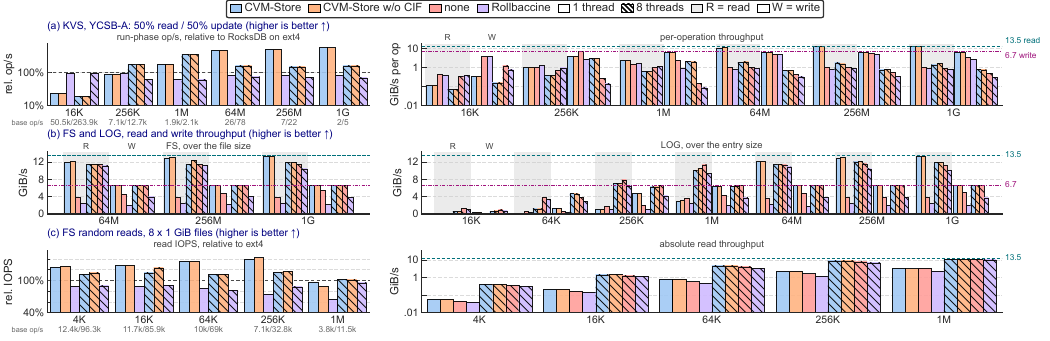}
    \end{center}
    \vspace{-1.5em}
    \caption{\revnoref[rev1]{Comparison of \projectname{}'s I/O characteristics.}}
    \label{fig:evaluation:io}
    \end{revfloat}
    \vspace{-2em}
\end{figure*}

\subsection{Limitations}
\label{sec:security:limitations}

\myparagraph{Metadata exposure}
\revnoref{As} \projectname{} encrypts at the abstraction rather than the block level~\cite{zadokCryptfsStackableVnode1998,grossAnalyzingAndroidsFileBased2019,brozAuthenticatedResilientDisk2018}, metadata such as file sizes, blob counts, and I/O access patterns remain visible to the untrusted host.
\begin{revblock}[rev1]{G.14}
Integrating oblivious execution~\cite{ahmadOBLIVIATEDataOblivious2018} could hide the access pattern with a performance overhead trade-off.
\end{revblock}

\myparagraph{Trusted counter service dependency}
\projectname{} relies on an external trusted counter service for freshness.
\revnoref{A configuration-controlled synchronization interval trades synchronization overhead for a larger rollback window:
data} written since the last synchronization point may be lost after an adversarial restart.

\myparagraph{Side-channel / physical attacks}
Side-channel and physical attacks are \rev[rev1]{G.14}{out of \projectname{}'s threat model (\autoref{subsec:overview:threat_model}).
\projectname{} relies on guest-side hardening and defenses~\cite{IntelTrustDomain}, and physical data center security~\cite{MoreInformationEncrypted}.
}

\myparagraph{Denial-of-service}
\rev[rev1]{G.14}{
Similar to the CVM threat model~\cite{amdAMDSEVSNPStrengthening2020}}, 
\projectname{} does not prevent an adversarial host, hypervisor, or storage device from dropping requests, delaying completions, exhausting resources, or refusing to provide the NVMe device.

\section{Implementation}
\label{sec:implementation}

We implement \projectname{}{} targeting AMD SEV-SNP, but our design is vendor-agnostic and applicable to other CVMs such as Intel TDX and Arm CCA.
\projectname{} is implemented as a userspace library with SPDK, totaling \revnoref[rev1]{$\approx$20,000 lines of code (LoC) for the core. Further, it provides} example clients, modifications to the Linux guest kernel, and extended SPDK~\cite{yang2017spdk}/DPDK~\cite{dpdk} components to support encrypted memory management, inspired by prior work~\cite{lamBoostingNetworkPerformance2024, yangFastSecureDPDK2023, liBifrostAnalysisOptimization2023, liBridgeFutureHighPerformance2024,thalheimRktioDirectStack2021}.

\myparagraph{Userspace I/O handling}
We re-enable userspace MMIO handling in SEV-SNP CVMs, which is disabled to prevent hypervisor coordination attacks~\cite{CVE202346813}, which attestation of the entire usersapce mitigates.
To enable userspace DMA memory allocation, we extend mmap with the \textit{VM\_CVM\_SHARED} flag, which decrypts pages on fault for DMA access and re-encrypts them on unmapping.

\myparagraph{DPDK integration}
SPDK utilizes DPDK~\cite{dpdk} for environment abstraction and memory management.
Since CVMs protect memory by default, DPDK's allocator must distinguish between regular and DMA-capable memory.
We modify the Environment Abstraction Layer (EAL) memory allocation APIs to propagate memory type and maintain separate \textit{memzones}, \textit{memseg} lists, and heaps per type ($\approx$1,500 LoC), and export APIs to manage distinct pools for encrypted and decrypted memory.
This approach reuses existing DPDK allocation infrastructure while enabling memory-type awareness, avoiding the alternative of external allocation which would forgo integration with DPDK's allocation logic.

\myparagraph{SPDK integration}
SPDK abstracts device handling and memory management via an environment layer.
In its abstraction layer, SPDK distinguishes between normal and DMA memory allocations, requiring changes in fewer than 50 LoC to integrate with \projectname{}'s encrypted memory management.
During development, we identified upstream SPDK code that did not properly use the allocation APIs and upstreamed fixes.

\myparagraph{\projectname{} library}
The I/O engine uses SPDK's Blobstore~\cite{spdk_blobstore} on top of NVMe \textit{bdev}, getting its DMA buffers through its own \texttt{iobuf} modules using SPDK's pools, which are allocated in CVM shared memory.
Authenticated encryption uses AEGIS-128X4 from libaegis~\cite{AegisaeadLibaegis2026}, and filesystem integrity is build on top of a Merkle tree from cryptsetup~\cite{CryptsetupCryptsetupGitLab2026}.
During development, we further identified that DMA buffers without 4K alignment slow down NVMe transfers, and CVMs without AES-NI and AVX512 enabled suffer a $\approx$50\% reduction in AES performance~\cite{EnableAESAVX}.

\myparagraph{Trusted counter}
\projectname{} implements trusted counters as a pluggable library \revnoref{that supports different latency and synchronization trade-offs and can be replaced} with a trusted third-party module.

\myparagraph{Attestation}
\projectname{}'s attestation follows the IETF RATS background check model~\cite{birkholzRemoteATtestationProcedureS2023}.
We provision SEV-SNP-specific reference values generated by \textit{go-gen-ref}~\cite{ramanJraman567Gogenref2025} along with AMD trust anchors~\cite{amdAMDSEVSNPStrengthening2020} into the verifier.
The CVM generates a key pair and includes the key hash with a verifier nonce in the attestation evidence.
\projectname{} uses ratsd~\cite{VeraisonRatsd2026} to collect CVM evidence via configfs-tsm~\cite{KernelorgDocDocumentation} in RATS CMW format with an Entity Attestation Token (EAT)~\cite{birkholzRATSConceptualMessages2025, lundbladeEntityAttestationToken2025a}, and forwards it to a relying party for verification.
This party uses Veraison~\cite{Veraison, VeraisonServices2026} to validate the evidence and receives an EAT Attestation Result (EAR)~\cite{fossatiEATAttestationResults2025}, seals the \projectname{} configuration against the attested public key, and transmits it to the CVM.
Deployment required backporting Linux kernel 7.0 patches~\cite{GITPULLKVM} for certificate chain provisioning, patching QEMU, and extending sev-snp-measure-go~\cite{VirteeSevsnpmeasurego2025} to support OVMF measurements; all applicable changes are sent upstream.

\section{Evaluation}
\renewcommand*{\subsectionautorefname}{\S}

\revnoref{
We evaluate \projectname{} across six dimensions:
I/O characteristics of its storage interfaces (\autoref{sec:evaluation:io}),
end-to-end performance (\autoref{sec:evaluation:e2e}),
read/write amplification (\autoref{sec:evaluation:rw}),
ablation study (\autoref{sec:evaluation:ablation}),
attestation and sealing (\autoref{sec:evaluation:attestation}), and
freshness and durability costs (\autoref{sec:evaluation:counter}).
}

\subsection{Experimental Setup}
\label{subsec:experimental_setup}

\myparagraph{Testbed}
We conduct our experiments on two AMD SEV-SNP servers with \revnoref{9334/9654} CPUs, and
\revnoref{a \qty{3.8}{\tera\byte} KIOXIA CM7-R NVMe.}
The host runs NixOS 25.11 with Linux 6.18, including patches for NVMe speed and backported SEV-SNP KVM certificate syscall~\cite{sunPATCHV402025,rothKVMSEVAdd2026, rothKVMIntroduceKVM_EXIT_SNP_REQ_CERTS2026}.
(C)VMs run Linux 6.17 with our mmap patch, \qty{64}{\giga\byte} RAM, \revnoref{64 vCPUs (32 vCPUs/\qty{256}{\giga\byte} for checkpointing), mirroring the host's NUMA topology for NVMe locality}.

\myparagraph{Variants}
\revnoref{We compare \projectname{} against ext4 and Rollbaccine~\cite{chuRollbaccineHerdImmunity2025}, a state-of-the-art anti-rollback block layer for TEEs.
We match their durability settings per experiment.}

\subsection{I/O Analysis}
\label{sec:evaluation:io}

\begin{revblock}[rev1]{G.06}
To isolate each interface's raw performance, we use synthetic workloads against ext4 with direct I/O and Rollbaccine at \(f\mathord{=}0\) (no replication, \texttt{O\_DIRECT}).
\projectname{} runs non-blocking, synchronizing its epoch only at shutdown, so neither pays per-operation synchronization.
We use small entry/file sizes of \qty{16}{\KB} to \qty{1}{\mega\byte} (\qty{128}{\giga\byte} in total) and large ones of \qty{64}{\mega\byte}, \qty{256}{\mega\byte}, and \qty{1}{\giga\byte} (\qty{256}{\giga\byte} in total), and run with 1 and 8 threads for single- and multi-core behavior. 
Dashed lines mark the raw NVMe bandwidth measured with fio (\qty{13.5}{\gibi\byte\per\second} read, \qty{6.7}{\gibi\byte\per\second} write).
The \textit{Filesystem} reads entire files sequentially, if not specified otherwise.
The \textit{KVS} runs the update-heavy YCSB workload A~\cite{ycsb} (50/50 read/update) against RocksDB with its write-ahead log disabled and BlobDB for values $\geq$\qty{4}{\mega\byte}.
The \textit{Shared Log} sequentially appends and retrieves data, simulating high-volume writes and recovery.

\projectname{}'s \textit{KVS} (\autoref{fig:evaluation:io} (a)) trades small-object throughput for large-object bandwidth.
It reaches \qtyrange{4.7}{4.9}{\gibi\byte\per\second} at \qty{256}{\mega\byte} and \qty{1}{\giga\byte} (per request, 94--98\% of the device's read and 99\% of its write bandwidth), 5.5--7.0$\times$ higher than RocksDB, which stalls below \qty{0.9}{\gibi\byte\per\second} on both ext4 and Rollbaccine, and overtakes it from \qty{1}{\mega\byte} upward (\qty{256}{\KB} with 8 threads).
Below that, \projectname{} is slower, as RocksDB is optimized for small objects.
On the \textit{Filesystem} (\autoref{fig:evaluation:io} (c)), a single \projectname{} thread leads both baselines on small random reads: at \qty{256}{\KB} it reads \qty{2.4}{\gibi\byte\per\second}, 1.4$\times$ higher than ext4 (\qty{1.7}{\gibi\byte\per\second}) and 1.9$\times$ higher than Rollbaccine (\qty{1.3}{\gibi\byte\per\second}), converging with ext4 by \qty{1}{\mega\byte}.
Eight threads lift \projectname{} to \qty{11.4}{\gibi\byte\per\second} at \qty{1}{\mega\byte}, still ahead of ext4 (\qty{11.2}{\gibi\byte\per\second}) and Rollbaccine (\qty{10.7}{\gibi\byte\per\second}).
For large sequential file reads (\qty{64}{\mega\byte}--\qty{1}{\giga\byte}) (\autoref{fig:evaluation:io} (b), left), \projectname{} splits each file into pipelined chunks. A single thread sustains \qtyrange{13.0}{13.3}{\gibi\byte\per\second} from \qty{256}{\mega\byte}, 96--99\% of the device, against 30\% for ext4 and 19\% for Rollbaccine.
On the \textit{Shared Log} (\autoref{fig:evaluation:io} (b), right), a single \projectname{} thread leads both baselines on appends at every entry size, 2.3--2.8$\times$ at \qty{16}{\KB} (\qty{397}{\mebi\byte\per\second}), while a fixed per-entry retrieval cost keeps small reads 2.2--3.5$\times$ below ext4 (\qty{80}{\mebi\byte\per\second} at \qty{16}{\KB}).
From \qty{1}{\mega\byte} and above, \projectname{}'s appends reach 95--99\% of the device's write bandwidth, 3$\times$ Rollbaccine (\qty{2.1}{\gibi\byte\per\second}), and sequential replay 97--98\% of its read bandwidth after the counter-cache build.
\end{revblock}

\begin{takeaway}
\rev[rev1]{G.06}{From \qty{64}{\mega\byte}, one \projectname{} core with CIF is within $\approx$1\% (write) and 1--19\% (read) of raw NVMe bandwidth, reaching \qty{13.3}{\gibi\byte\per\second} read and \qty{6.6}{\gibi\byte\per\second} write; single-threaded ext4 and Rollbaccine reach 18--30\% on reads and need eight threads to compensate bounce buffer and CIF overhead.}
\end{takeaway}

\subsection{End-to-end Performance}
\label{sec:evaluation:e2e}

\begin{figure}[t]
    \centering
    \begin{revfloat}[rev1]{G.03}
    \includegraphics[width=\linewidth]{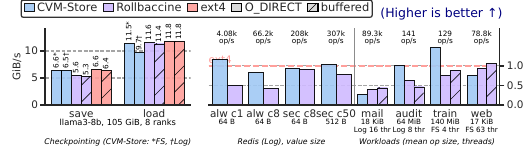}
    \vspace{-1.8em}
    \caption{
    \revnoref{
    End-to-end evaluation results.
    }
    }
    \label{fig:evalutation:e2e-performance}
    \end{revfloat}
    \vspace{-0.9em}
\end{figure}

\begin{revblock}[rev1]{G.03}
We evaluate the \projectname{}'s end-to-end application performance with the following benchmarks: ML training checkpointing (\qty{105}{\giga\byte} Llama3-8B model) from MLPerf-Storage~\cite{balmauCharacterizingMachineLearning2022}, 
Redis with AOF (Append Only File) persistence~\cite{RedisPersistence}, 
and a Filebench~\cite{tarasovFilebenchFlexibleFramework2016} inspired client that replays representative file-system operations.
We compare \projectname{} in blocking mode, closing an epoch per fsync or barrier, with ext4 and Rollbaccine at \(f\mathord{=}1\) with local metadata replication, so both are durable per sync.
\autoref{fig:evalutation:e2e-performance} presents the results.

In the checkpointing workload, \projectname{}'s \textit{Filesystem} and \textit{Shared Log} reach \qtyrange{6.53}{6.55}{\giga\byte\per\second} on saves, matching ext4 and outperforming Rollbaccine (\qty{5.61}{\giga\byte\per\second}) by 1.17$\times$.
Loading matches both baselines within 2.3\% via the \textit{Filesystem} (\qty{11.5}{\giga\byte\per\second}). The \textit{Shared Log} trails by 15\% (\qty{9.74}{\giga\byte\per\second}) because the retrieval of checkpoint entries is partially serialized on reads, not achieving full bandwidth.

On Redis, \projectname{} outperforms Rollbaccine by 1.95--2.27$\times$ under per-operation durability, due to faster per-operation sync.
Under once-per-second batching, it overtakes Rollbaccine by 1.32$\times$ at 50 clients (\qty{325}{\kilo\ops\per\second}).
Against ext4, it is either on par or trailing by 6--15\%.
\projectname{} is integrated with Redis on its own userspace SPDK core with I/O batching, while the others go through the kernel's I/O stack.

Across four Filebench-style workloads, \projectname{}'s performance matches the expected characteristics from \autoref{sec:evaluation:io}.
The \textit{train}, modeled after 3D~U-Net, and the fsync-barriered \textit{audit} transfer megabytes per operation. In \textit{train}, \projectname{} reads \qty{9.28}{\gibi\byte\per\second} (1.48$\times$ ext4, 1.94$\times$ Rollbaccine), while in \textit{audit} it sustains \qty{6.63}{\gibi\byte\per\second} (1.03$\times$ ext4, 1.63$\times$ Rollbaccine).
At the small-object and high-thread scale, which \projectname{}'s design deprioritizes, it trails ext4: \textit{web} reaches 0.76$\times$ and \textit{mail} is its worst case at 0.29$\times$.
In these workloads, small operations cannot amortize the per-operation cost caused by partial serialization for metadata and SPDK Blobstore handling.
\end{revblock}

\begin{takeaway}
\rev[rev1]{G.03}{
On data-heavy workloads, \projectname{}'s I/O characteristics carry through to full applications: it matches or outperforms both baselines on its targeted large-object workloads, trailing only on small-object and high thread count workloads.
}
\end{takeaway}

\subsection{Read/Write Amplification}
\label{sec:evaluation:rw}
\begin{table}[t]
    \centering
    \footnotesize
    \revnoref{
    \begin{tabular*}{\linewidth}{@{\hspace{2pt}\extracolsep{\fill}}lcccc|c@{\hspace{2pt}}}
        \toprule
        Workload (R/W)  & SPDK  & MS  & IO Meta  & MT & Total (\textperthousand)  \\
        \midrule
        KVS \qty{64}{\mega\byte} & 80/70 & 0/0.2 & 20/53 & 0/0 & 0.10/0.13 \\
        KVS \qty{4}{\giga\byte} & 1.3/1.6 & 0/0 & 0.2/3.0 & 0/0 & 0.0015/0.0044 \\
        Log \qty{64}{\mega\byte} & 30/70 & 0/90 & 20/21 & 0/0 & 0.05/0.18 \\
        Log \qty{4}{\giga\byte} & 0/1.5 & 0/8.0 & 0.2/0.3 & 0/0 & 0.0002/0.0097 \\
        FS \qty{64}{\mega\byte} & 60/190 & 0/100 & 7.6/8.3 & 20/20 & 0.09/0.32 \\
        FS \qty{4}{\giga\byte} & 0/3.7 & 0/8.2 & 0.1/0.2 & 20/20 & 0.02/0.03 \\
        \bottomrule
    \end{tabular*}
    }
    \vspace{-0.5em}
    \caption{\revnoref{Breakdown of read/write amplification. SPDK blob meta (SPDK), metadata sync (MS), I/O engine meta (IO Meta), and Merkle tree (MT) in $10^{-4}$\%; Total in \textperthousand.}}
    \label{tab:evaluation:rw-amplification}
    \vspace{-0.7em}
\end{table}

To measure \projectname{}'s read/write amplification \revnoref{with \qty{100}{\milli\second} synchronization epoch,} we trace API-level operations and compare application-visible I/O against actual bytes written through the SPDK bdev layer, capturing overhead from SPDK metadata, \projectname{} metadata, and CIF structures.
We do not measure physical write amplification on the NVMe, as it is NVMe- and internal-state-specific.

\autoref{tab:evaluation:rw-amplification} presents the results.
\revnoref{SPDK allocates a single \qty{4}{\kibi\byte} metadata page per blob, and \projectname{} adds one further block for AEAD MACs, nonces, and sizes. 
The rest is for global metadata and block size padding.
KVS and shared log store an additional encrypted metadata block for their freshness guarantees, while the filesystem stores a Merkle tree, one leaf per chunk ($<$0.002\%, logarithmic depth).}
\revnoref{SPDK's fragmentation-free allocation keeps amplification across workloads minimal.}

\begin{takeaway}
\projectname{}'s abstractions incur minimal read/write amplification for their metadata and operations.
\end{takeaway}

\subsection{\texorpdfstring{\revnoref{Ablation Study}}{Ablation Study}}
\label{sec:evaluation:ablation}

\begin{figure}
    \centering
    \begin{revfloat}[rev1]{G.07}
    \includegraphics[width=\linewidth]{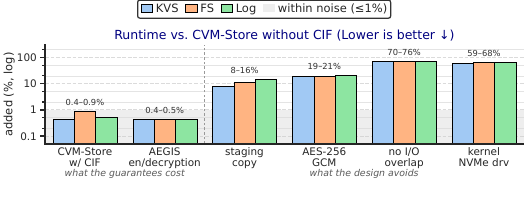}
    \vspace{-2.4em}
    \caption{\rev[rev1]{G.07}{Ablation of \projectname{}'s design decisions.
    }}
    \label{fig:evaluation:ablation}
    \end{revfloat}
    \vspace{-0.9em}
\end{figure}

\begin{revblock}[rev1]{G.07}
To justify \projectname{}'s individual design decisions, we run an ablation study with a synthetic workload that reads and writes \qty{256}{\giga\byte} data.
We compare the following configurations: \projectname{} with CIF, \projectname{} without CIF, and \projectname{} variants in which one mechanism is disabled.
\autoref{fig:evaluation:ablation} reports the runtime normalized to \projectname{} without CIF.

First, full CIF protection (``\projectname{} w/ CIF'') adds 0.4--0.9\% of runtime.
Merkle tree, metadata, and trusted counter operations each contribute less than 0.3\% of runtime in traces of the same runs.
``AEGIS en/decryption'' solely shows the encryption cost: the crypto pipeline runs in parallel with I/O and spans 51--56\% of total runtime, of which only 0.4--0.5\% is visible end-to-end, as the remainder overlaps with in-flight DMA.

The remaining configurations isolate the contribution of each mechanism.
Simulating only the additional copy through a staging buffer (``staging copy'') adds 8--16\%.
Substituting AES-256-GCM for AEGIS-128X4 (``AES-256-GCM'') adds 19--21\%, making the I/O cipher-bound instead of device-bound.
Serializing encryption and I/O instead of pipelining them (``no I/O overlap'') inflates runtime by 70--76\%.
Replacing the hardened kernel-bypass data path with the guest's NVMe driver (``kernel NVMe drv'') incurs a 59--68\% overhead, since every request then traverses the kernel and a bounce buffer.
\end{revblock}

\begin{takeaway}
\rev[rev1]{G.07}{
\projectname{} provides CIF guarantees at 0.4--0.9\% runtime overhead by interleaving cryptographic operations with in-flight DMA on a hardened kernel-bypass data path.
Naively implementing any one of these parts instead could cost up to 76\% runtime overhead.
}
\end{takeaway}

\begin{table}[t]
    \centering
    \footnotesize
    \begin{revfloat}[rev1]{G.12}
\begin{tabular*}{\linewidth}
    {@{}l@{\hspace{4pt}}c@{\hspace{4pt}}c@{\hspace{4pt}}c@{\hspace{4pt}}c@{\hspace{4pt}}|r@{}}
    \toprule
    \multirow{2}{*}{} & Evidence & Evidence & Attestation & Seal/ & \multicolumn{1}{c}{\multirow{2}{*}{Total}} \\
                               & collection & appraisal & appraisal & Unseal & \\
    \midrule
    CVM & 26.57 & — & — & 24.65 & \qty{188.35}{\milli\second} \\
    Trusted party & — & 65.95 & 0.76 & 20.73 & \qty{87.43}{\milli\second} \\
    \bottomrule
\end{tabular*}
    \vspace{-0.5em}
    \caption{\revnoref[rev1]{Breakdown of attestation and provisioning time.}}
    \label{tab:evaluation:attestation}
    \end{revfloat}
    \vspace{-0.5em}
\end{table}

\begin{table}[t]
    \centering
    \footnotesize
    \revnoref{
    \begin{tabular*}{\linewidth}{@{\hspace{6pt}\extracolsep{\fill}}lccc@{\hspace{6pt}}}
        \toprule
        Data Size & KVS (s) & Filesystem (s) & Shared log (s) \\
            &  S / M / C & S / M / C & S / M / C \\
            \midrule
            \qty{100}{\mega\byte} & 0.03 / 0.12 / {--} & 0.06 / 0.25 / 0.13 & 0.03 / 0.05 / 0.08 \\
            \qty{1}{\giga\byte} & 0.16 / 0.32 / {--} & 0.16 / 0.49 / 0.14 & 0.15 / 0.23 / 0.07 \\
            \qty{10}{\giga\byte} & 1.5 / 1.3 / {--} & 1.5 / 1.3 / 0.15 & 1.5 / 0.9 / 0.08 \\
            \qty{100}{\giga\byte} & 15.5 / 7.9 / {--} & 15.1 / 8.1 / 0.14 & 15.0 / 8.8 / 0.08 \\
            \qty{1000}{\giga\byte} & 150.1 / 75.2 / {--} & 150.2 / 76.4 / 0.15 & 150.2 / 76.1 / 0.09 \\
            \bottomrule
    \end{tabular*}
    }
    \vspace{-0.6em}
    \caption{\revnoref[rev1]{Time (in seconds) to seal (S) new data to \projectname{}, to unseal \& load abstraction metadata (M), and to check consistency (C).}}
    \label{tab:evaluation:api-recovery}
    \vspace{-1em}
\end{table}

\subsection{Attestation and Sealing}
\label{sec:evaluation:attestation}

\revnoref{%
We evaluate \projectname{}'s attestation and sealing mechanisms using a Veraison~\cite{Veraison} verifier with experimental
SEV-SNP support, which returns a standardized EAT Attestation Result
(EAR)~\cite{fossatiEATAttestationResults2025}.
\projectname{} collects CVM attestation evidence, the trusted party submits them to the verifier, and successful appraisal releases secrets (\qty{1}{\mebi\byte}).
For provisioning, \projectname{} runs externally to seal data before providing the symmetric key and trusted counter values to the CVM.
For recovery, data is split across 1024 entries/files.
The filesystem consistency check validates the superblock metadata; the shared log checks its counter chain on log opening; and the KVS folds its check into loading, rebuilding its in-memory key lookup by reading all entries.
We measure CVM-side and trusted-party operations separately and report the average of three runs over a local network.
This one-time startup cost is \qty{188.35}{\milli\second} on the CVM and \qty{87.43}{\milli\second} on the trusted party (\autoref{tab:evaluation:attestation}).
Sealing and unsealing then scale linearly with the data, from \qtyrange{0.03}{0.06}{\second} at \qty{100}{\mega\byte} to \qty{150.2}{\second} at \qty{1}{\tera\byte} (\autoref{tab:evaluation:api-recovery}).
Recovery, in contrast, is asymmetric: the filesystem and shared log validate metadata only (\qtyrange{0.07}{0.15}{\second}), while the KVS combines the check as part of loading, reading every entry on rebuild, which scales with the data size.
\ifextended Appendix~\ref{app:lifecycle} details the instance lifecycle.\fi%
}

\begin{takeaway}
\projectname{} incurs a one-time overhead (\qty{188}{\milli\second}) for its attestation process and enables external data sealing for subsequent use, with standardized attestation results supporting future multi-party provisioning scenarios.
\end{takeaway}

\subsection{Trusted Counter}
\label{sec:evaluation:counter}

\begin{figure}[t]
    \centering
    \begin{revfloat}[rev1]{G.05, G.11}
    \includegraphics[width=\linewidth,trim=0 0 0 6.8,clip]{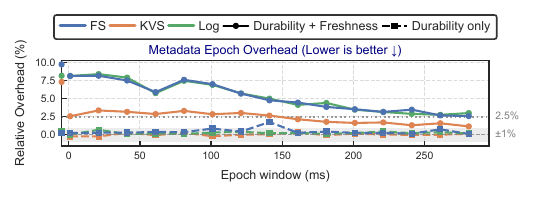}
    \vspace{-2.4em}
    \caption{\rev[rev1]{G.05, G.11}{Blocking-mode epoch synchronization relative overhead with and without freshness (\qty{5}{\milli\second} RTT).}}
    \label{fig:evaluation:trustedcounter}
    \end{revfloat}
    \vspace{-1.03em}
\end{figure}

\begin{revblock}[rev1]{G.05, G.11}%
To evaluate trusted counter synchronization overhead in blocking mode, we inject a synthetic \qty{5}{\milli\second} delay, modeling network Round Trip Time (RTT) to a remote service~\cite{angelNimbleRollbackProtection2023} using the same workloads as in the ablation study.
We vary the epoch length from \qty{1}{\milli\second} to \qty{281}{\milli\second} in \qty{20}{\milli\second} increments, and compare durability alone against durability coupled with freshness.

\autoref{fig:evaluation:trustedcounter} shows the relative overhead, compared to only durability at shutdown.
The markers on the y-axis show the worst case of synchronizing every operation.
Durability alone persists metadata without contacting the trusted party, keeping its overhead within $\pm$1\% across most workloads and epoch lengths.
Coupling durability with freshness adds a round trip to the trusted party at every epoch boundary.
At the worst case, this overhead is at 7.3--9.7\% for all three workloads.
The filesystem and shared log stay near this worst case at a \qty{1}{\milli\second} epoch ($\approx$8\%), close to 2.5\% at the widest epochs.
KVS overhead is lower (1--3\% over all epochs), as the workload mixes reads and writes and some epochs do not entail synchronization.
\end{revblock}

\begin{takeaway}
\rev[rev1]{G.05, G.11}{%
In \projectname{}, applications can trade freshness for performance by widening the epoch, taking filesystem and shared log overhead from 8--10\% at per-operation synchronization down to close to 2.5\% at the widest epochs.
}
\end{takeaway}

\section{Related Work}

\myparagraph{\revnoref[rev1]{Secure storage for VMs and TEEs}}
\rev[rev1]{G.01}{
We discuss related work on secure storage systems for TEEs in \autoref{sec:motivation}.
}
\revnoref[rev1]{Additionally, TEE-compatible designs target secure storage in single-node~\cite{xuMlsDiskTrustedBlock2026, chakrabortiDmxProtectingVolumelevel2017, tianAtomicDiskSecureVirtual2025, ahmadOBLIVIATEDataOblivious2018, dickensiiiStrongBoxConfidentialityIntegrity2018, weinholdJVPFSAddingRobustness2011} and distributed settings~\cite{bailleuAvocadoSecureInMemory2021, mahajanDepotCloudStorage2011, popaEnablingSecurityCloud2011}, as well as secure databases and key-value stores~\cite{popaCryptDBProtectingConfidentiality2011, papadimitriouBigDataAnalytics2016, tuProcessingAnalyticalQueries2013, kimLogStructuredMergeTreeaware2022, bailleuSPEICHERSecuringLSMbased2019, kimShieldStoreShieldedInmemory2019a, messadiPrecursorFastClientcentric2021, priebeEnclaveDBSecureDatabase2018, sunBuildingEnclavenativeStorage2021}.
Generic kernel-level mechanisms (dm-crypt, dm-integrity, dm-verity) provide block-level security without CVM-specific optimizations.
Building on these mechanisms, the Confidential Data Hub~\cite{confidentialcontainersConfidentialContainersSecure} provides secure storage for confidential containers, using LUKS~\cite{CryptsetupCryptsetupGitLab2026} for block device encryption or Gocryptfs~\cite{gocryptfs} for object storage.}
However, \revnoref{these backends rely on the kernel I/O stack,} and neither provides full integrity or freshness guarantees.
\revnoref[rev1]{In contrast,} \projectname{} provides optimized storage abstractions that support diverse access patterns within CVMs while maintaining strong CIF guarantees.

\myparagraph{Userspace I/O in confidential computing}
Userspace I/O frameworks (e.g., DPDK~\cite{dpdk}, SPDK~\cite{yang2017spdk}) reduce context switching overhead by directly interacting with devices~\cite{thalheimRktioDirectStack2021, dpdk}.
\revnoref[rev1]{SPDK has been combined with SGX~\cite{thalheimRktioDirectStack2021}, but not adapted for CVMs, due to limited device passthrough and DMA memory support in CVM userspace.
In the network domain,} Bifrost~\cite{liBifrostAnalysisOptimization2023} leverages the fact that application-level protocols (e.g., TLS) already provide end-to-end confidentiality, thereby avoiding unnecessary encryption at the I/O layer,
\revnoref[rev1]{and} other approaches demonstrate that careful software optimization can achieve near-optimal I/O performance~\cite{liBridgeFutureHighPerformance2024}.
\projectname{} builds upon these insights to design a secure userspace I/O engine tailored for storage workloads in CVMs.

\myparagraph{Trusted counters}
CVMs do not natively support trusted counters.
AMD provides SecureTSC to protect time-source values from hypervisor manipulation, but it introduces new attack vectors~\cite{juffingerNotSecureTSC2025, amdAMDSecurityBulletin2024} and cannot be used as a replacement. 
Therefore, remote trusted services have been proposed~\cite{angelNimbleRollbackProtection2023,martinADAMCSAdvancedAsynchronous2021,mateticROTERollbackProtection2017} to restore the trusted counter functionality and build replication on top.
\projectname{} addresses this by exposing a pluggable trusted counter interface through which it synchronizes with an external trusted party to provide its freshness guarantees.

\myparagraph{Attestation}
\revnoref{While CVM attestation performance has already been studied~\cite{misonoConfidentialVMsExplained2024}, supply chain challenges remain for endorsements and reference values~\cite{glazeMeaningfulAttestationsSupply2025}.}
RATS~\cite{birkholzRemoteATtestationProcedureS2023} provides standardized frameworks used by Trustee KBS~\cite{Trustee2025} and Veraison~\cite{Veraison}.
\projectname{} contributed to Veraison to attest evidence from the CVM before provisioning cryptographic materials.

\myparagraph{Authenticated data structures}
Recent work proposes skewed Merkle trees with shorter paths for frequently accessed addresses~\cite{szeferFastHardwareMemory2014} and Bonsai Merkle Forests that split integrity trees to reduce overhead~\cite{freijBonsaiMerkleForests2021}.
Additional optimizations adapt the tree layout at runtime~\cite{rakshitASSUREAuthenticationScheme2017,thomasMidsummerNightsTree2024}.
\projectname{} only requires a simple Merkle tree for filesystem integrity verification, while avoiding the complexity of dynamic restructuring through its immutable file design.

\section{Conclusion}

\revnoref{CVMs protect data in use, but leave a \emph{storage gap}: no prior CVM storage system provides CIF while matching modern NVMe performance, forcing stateful workloads to trade security for performance.
\projectname{} closes this gap by hardening the userspace I/O trust boundary, interleaving cryptographic operations with in-flight DMA, aligning integrity structures with each abstraction's access pattern,
and decoupling freshness into an externally epoch-batched counter service.
With full CIF, \projectname{} stays within $\approx$1\% (write) and 1--19\% (read) of raw NVMe bandwidth on large objects, outperforming unencrypted CVM kernel I/O and rollback-protected storage in terms of runtime overhead, while applications can tune the freshness cost through the epoch length configuration.}

\myparagraph{Artifact and appendix}
\projectname{}'s source code is available at \url{https://github.com/TUM-DSE/CVM-Store-NDSS27}.
\revnoref{\ifextended
The \hyperref[sec:appendix:counter]{Appendix} details the trusted counter service, the instance lifecycle, and the SPDK hardening.
\else
The \hyperref[app:spdk]{Appendix} details the SPDK hardening, and \extversion{} additionally covers the trusted counter service and the instance lifecycle.
\fi}

\section*{Acknowledgment}
This work was supported in part by an ERC Starting Grant (ID: 101077577) and
the Chips Joint Undertaking (JU), European Union (EU) HORIZON-JU-IA, under
grant agreement No. 101140087 (SMARTY).
We also thank the Veraison and SPDK open-source projects, whose work
\projectname{} builds upon, for their input and help contributing fixes upstream.

\section*{Ethics Considerations}
This paper does not introduce any new vulnerability and does not involve human subjects research.
Our work aims to strengthen security in confidential cloud computing, and we therefore believe this work raises no ethical concerns.

\ifextended
  \bibliographystyle{plainurl}
\else
  \bibliographystyle{abbrvurl}
\fi
\bibliography{for-paper, extra_refs, revision-citiations}

@inproceedings{yang2017spdk,
  title={SPDK: A development kit to build high performance storage applications},
  author={Yang, Ziye and Harris, James R and Walker, Benjamin and Verkamp, Daniel and Liu, Changpeng and Chang, Cunyin and Cao, Gang and Stern, Jonathan and Verma, Vishal and Paul, Luse E},
  booktitle={2017 IEEE International Conference on Cloud Computing Technology and Science (CloudCom)},
  pages={154--161},
  year={2017},
  organization={IEEE},
  doi={10.1109/CloudCom.2017.14}}

@misc{azure-confidential-computing,
    author = {Microsoft},
    title = "{Azure Confidential Computing}",
    url = {https://azure.microsoft.com/en-us/solutions/confidential-compute},
    Xhowpublished = {\url{https://azure.microsoft.com/en-us/solutions/confidential-compute}},
    note = {Last accessed: March, 2025},
    year = {2019},
}

@misc{gcp-confidential-computing,
    author = {Google},
    title = "{Google Cloud Confidential Computing}",
    url = {https://cloud.google.com/confidential-computing/},
    Xhowpublished = {\url{https://cloud.google.com/confidential-computing/}},
    note = {Last accessed: March, 2025},
    year = {2019},
}

@misc{ibm-confidential-computing,
    author = {IBM},
    title = "{Confidential computing on IBM Cloud}",
    url = {https://www.ibm.com/cloud/confidential-computing},
    Xhowpublished = {\url{https://www.ibm.com/cloud/confidential-computing}},
    note = {Last accessed: March, 2025},
    year = {2019},
}

@misc{tdx,
    author = {Intel},
    title = "{Intel® Trust Domain Extensions (Intel TDX)}",
    url = {https://www.intel.com/content/www/us/en/developer/articles/technical/intel-trust-domain-extensions.html},
    note = {Last accessed: March, 2025},
    year = {[n.\,d.]},
}

@misc{cca,
    author = {ARM},
    title = "{Arm Confidential Compute Architecture}",
    url = {https://www.arm.com/architecture/security-features/arm-confidential-compute-architecture},
    Xhowpublished = {\url{https://www.arm.com/architecture/security-features/arm-confidential-compute-architecture}},
    note = {Last accessed: March, 2025},
    year = {[n.\,d.]},
}

@misc{bounce_buffer,
    author = {Marta Rybczy\'{n}ska},
    title = "{Bounce Buffers for Untrusted Devices  [LWN.net]}",
    url = {https://lwn.net/Articles/786558/},
    note = {Last accessed: March, 2025},
    year = {2019},
}

@misc{ycsb,
title = {{YCSB}},
howpublished ={\url{https://github.com/brianfrankcooper/YCSB}},
note = {Last accessed: May, 2025},
author = {Brian Frank Cooper},
year = {2025}
}

@misc{dpdk,
title = {{DPDK}},
howpublished ={\url{https://www.dpdk.org/}},
note = {Last accessed: February, 2026},
author = {{DPDK Project}},
year = {2026}
}

@misc{pcie-ide-tdisp,
title = {{IDE and TDISP: An Overview of PCIe® Technology Security Features}},
howpublished ={\url{https://pcisig.com/blog/ide-and-tdisp-overview-pcie%C2%AE-technology-security-features}},
note = {Last accessed: February, 2026},
author = {Lou Ternullo},
year = {2026}
}

@misc{fde,
title = {{What is Full Disk Encryption (FDE) \& How It Works}},
howpublished ={\url{https://www.bitdefender.com/en-us/business/infozone/what-is-full-disk-encryption-fde}},
note = {Last accessed: February, 2026},
author = {{BitDefender}},
year = {2026}
}

@misc{nvme-spec,
title = {{NVM Express® Base Specification}},
howpublished ={\url{https://nvmexpress.org/specification/nvm-express-base-specification/}},
note = {Last accessed: February, 2026},
author = {{NVM Express®}},
year = {2026}
}

@misc{spdk_blobstore,
title = {{Blobstore Programmer's Guide}},
howpublished ={\url{https://spdk.io/doc/blob.html}},
note = {Last accessed: February, 2026},
author = {{SPDK community}},
year = {2026}
}

@misc{vfio,
title = {{VFIO - “Virtual Function I/O”}},
howpublished ={\url{https://docs.kernel.org/driver-api/vfio.html}},
note = {Last accessed: February, 2026},
author = {{The kernel development community.}},
year = {2026}
}

@misc{iommufd,
title = {{IOMMUFD}},
howpublished ={\url{https://docs.kernel.org/userspace-api/iommufd.html}},
note = {Last accessed: February, 2026},
author = {Jason Gunthorpe and Kevin Tian},
year = {2026}
}

@misc{dm-crypt,
title = {{dm-crypt}},
howpublished ={\url{https://docs.kernel.org/admin-guide/device-mapper/dm-crypt.html}},
note = {Last accessed: February, 2026},
author = {{The kernel development community.}},
year = {2026}
}

@article{cheng_tdxsurvey_2023,
author = {Cheng, Pau-Chen and Ozga, Wojciech and Valdez, Enriquillo and Ahmed, Salman and Gu, Zhongshu and Jamjoom, Hani and Franke, Hubertus and Bottomley, James},
title = "{Intel TDX Demystified: A Top-Down Approach}",
year = {2024},
Xissue_date = {September 2024},
publisher = {Association for Computing Machinery},
Xaddress = {New York, NY, USA},
volume = {56},
number = {9},
Xissn = {0360-0300},
url = {https://doi.org/10.1145/3652597},
doi = {10.1145/3652597},
journal = {ACM Comput. Surv.},
Xmonth = apr,
articleno = {238},
Xnumpages = {33},
Xkeywords = {Confidential computing, trusted execution environment}
}

@misc{tdx_google_security_review,
    author = {Erdem Aktas and Cfir Cohen and Josh Eads and James Forshaw and Felix Wilhelm},
    title = "{Intel Trust Domain Extensions (TDX) Security Review April 2023}",
    url = {https://services.google.com/fh/files/misc/intel_tdx_-_full_report_041423.pdf},
    note = {Last accessed: March, 2025},
    year = {2023},
}

@misc{linux_coco_threat_model,
    author = {Elena Reshetova and Carlos Bilbao},
    title = "{Confidential Computing in Linux for x86 virtualization}",
    url = {https://docs.kernel.org/security/snp-tdx-threat-model.html},
    note = {Last accessed: March, 2025},
    year = {2023},
}

@INPROCEEDINGS{alam_cc-side-channel_2023,
  author={Alam, A K M Mubashwir and Chen, Keke},
  Xbooktitle={2023 IEEE 16th International Conference on Cloud Computing (CLOUD)},
  booktitle={Proceedings of the IEEE 16th International Conference on Cloud Computing}, 
  title="{Making Your Program Oblivious: A Comparative Study for Side-channel-Safe Confidential Computing}", 
  year={2023},
  volume={},
  number={},
  publisher={Institute of Electrical and Electronics Engineers},
  Xpages={282-289},
  doi={10.1109/CLOUD60044.2023.00040}
}

@misc{intel_secure_coding_guideline,
    author = {Intel},
    title = "{Guidelines for Mitigating Timing Side Channels Against Cryptographic Implementations}",
    url = {https://www.intel.com/content/www/us/en/developer/articles/technical/software-security-guidance/secure-coding/mitigate-timing-side-channel-crypto-implementation.html},
    note = {Last accessed: July, 2024},
    year = {2022},
}

@inproceedings {wichelmann_cipherfix_2023,
    author = {Jan Wichelmann and Anna P{\"a}tschke and Luca Wilke and Thomas Eisenbarth},
    title = "{Cipherfix: Mitigating Ciphertext {Side-Channel} Attacks in Software}",
    booktitle = {Proceedings of the 32nd USENIX Security Symposium},
    Xbooktitle = {32nd USENIX Security Symposium (USENIX Security 23)},
    year = {2023},
    isbn = {978-1-939133-37-3},
    Xaddress = {Anaheim, CA},
    Xpages = {6789--6806},
    Xurl = {https://www.usenix.org/conference/usenixsecurity23/presentation/wichelmann},
    publisher = {USENIX Association},
    Xmonth = aug
}

@misc{sev-cipher-sidechannel-mitigation,
    author = {AMD},
    title = "{TECHNICAL GUIDANCE FOR MITIGATING EFFECTS OF CIPHERTEXT VISIBILITY UNDER AMD SEV REVISION 5.10.22}",
    url = {https://www.amd.com/system/files/documents/221404394-a_security_wp_final.pdf},
    note = {Last accessed: July, 2024},
    year = {2022},
}

@inproceedings {deng_cipherh_2023,
    author = {Sen Deng and Mengyuan Li and Yining Tang and Shuai Wang and Shoumeng Yan and Yinqian Zhang},
    title = "{{CipherH}: Automated Detection of Ciphertext Side-channel Vulnerabilities in Cryptographic Implementations}",
    Xbooktitle = {32nd USENIX Security Symposium (USENIX Security 23)},
    booktitle = {Proceedings of the 32nd USENIX Security Symposium},
    year = {2023},
    isbn = {978-1-939133-37-3},
    Xaddress = {Anaheim, CA},
    Xpages = {6843--6860},
    Xurl = {https://www.usenix.org/conference/usenixsecurity23/presentation/deng-sen},
    publisher = {USENIX Association},
    Xmonth = aug
}

@misc{firmware-sidechannel-mitigation,
    author = {Intel},
    title = "{Host Firmware Speculative Execution Side Channel Mitigation}",
    url = {https://www.intel.com/content/www/us/en/developer/articles/technical/software-security-guidance/technical-documentation/host-firmware-speculative-side-channel-mitigation.html},
    note = {Last accessed: July, 2024},
    year = {2018},
}

@inproceedings{wang_pwrleak_2023,
    author = {Wang, Wubing and Li, Mengyuan and Zhang, Yinqian and Lin, Zhiqiang},
    title = "{PwrLeak: Exploiting Power Reporting Interface for Side-Channel Attacks on AMD SEV}",
    year = {2023},
    Xisbn = {978-3-031-35503-5},
    publisher = {Springer-Verlag},
    Xaddress = {Berlin, Heidelberg},
    Xurl = {https://doi.org/10.1007/978-3-031-35504-2_3},
    Xdoi = {10.1007/978-3-031-35504-2_3},
    Xbooktitle = {Detection of Intrusions and Malware, and Vulnerability Assessment: 20th International Conference, DIMVA 2023, Hamburg, Germany, July 12–14, 2023, Proceedings},
    booktitle = {Proceedings of the 20th International Conference on Detection of Intrusions and Malware, and Vulnerability Assessment},
    Xpages = {46–66},
    Xnumpages = {21},
    Xlocation = {Hamburg, Germany}
}

@misc{confidential_computing,
    author={{Confidential Computing Consortium}},
    title="{Confidential Computing: Hardware-Based Trusted Execution for Applications and Data: November 2022, V1.3}",
    url={https://confidentialcomputing.io/wp-content/uploads/sites/10/2023/03/CCC_outreach_whitepaper_updated_November_2022.pdf},
    note = {Last accessed: September, 2025},
    year={2022},
}

@INPROCEEDINGS{rollback_parno,
  author={Parno, Bryan and Lorch, Jacob R. and Douceur, John R. and Mickens, James and McCune, Jonathan M.},
  booktitle={2011 IEEE Symposium on Security and Privacy}, 
  title={Memoir: Practical State Continuity for Protected Modules}, 
  year={2011},
  volume={},
  number={},
  pages={379-394},
  doi={10.1109/SP.2011.38}}

@misc{qemu,
    author={{QEMU}},
    title="{QEMU: A generic and open source machine emulator and virtualizer}",
    url={https://www.qemu.org/},
    note = {Last accessed: February, 2026},
    year={2026},
}

@article{anchor,
author = {Stavrakakis, Dimitrios and Giantsidi, Dimitra and Bailleu, Maurice and S\"{a}ndig, Philip and Issa, Shady and Bhatotia, Pramod},
title = {Anchor: A Library for Building Secure Persistent Memory Systems},
year = {2023},
issue_date = {December 2023},
publisher = {Association for Computing Machinery},
address = {New York, NY, USA},
volume = {1},
number = {4},
url = {https://doi.org/10.1145/3626718},
doi = {10.1145/3626718},
journal = {Proc. ACM Manag. Data},
month = dec,
articleno = {231},
numpages = {31}
}

@misc{gocryptfs,
 author = {rfjakob},
 title= {{gocryptfs: an encrypted overlay filesystem written in Go}},
 howpublished ={\url{https://github.com/rfjakob/gocryptfs}},
 note = {Last accessed: February, 2026}
}

@inproceedings{qiu-cvmdb-2024,
author = {Qiu, Lina and Taft, Rebecca and Shraer, Alexander and Kollios, George},
title = {{The Price of Privacy: A Performance Study of Confidential Virtual Machines for Database Systems}},
year = {2024},
Xisbn = {9798400706677},
publisher = {Association for Computing Machinery},
Xaddress = {New York, NY, USA},
url = {https://doi.org/10.1145/3662010.3663440},
Xdoi = {10.1145/3662010.3663440},
booktitle = {Proceedings of the 20th International Workshop on Data Management on New Hardware},
Xarticleno = {2},
Xnumpages = {8},
Xlocation = {Santiago, AA, Chile},
Xseries = {DaMoN '24}
}

@misc{cvmstoreExtended,
  title        = {{{CVM-Store}}: {{A}} {{Secure}} {{Storage}} {{System}} for {{Confidential}} {{Virtual}} {{Machines}} ({{Extended}} {{Version}})},
  author       = {Sommer, Thore and Misono, Masanori and Stavrakakis, Dimitrios and Bhatotia, Pramod},
  year         = {2027},
  howpublished = {arXiv preprint arXiv:XXXX.XXXXX},
  url          = {https://arxiv.org/abs/XXXX.XXXXX},
  note         = {PLACEHOLDER -- replace XXXX.XXXXX with the arXiv ID},
}

@misc{AegisaeadLibaegis2026,
  title = {Aegis-Aead/Libaegis},
  year = 2026,
  month = feb,
  copyright = {MIT},
  howpublished = {The AEGIS authenticated encryption algorithms},
  note = {Last accessed: February 2026}
}

@misc{AegisaeadOpenssl2026,
  title = {Aegis-Aead/Openssl},
  year = 2026,
  month = jan,
  copyright = {Apache-2.0},
  howpublished = {The AEGIS authenticated encryption algorithms},
  note = {Last accessed: March 2026}
}

@inproceedings{ahmadOBLIVIATEDataOblivious2018,
  title = {{{OBLIVIATE}}: {{A Data Oblivious Filesystem}} for {{Intel SGX}}},
  shorttitle = {{{OBLIVIATE}}},
  booktitle = {Proceedings 2018 {{Network}} and {{Distributed System Security Symposium}}},
  author = {Ahmad, Adil and Kim, Kyungtae and Sarfaraz, Muhammad Ihsanulhaq and Lee, Byoungyoung},
  year = 2018,
  publisher = {Internet Society},
  address = {San Diego, CA},
  doi = {10.14722/ndss.2018.23284},
  urldate = {2025-07-14},
  langid = {english},
}

@misc{amdAMDSecurityBulletin2024,
  title = {{{AMD Security Bulletin AMD-SB-3007}}},
  author = {{AMD}},
  year = 2024,
  month = feb,
  journal = {AMD},
  howpublished = {\url{https://www.amd.com/en/resources/product-security/bulletin/amd-sb-3007.html}},
  langid = {english},
  note = {Last accessed: July 2025}
}

@techreport{amdAMDSEVSNPStrengthening2020,
  title = {{{AMD SEV-SNP}}: {{Strengthening VM Isolation}} with {{Integrity Protection}} and {{More}}},
  author = {{AMD}},
  year = 2020,
  month = jan,
  institution = {AMD},
  urldate = {2025-07-29},
  langid = {english},
}

@misc{amdVersionedChipEndorsement2025,
  title = {Versioned {{Chip Endorsement Key}} ({{VCEK}}) {{Certificate}} and {{KDS Interface Specification}}},
  author = {{AMD}},
  year = 2025,
  month = jan,
  urldate = {2025-07-29},
  langid = {english},
}

@inproceedings{angelNimbleRollbackProtection2023,
  title = {Nimble: {{Rollback Protection}} for {{Confidential Cloud Services}}},
  shorttitle = {Nimble},
  booktitle = {17th {{USENIX Symposium}} on {{Operating Systems Design}} and {{Implementation}} ({{OSDI}} 23)},
  author = {Angel, Sebastian and Basu, Aditya and Cui, Weidong and Jaeger, Trent and Lau, Stella and Setty, Srinath and Singanamalla, Sudheesh},
  year = 2023,
  pages = {193--208},
  urldate = {2025-07-29},
  isbn = {978-1-939133-34-2},
  langid = {english},
}

@inproceedings{bailleuAvocadoSecureInMemory2021,
  title = {Avocado: {{A Secure In-Memory Distributed Storage System}}},
  shorttitle = {Avocado},
  booktitle = {2021 {{USENIX Annual Technical Conference}} ({{USENIX ATC}} 21)},
  author = {Bailleu, Maurice and Giantsidi, Dimitra and Gavrielatos, Vasilis and Quoc, Do Le and Nagarajan, Vijay and Bhatotia, Pramod},
  year = 2021,
  pages = {65--79},
  urldate = {2025-07-29},
  isbn = {978-1-939133-23-6},
  langid = {english},
}

@article{bailleuSPEICHERSecuringLSMbased2019,
  title = {{{SPEICHER}}: {{Securing LSM-based Key-Value Stores}} Using {{Shielded Execution}}},
  author = {Bailleu, Maurice and Thalheim, J{\"o}rg and Bhatotia, Pramod and Fetzer, Christof and Honda, Michio and Vaswani, Kapil},
  year = 2019,
  month = feb,
  journal = {17th USENIX Conference on File and Storage Technologies (FAST 19)},
  pages = {173--190},
  issn = {978-1-939133-09-0},
  langid = {english},
}

@inproceedings{batteringramsp26,
  title = {Battering {{RAM}}: {{Low-Cost Interposer Attacks}} on {{Confidential Computing}} via {{Dynamic Memory Aliasing}}},
  booktitle = {47th {{IEEE Symposium}} on {{Security}} and {{Privacy}} ({{IEEE S}}\&{{P}} '26)},
  author = {Meulemeester, Jesse De and Oswald, David and Verbauwhede, Ingrid and Van Bulck, Jo},
  year = 2026,
  month = may,
}

@techreport{birkholzRATSConceptualMessages2025,
  type = {Internet {{Draft}}},
  title = {{{RATS Conceptual Messages Wrapper}} ({{CMW}})},
  author = {Birkholz, Henk and Smith, Ned and Fossati, Thomas and Tschofenig, Hannes and Glaze, Dionna},
  year = 2025,
  month = dec,
  number = {draft-ietf-rats-msg-wrap-23},
  institution = {Internet Engineering Task Force},
  urldate = {2026-02-19},
  note = {Work in Progress}
}

@techreport{birkholzRemoteATtestationProcedureS2023,
  type = {Request for {{Comments}}},
  title = {Remote {{ATtestation procedureS}} ({{RATS}}) {{Architecture}}},
  author = {Birkholz, Henk and Thaler, Dave and Richardson, Michael and Smith, Ned and Pan, Wei},
  year = 2023,
  month = jan,
  number = {RFC 9334},
  institution = {Internet Engineering Task Force},
  doi = {10.17487/RFC9334},
  urldate = {2023-03-07},
}

@phdthesis{brozAuthenticatedResilientDisk2018,
  type = {Doctoral {{Thesis}}},
  title = {Authenticated and {{Resilient Disk Encryption}}},
  author = {Bro{\v z}, Milan},
  year = 2018,
  urldate = {2026-03-16},
  langid = {english},
}

@inproceedings{burkeScalableIntegrityChecking2025,
  title = {On {{Scalable Integrity Checking}} for {{Secure Cloud Disks}}},
  booktitle = {23rd {{USENIX Conference}} on {{File}} and {{Storage Technologies}} ({{FAST}} 25)},
  author = {Burke, Quinn and Sheatsley, Ryan and King, Rachel and Hines, Owen and Swift, Michael and McDaniel, Patrick},
  year = 2025,
  pages = {391--405},
  urldate = {2026-03-02},
  isbn = {978-1-939133-45-8},
  langid = {english},
}

@inproceedings{chakrabortiDmxProtectingVolumelevel2017,
  title = {Dm-x: {{Protecting Volume-level Integrity}} for {{Cloud Volumes}} and {{Local Block Devices}}},
  shorttitle = {Dm-x},
  booktitle = {Proceedings of the 8th {{Asia-Pacific Workshop}} on {{Systems}}},
  author = {Chakraborti, Anrin and Jain, Bhushan and Kasiak, Jan and Zhang, Tao and Porter, Donald and Sion, Radu},
  year = 2017,
  month = sep,
  pages = {1--7},
  publisher = {ACM},
  address = {Mumbai India},
  doi = {10.1145/3124680.3124732},
  urldate = {2026-03-06},
  isbn = {978-1-4503-5197-3},
  langid = {english},
}

@misc{chuRollbaccineHerdImmunity2025,
  title = {Rollbaccine : {{Herd Immunity}} against {{Storage Rollback Attacks}} in {{TEEs}} [{{Technical Report}}]},
  shorttitle = {Rollbaccine},
  author = {Chu, David and Balasubramanian, Aditya and Bao, Dee and Crooks, Natacha and Howard, Heidi and Katahanas, Lucky E. and Ponnapalli, Soujanya},
  year = 2025,
  month = may,
  number = {arXiv:2505.04014},
  eprint = {2505.04014},
  primaryclass = {cs},
  publisher = {arXiv},
  doi = {10.48550/arXiv.2505.04014},
  urldate = {2025-07-08},
  archiveprefix = {arXiv},
  langid = {english},
}

@misc{confidentialcontainersConfidentialContainersSecure,
  title = {Confidential {{Containers}} - {{Secure Storage Documentation}}},
  author = {{Confidential Containers}},
  journal = {guest-components/confidential-data-hub/docs/SECURE\_STORAGE.md at main {$\cdot$} confidential-containers/guest-components},
  howpublished = {\url{https://github.com/confidential-containers/guest-components/blob/main/confidential-data-hub/docs/SECURE_STORAGE.md}},
  note = {Last accessed: July 2025}
}

@misc{CryptsetupCryptsetupGitLab2026,
  title = {{cryptsetup / cryptsetup {$\cdot$} GitLab}},
  year = 2026,
  month = feb,
  journal = {GitLab},
  howpublished = {\url{https://gitlab.com/cryptsetup/cryptsetup}},
  langid = {ngerman},
  note = {Last accessed: February 2026}
}

@misc{CVE202346813,
  title = {{{CVE-2023-46813}}},
  howpublished = {\url{https://www.cve.org/CVERecord?id=CVE-2023-46813}},
  note = {Last accessed: October 2025}
}

@misc{denisAddingMoreParallelism2023,
  title = {Adding More Parallelism to the {{AEGIS}} Authenticated Encryption Algorithms},
  author = {Denis, Frank},
  year = 2023,
  number = {2023/523},
  urldate = {2026-03-10},
}

@inproceedings{dickensiiiStrongBoxConfidentialityIntegrity2018,
  title = {{{StrongBox}}: {{Confidentiality}}, {{Integrity}}, and {{Performance}} Using {{Stream Ciphers}} for {{Full Drive Encryption}}},
  shorttitle = {{{StrongBox}}},
  booktitle = {Proceedings of the {{Twenty-Third International Conference}} on {{Architectural Support}} for {{Programming Languages}} and {{Operating Systems}}},
  author = {Dickens, {III}, Bernard and Gunawi, Haryadi S. and Feldman, Ariel J. and Hoffmann, Henry},
  year = 2018,
  month = mar,
  pages = {708--721},
  publisher = {ACM},
  address = {Williamsburg VA USA},
  doi = {10.1145/3173162.3173183},
  urldate = {2025-07-14},
  copyright = {https://www.acm.org/publications/policies/copyright\_policy\#Background},
  langid = {english},
}

@misc{EnableAESAVX,
  title = {Enable {{AES}} and {{AVX}} Instructions for {{SEV-SNP CVMs}} {$\cdot$} {{Issue}} \#12382 {$\cdot$} Kata-Containers/Kata-Containers},
  journal = {GitHub},
  howpublished = {\url{https://github.com/kata-containers/kata-containers/issues/12382}},
  langid = {english},
  note = {Last accessed: February 2026}
}

@misc{FioFlexibleTester,
  title = {Fio - {{Flexible I}}/{{O}} Tester Rev. 3.38 --- Fio 3.38-38-G3ee5 Documentation},
  howpublished = {\url{https://fio.readthedocs.io/en/latest/fio_doc.html}},
  note = {Last accessed: July 2025}
}

@techreport{fossatiEATAttestationResults2025,
  type = {Internet {{Draft}}},
  title = {{{EAT Attestation Results}}},
  author = {Fossati, Thomas and Voit, Eric and Trofimov, Sergei and Birkholz, Henk},
  year = 2025,
  month = apr,
  number = {draft-ietf-rats-ear-00},
  institution = {Internet Engineering Task Force},
  urldate = {2025-07-14},
  note = {Work in Progress}
}

@inproceedings{freijBonsaiMerkleForests2021,
  title = {Bonsai {{Merkle Forests}}: {{Efficiently Achieving Crash Consistency}} in {{Secure Persistent Memory}}},
  shorttitle = {Bonsai {{Merkle Forests}}},
  booktitle = {{{MICRO-54}}: 54th {{Annual IEEE}}/{{ACM International Symposium}} on {{Microarchitecture}}},
  author = {Freij, Alexander and Zhou, Huiyang and Solihin, Yan},
  year = 2021,
  month = oct,
  series = {{{MICRO}} '21},
  pages = {1227--1240},
  publisher = {Association for Computing Machinery},
  address = {New York, NY, USA},
  doi = {10.1145/3466752.3480067},
  urldate = {2025-07-21},
  isbn = {978-1-4503-8557-2},
}

@misc{GITPULLKVM,
  title = {[{{GIT PULL}}] {{KVM}} Changes for {{Linux}} 7.0 - {{Paolo Bonzini}}},
  howpublished = {\url{https://lore.kernel.org/all/20260213162059.22230-1-pbonzini@redhat.com/}},
  note = {Last accessed: February 2026}
}

@misc{glazeMeaningfulAttestationsSupply2025,
  title = {Meaningful {{Attestations}} from {{Supply Chain}} to the {{World}} by {{Dionna Glaze}}},
  author = {Glaze, Dionna},
  year = 2025,
  month = apr,
  urldate = {2025-07-14}
}

@inproceedings{grossAnalyzingAndroidsFileBased2019,
  title = {Analyzing {{Android}}'s {{File-Based Encryption}}: {{Information Leakage}} through {{Unencrypted Metadata}}},
  shorttitle = {Analyzing {{Android}}'s {{File-Based Encryption}}},
  booktitle = {Proceedings of the 14th {{International Conference}} on {{Availability}}, {{Reliability}} and {{Security}}},
  author = {Gro{\ss}, Tobias and Ahmadova, Matanat and M{\"u}ller, Tilo},
  year = 2019,
  month = aug,
  pages = {1--7},
  publisher = {ACM},
  address = {Canterbury CA United Kingdom},
  doi = {10.1145/3339252.3340340},
  urldate = {2026-02-26},
  isbn = {978-1-4503-7164-3},
  langid = {english},
}

@inproceedings{hetzeltAnalyzingDeviceInterfaces2021,
  title = {{{VIA}}: {{Analyzing Device Interfaces}} of {{Protected Virtual Machines}}},
  shorttitle = {{{VIA}}},
  booktitle = {Proceedings of the 37th {{Annual Computer Security Applications Conference}}},
  author = {Hetzelt, Felicitas and Radev, Martin and Buhren, Robert and Morbitzer, Mathias and Seifert, Jean-Pierre},
  year = 2021,
  month = dec,
  series = {{{ACSAC}} '21},
  pages = {273--284},
  publisher = {Association for Computing Machinery},
  address = {New York, NY, USA},
  doi = {10.1145/3485832.3488011},
  urldate = {2026-05-05},
  isbn = {978-1-4503-8579-4},
}

@inproceedings{juffingerNotSecureTSC2025,
  title = {Not {{So Secure TSC}}},
  booktitle = {Applied {{Cryptography}} and {{Network Security}}},
  author = {Juffinger, Jonas and Neela, Sudheendra Raghav and Gruss, Daniel},
  editor = {Fischlin, Marc and Moonsamy, Veelasha},
  year = 2025,
  pages = {3--23},
  publisher = {Springer Nature Switzerland},
  address = {Cham},
  doi = {10.1007/978-3-031-95764-2_1},
  isbn = {978-3-031-95764-2},
  langid = {english},
}

@inproceedings{kangIceClaveTrustedExecution2021,
  title = {{{IceClave}}: {{A Trusted Execution Environment}} for {{In-Storage Computing}}},
  shorttitle = {{{IceClave}}},
  booktitle = {{{MICRO-54}}: 54th {{Annual IEEE}}/{{ACM International Symposium}} on {{Microarchitecture}}},
  author = {Kang, Luyi and Xue, Yuqi and Jia, Weiwei and Wang, Xiaohao and Kim, Jongryool and Youn, Changhwan and Kang, Myeong Joon and Lim, Hyung Jin and Jacob, Bruce and Huang, Jian},
  year = 2021,
  month = oct,
  series = {{{MICRO}} '21},
  pages = {199--211},
  publisher = {Association for Computing Machinery},
  address = {New York, NY, USA},
  doi = {10.1145/3466752.3480109},
  urldate = {2026-03-16},
  isbn = {978-1-4503-8557-2},
}

@misc{KernelorgDocDocumentation,
  title = {Kernel.Org/Doc/{{Documentation}}/{{ABI}}/Testing/Configfs-Tsm},
  howpublished = {\url{https://www.kernel.org/doc/Documentation/ABI/testing/configfs-tsm}},
  note = {Last accessed: February 2026}
}

@inproceedings{kimLogStructuredMergeTreeaware2022,
  title = {A {{Log-Structured Merge Tree-aware Message Authentication Scheme}} for {{Persistent Key-Value Stores}}},
  booktitle = {20th {{USENIX Conference}} on {{File}} and {{Storage Technologies}} ({{FAST}} 22)},
  author = {Kim, Igjae and Kim, J. Hyun and Chung, Minu and Moon, Hyungon and Noh, Sam H.},
  year = 2022,
  pages = {363--380},
  urldate = {2025-05-30},
  isbn = {978-1-939133-26-7},
  langid = {english},
}

@inproceedings{kimShieldStoreShieldedInmemory2019a,
  title = {{{ShieldStore}}: {{Shielded In-memory Key-value Storage}} with {{SGX}}},
  shorttitle = {{{ShieldStore}}},
  booktitle = {Proceedings of the {{Fourteenth EuroSys Conference}} 2019},
  author = {Kim, Taehoon and Park, Joongun and Woo, Jaewook and Jeon, Seungheun and Huh, Jaehyuk},
  year = 2019,
  month = mar,
  series = {{{EuroSys}} '19},
  pages = {1--15},
  publisher = {Association for Computing Machinery},
  address = {New York, NY, USA},
  doi = {10.1145/3302424.3303951},
  urldate = {2025-07-13},
  isbn = {978-1-4503-6281-8},
}

@misc{lamBoostingNetworkPerformance2024,
  title = {Boosting the {{Network Performance}} of {{Confidential VM Using Userspace Stack}}},
  author = {Lam, William},
  year = 2024,
  month = sep,
  address = {Montreal, QC, Canada},
  urldate = {2026-02-26}
}

@inproceedings{liBifrostAnalysisOptimization2023,
  title = {Bifrost: {{Analysis}} and {{Optimization}} of {{Network I}}/{{O Tax}} in {{Confidential Virtual Machines}}},
  shorttitle = {Bifrost},
  booktitle = {2023 {{USENIX Annual Technical Conference}} ({{USENIX ATC}} 23)},
  author = {Li, Dingji and Mi, Zeyu and Ji, Chenhui and Tan, Yifan and Zang, Binyu and Guan, Haibing and Chen, Haibo},
  year = 2023,
  pages = {1--15},
  urldate = {2025-07-29},
  isbn = {978-1-939133-35-9},
  langid = {english},
}

@misc{liBridgeFutureHighPerformance2024,
  title = {Bridge the {{Future}}: {{High-Performance Networks}} in {{Confidential VMs}} without {{Trusted I}}/{{O}} Devices},
  shorttitle = {Bridge the {{Future}}},
  author = {Li, Mengyuan and Srivastava, Shashvat and Yan, Mengjia},
  year = 2024,
  month = mar,
  number = {arXiv:2403.03360},
  eprint = {2403.03360},
  primaryclass = {cs},
  publisher = {arXiv},
  doi = {10.48550/arXiv.2403.03360},
  urldate = {2026-02-26},
  archiveprefix = {arXiv},
}

@inproceedings{linLargescaleDataSet2020,
  title = {A {{Large-scale Data Set}} and an {{Empirical Study}} of {{Docker Images Hosted}} on {{Docker Hub}}},
  booktitle = {2020 {{IEEE International Conference}} on {{Software Maintenance}} and {{Evolution}} ({{ICSME}})},
  author = {Lin, Changyuan and Nadi, Sarah and Khazaei, Hamzeh},
  year = 2020,
  month = sep,
  pages = {371--381},
  issn = {2576-3148},
  doi = {10.1109/ICSME46990.2020.00043},
  urldate = {2026-03-06},
}

@techreport{lundbladeEntityAttestationToken2025a,
  type = {Request for {{Comments}}},
  title = {The {{Entity Attestation Token}} ({{EAT}})},
  author = {Lundblade, Laurence and Mandyam, Giridhar and O'Donoghue, Jeremy and Wallace, Carl},
  year = 2025,
  month = apr,
  number = {RFC 9711},
  institution = {Internet Engineering Task Force},
  doi = {10.17487/RFC9711},
  urldate = {2026-02-19},
}

@inproceedings{luWiscKeySeparatingKeys2016,
  title = {{{WiscKey}}: {{Separating Keys}} from {{Values}} in {{SSD-conscious Storage}}},
  shorttitle = {{{WiscKey}}},
  booktitle = {14th {{USENIX Conference}} on {{File}} and {{Storage Technologies}} ({{FAST}} 16)},
  author = {Lu, Lanyue and Pillai, Thanumalayan Sankaranarayana and {Arpaci-Dusseau}, Andrea C. and {Arpaci-Dusseau}, Remzi H.},
  year = 2016,
  pages = {133},
  urldate = {2026-03-06},
  isbn = {978-1-931971-28-7},
  langid = {english},
}

@article{mahajanDepotCloudStorage2011,
  title = {Depot: {{Cloud Storage}} with {{Minimal Trust}}},
  shorttitle = {Depot},
  author = {Mahajan, Prince and Setty, Srinath and Lee, Sangmin and Clement, Allen and Alvisi, Lorenzo and Dahlin, Mike and Walfish, Michael},
  year = 2011,
  month = dec,
  journal = {ACM Transactions on Computer Systems},
  volume = {29},
  number = {4},
  pages = {1--38},
  publisher = {Association for Computing Machinery (ACM)},
  issn = {0734-2071, 1557-7333},
  doi = {10.1145/2063509.2063512},
  urldate = {2025-07-23},
  copyright = {https://www.acm.org/publications/policies/copyright\_policy\#Background},
  langid = {english},
}

@misc{mahhoukMoeMahhoukGraminesealingkeyprovider2025,
  title = {{{MoeMahhouk}}/Gramine-Sealing-Key-Provider},
  author = {Mahhouk, Moe},
  year = 2025,
  month = apr,
  copyright = {MIT},
  note = {Last accessed: July 2025}
}

@inproceedings{martinADAMCSAdvancedAsynchronous2021,
  title = {{{ADAM-CS}}: {{Advanced Asynchronous Monotonic Counter Service}}},
  shorttitle = {{{ADAM-CS}}},
  booktitle = {2021 51st {{Annual IEEE}}/{{IFIP International Conference}} on {{Dependable Systems}} and {{Networks}} ({{DSN}})},
  author = {Martin, Andr{\'e} and Lian, Cong and Gregor, Franz and Krahn, Robert and Schiavoni, Valerio and Felber, Pascal and Fetzer, Christof},
  year = 2021,
  month = jun,
  pages = {426--437},
  issn = {2158-3927},
  doi = {10.1109/DSN48987.2021.00053},
  urldate = {2025-07-18},
}

@inproceedings{mateticROTERollbackProtection2017,
  title = {{{ROTE}}: {{Rollback Protection}} for {{Trusted Execution}}},
  shorttitle = {\textbraceleft{{ROTE}}\textbraceright},
  booktitle = {26th {{USENIX Security Symposium}} ({{USENIX Security}} 17)},
  author = {Matetic, Sinisa and Ahmed, Mansoor and Kostiainen, Kari and Dhar, Aritra and Sommer, David and Gervais, Arthur and Juels, Ari and Capkun, Srdjan},
  year = 2017,
  pages = {1289},
  urldate = {2025-07-18},
  isbn = {978-1-931971-40-9},
  langid = {english},
}

@inproceedings{messadiPrecursorFastClientcentric2021,
  title = {Precursor: A Fast, Client-Centric and Trusted Key-Value Store Using {{RDMA}} and {{Intel SGX}}},
  shorttitle = {Precursor},
  booktitle = {Proceedings of the 22nd {{International Middleware Conference}}},
  author = {Messadi, Ines and Neumann, Shivananda and Weichbrodt, Nico and Almstedt, Lennart and Mahhouk, Mohammad and Kapitza, R{\"u}diger},
  year = 2021,
  month = nov,
  pages = {1--13},
  publisher = {ACM},
  address = {Qu{\'e}bec City, Canada},
  doi = {10.1145/3464298.3476129},
  urldate = {2025-07-23},
  copyright = {https://www.acm.org/publications/policies/copyright\_policy\#Background},
  langid = {english},
}

@inproceedings{michaelForensicValidityApproximated2020,
  title = {On the {{Forensic Validity}} of {{Approximated Audit Logs}}},
  booktitle = {Annual {{Computer Security Applications Conference}}},
  author = {Michael, Noor and Mink, Jaron and Liu, Jason and Gaur, Sneha and Hassan, Wajih Ul and Bates, Adam},
  year = 2020,
  month = dec,
  pages = {189--202},
  publisher = {ACM},
  address = {Austin USA},
  doi = {10.1145/3427228.3427272},
  urldate = {2026-03-06},
  isbn = {978-1-4503-8858-0},
  langid = {english},
}

@article{misonoConfidentialVMsExplained2024,
  title = {Confidential {{VMs Explained}}: {{An Empirical Analysis}} of {{AMD SEV-SNP}} and {{Intel TDX}}},
  shorttitle = {Confidential {{VMs Explained}}},
  author = {Misono, Masanori and Stavrakakis, Dimitrios and Santos, Nuno and Bhatotia, Pramod},
  year = 2024,
  month = dec,
  journal = {Proceedings of the ACM on Measurement and Analysis of Computing Systems},
  volume = {8},
  number = {3},
  pages = {1--42},
  publisher = {Association for Computing Machinery (ACM)},
  issn = {2476-1249},
  doi = {10.1145/3700418},
  urldate = {2025-07-11},
  copyright = {https://creativecommons.org/licenses/by/4.0/},
  langid = {english},
}

@inproceedings{mohanVelaVirtualizedLLM2025,
  title = {Vela: {{A Virtualized LLM Training System}} with {{GPU Direct RoCE}}},
  shorttitle = {Vela},
  booktitle = {Proceedings of the 30th {{ACM International Conference}} on {{Architectural Support}} for {{Programming Languages}} and {{Operating Systems}}, {{Volume}} 2},
  author = {Mohan, Apoorve and Walkup, Robert and Karacali, Bengi and Chen, Ming-hung and Kayi, Abdullah and Schour, Liran and Salaria, Shweta and Wen, Sophia and Chung, I-hsin and Alim, Abdul and Evangelinos, Constantinos and Luo, Lixiang and Dombrowa, Marc and Schares, Laurent and Sydney, Ali and Maniotis, Pavlos and Koteshwara, Sandhya and Tang, Brent and Belog, Joel and Odaira, Rei and Tarasov, Vasily and Gampel, Eran and Thorstensen, Drew and Gershon, Talia and Seelam, Seetharami},
  year = 2025,
  month = mar,
  pages = {1348--1364},
  publisher = {ACM},
  address = {Rotterdam Netherlands},
  doi = {10.1145/3676641.3716280},
  urldate = {2025-04-28},
  isbn = {979-8-4007-1079-7},
  langid = {english},
}

@inproceedings{nguyenWhyFilesIf2024,
  title = {Why {{Files If You Have}} a {{DBMS}}?},
  booktitle = {2024 {{IEEE}} 40th {{International Conference}} on {{Data Engineering}} ({{ICDE}})},
  author = {Nguyen, Lam-Duy and Leis, Viktor},
  year = 2024,
  month = may,
  pages = {3878--3892},
  publisher = {IEEE},
  address = {Utrecht, Netherlands},
  doi = {10.1109/ICDE60146.2024.00297},
  urldate = {2026-03-06},
  copyright = {https://doi.org/10.15223/policy-029},
  isbn = {979-8-3503-1715-2},
  langid = {english},
}

@inproceedings{papadimitriouBigDataAnalytics2016,
  title = {Big {{Data Analytics}} over {{Encrypted Datasets}} with {{Seabed}}},
  booktitle = {12th {{USENIX Symposium}} on {{Operating Systems Design}} and {{Implementation}} ({{OSDI}} 16)},
  author = {Papadimitriou, Antonis and Bhagwan, Ranjita and Chandran, Nishanth and Ramjee, Ramachandran and Haeberlen, Andreas and Singh, Harmeet and Modi, Abhishek and Badrinarayanan, Saikrishna},
  year = 2016,
  pages = {587--602},
  urldate = {2025-07-29},
  isbn = {978-1-931971-33-1},
  langid = {english},
}

@misc{pci-sigPCIExpressBase2024,
  title = {{{PCI Express}}\textregistered{} {{Base Specification Revision}} 6.3},
  author = {{PCI-SIG}},
  year = 2024,
  month = dec,
  langid = {english},
}

@inproceedings{popaCryptDBProtectingConfidentiality2011,
  title = {{{CryptDB}}: Protecting Confidentiality with Encrypted Query Processing},
  shorttitle = {{{CryptDB}}},
  booktitle = {Proceedings of the {{Twenty-Third ACM Symposium}} on {{Operating Systems Principles}}},
  author = {Popa, Raluca Ada and Redfield, Catherine M. S. and Zeldovich, Nickolai and Balakrishnan, Hari},
  year = 2011,
  month = oct,
  pages = {85--100},
  publisher = {ACM},
  address = {Cascais Portugal},
  doi = {10.1145/2043556.2043566},
  urldate = {2025-07-18},
  copyright = {https://www.acm.org/publications/policies/copyright\_policy\#Background}
}

@inproceedings{popaEnablingSecurityCloud2011,
  title = {Enabling {{Security}} in {{Cloud Storage SLAs}} with {{CloudProof}}},
  booktitle = {2011 {{USENIX Annual Technical Conference}} ({{USENIX ATC}} 11)},
  author = {Popa, Raluca Ada and Lorch, Jacob R. and Molnar, David and Wang, Helen J. and Zhuang, Li},
  year = 2011,
  urldate = {2025-07-29},
  langid = {english},
}

@inproceedings{priebeEnclaveDBSecureDatabase2018,
  title = {{{EnclaveDB}}: {{A Secure Database Using SGX}}},
  shorttitle = {{{EnclaveDB}}},
  booktitle = {2018 {{IEEE Symposium}} on {{Security}} and {{Privacy}} ({{SP}})},
  author = {Priebe, Christian and Vaswani, Kapil and Costa, Manuel},
  year = 2018,
  month = may,
  pages = {264--278},
  publisher = {IEEE},
  address = {San Francisco, CA},
  doi = {10.1109/sp.2018.00025},
  urldate = {2025-07-18},
}

@inproceedings{quocSecureTFSecureTensorFlow2020,
  title = {{{secureTF}}: {{A Secure TensorFlow Framework}}},
  shorttitle = {{{secureTF}}},
  booktitle = {Proceedings of the 21st {{International Middleware Conference}}},
  author = {Quoc, Do Le and Gregor, Franz and Arnautov, Sergei and Kunkel, Roland and Bhatotia, Pramod and Fetzer, Christof},
  year = 2020,
  month = dec,
  series = {Middleware '20},
  pages = {44--59},
  publisher = {Association for Computing Machinery},
  address = {New York, NY, USA},
  doi = {10.1145/3423211.3425687},
  urldate = {2026-05-06},
  isbn = {978-1-4503-8153-6},
}

@inproceedings{rakshitASSUREAuthenticationScheme2017,
  title = {{{ASSURE}}: {{Authentication Scheme}} for {{SecURE}} Energy Efficient Non-Volatile Memories},
  shorttitle = {{{ASSURE}}},
  booktitle = {2017 54th {{ACM}}/{{EDAC}}/{{IEEE Design Automation Conference}} ({{DAC}})},
  author = {Rakshit, Joydeep and Mohanram, Kartik},
  year = 2017,
  month = jun,
  pages = {1--6},
  doi = {10.1145/3061639.3062205},
  urldate = {2025-07-22},
}

@misc{ramanJraman567Gogenref2025,
  title = {Jraman567/Go-Gen-Ref},
  author = {Raman, Jag},
  year = 2025,
  month = mar,
  note = {Last accessed: February 2026}
}

@misc{rothKVMIntroduceKVM_EXIT_SNP_REQ_CERTS2026,
  title = {{{KVM}}: {{Introduce KVM}}\_{{EXIT}}\_{{SNP}}\_{{REQ}}\_{{CERTS}} for {{SNP}} Certificate-Fetching},
  author = {Roth, Michael},
  year = 2026,
  month = jan,
  note = {Last accessed: February 2026}
}

@misc{rothKVMSEVAdd2026,
  title = {{{KVM}}: {{SEV}}: {{Add KVM}}\_{{SEV}}\_{{SNP}}\_{{ENABLE}}\_{{REQ}}\_{{CERTS}} Command},
  author = {Roth, Michael},
  year = 2026,
  month = jan,
  note = {Last accessed: February 2026}
}

@misc{segallUsingTPMMachine2012,
  title = {Using the {{TPM}}: {{Machine Authentication}} and {{Attestation}}},
  author = {Segall, Ariel},
  year = 2012,
  month = oct,
  urldate = {2026-02-26},
  langid = {english},
}

@article{sunBuildingEnclavenativeStorage2021,
  title = {Building Enclave-Native Storage Engines for Practical Encrypted Databases},
  author = {Sun, Yuanyuan and Wang, Sheng and Li, Huorong and Li, Feifei},
  year = 2021,
  month = feb,
  journal = {Proceedings of the VLDB Endowment},
  volume = {14},
  number = {6},
  pages = {1019--1032},
  publisher = {Association for Computing Machinery (ACM)},
  issn = {2150-8097},
  doi = {10.14778/3447689.3447705},
  urldate = {2025-07-23},
  langid = {english},
}

@misc{sunPATCHV402025,
  title = {[{{PATCH}} v4 0/2] {{PCI}}: {{Fix}} the Issue of Failed Speed Limit Lifting},
  author = {Sun, Jiwei},
  year = 2025,
  month = jan,
  note = {Last accessed: February 2026}
}

@inproceedings{szeferFastHardwareMemory2014,
  title = {Towards Fast Hardware Memory Integrity Checking with Skewed {{Merkle}} Trees},
  booktitle = {Proceedings of the {{Third Workshop}} on {{Hardware}} and {{Architectural Support}} for {{Security}} and {{Privacy}}},
  author = {Szefer, Jakub and Biedermann, Sebastian},
  year = 2014,
  month = jun,
  series = {{{HASP}} '14},
  pages = {1--8},
  publisher = {Association for Computing Machinery},
  address = {New York, NY, USA},
  doi = {10.1145/2611765.2611774},
  urldate = {2025-07-21},
  isbn = {978-1-4503-2777-0}
}

@inproceedings{tee-fail,
  title = {{{TEE}}.Fail: {{Breaking Trusted Execution Environments}} via {{DDR5 Memory Bus Interposition}}},
  booktitle = {47th {{IEEE Symposium}} on {{Security}} and {{Privacy}} ({{IEEE S}}\&{{P}} '26)},
  author = {Chuang, Jalen and Seto, Alex and Berrios, Nicolas and van Schaik, Stephan and Garman, Christina and Genkin, Daniel},
  year = 2026,
  publisher = {IEEE Computer Society},
}

@inproceedings{thalheimRktioDirectStack2021,
  title = {Rkt-Io: A Direct {{I}}/{{O}} Stack for Shielded Execution},
  shorttitle = {Rkt-Io},
  booktitle = {Proceedings of the {{Sixteenth European Conference}} on {{Computer Systems}}},
  author = {Thalheim, J{\"o}rg and Unnibhavi, Harshavardhan and Priebe, Christian and Bhatotia, Pramod and Pietzuch, Peter},
  year = 2021,
  month = apr,
  series = {{{EuroSys}} '21},
  pages = {490--506},
  publisher = {Association for Computing Machinery},
  address = {New York, NY, USA},
  doi = {10.1145/3447786.3456255},
  urldate = {2026-03-06},
  isbn = {978-1-4503-8334-9}
}

@inproceedings{thomasMidsummerNightsTree2024,
  title = {A {{Midsummer Night}}'s {{Tree}}: {{Efficient}} and {{High Performance Secure SCM}}},
  shorttitle = {A {{Midsummer Night}}'s {{Tree}}},
  booktitle = {Proceedings of the 29th {{ACM International Conference}} on {{Architectural Support}} for {{Programming Languages}} and {{Operating Systems}}, {{Volume}} 3},
  author = {Thomas, Samuel and Workneh, Kidus and McCarty, Jac and Izraelevitz, Joseph and Lehman, Tamara and Bahar, R. Iris},
  year = 2024,
  month = apr,
  series = {{{ASPLOS}} '24},
  volume = {3},
  pages = {22--37},
  publisher = {Association for Computing Machinery},
  address = {New York, NY, USA},
  doi = {10.1145/3620666.3651354},
  urldate = {2025-07-21},
  isbn = {979-8-4007-0386-7},
}

@inproceedings{tianAtomicDiskSecureVirtual2025,
  title = {{{AtomicDisk}}: {{A Secure Virtual Disk}} for {{TEEs}} against {{Eviction Attacks}}},
  shorttitle = {{{AtomicDisk}}},
  booktitle = {23rd {{USENIX Conference}} on {{File}} and {{Storage Technologies}} ({{FAST}} 25)},
  author = {Tian, Hongliang and Yu, Xinyi and Song, Shaowei and Chen, Qingsong and Zhang, Zhihao and Wang, Shiyu and Liu, Weijie and Xu, Erci and Yan, Shoumeng and Zhang, Yiming},
  year = 2025,
  pages = {449--459},
  urldate = {2025-08-22},
  isbn = {978-1-939133-45-8},
  langid = {english},
}

@techreport{trustedcomputinggroupTCGStorageSecurity2025,
  title = {{{TCG Storage Security Subsystem Class}} ({{SSC}}): {{Opal Version}} 2.30},
  author = {{Trusted Computing Group}},
  year = 2025,
  month = jan,
  institution = {Trusted Computing Group},
  langid = {english},
}

@misc{Trustee2025,
  title = {Trustee},
  year = 2025,
  month = jul,
  copyright = {Apache-2.0},
  howpublished = {Confidential Containers},
  note = {Last accessed: July 2025}
}

@article{tuProcessingAnalyticalQueries2013,
  title = {Processing Analytical Queries over Encrypted Data},
  author = {Tu, Stephen and Kaashoek, M. Frans and Madden, Samuel and Zeldovich, Nickolai},
  year = 2013,
  month = mar,
  journal = {Proceedings of the VLDB Endowment},
  volume = {6},
  number = {5},
  pages = {289--300},
  publisher = {Association for Computing Machinery (ACM)},
  issn = {2150-8097},
  doi = {10.14778/2535573.2488336},
  urldate = {2025-07-18},
  langid = {english},
}

@misc{Veraison,
  title = {Veraison},
  journal = {Veraison},
  howpublished = {\url{https://github.com/veraison}},
  langid = {english},
  note = {Last accessed: July 2025}
}

@misc{VeraisonRatsd2026,
  title = {Veraison/Ratsd},
  year = 2026,
  month = jan,
  copyright = {Apache-2.0},
  howpublished = {Veraison},
  note = {Last accessed: February 2026}
}

@misc{VeraisonServices2026,
  title = {Veraison/Services},
  year = 2026,
  month = feb,
  copyright = {Apache-2.0},
  howpublished = {Veraison},
  note = {Last accessed: February 2026}
}

@misc{VirteeSevsnpmeasurego2025,
  title = {Virtee/Sev-Snp-Measure-Go},
  year = 2025,
  month = nov,
  copyright = {Apache-2.0},
  howpublished = {VirTEE},
  note = {Last accessed: February 2026}
}

@misc{wangFastPersistAcceleratingModel2024,
  title = {{{FastPersist}}: {{Accelerating Model Checkpointing}} in {{Deep Learning}}},
  shorttitle = {{{FastPersist}}},
  author = {Wang, Guanhua and Ruwase, Olatunji and Xie, Bing and He, Yuxiong},
  year = 2024,
  month = jun,
  number = {arXiv:2406.13768},
  eprint = {2406.13768},
  primaryclass = {cs},
  publisher = {arXiv},
  doi = {10.48550/arXiv.2406.13768},
  urldate = {2025-06-06},
  archiveprefix = {arXiv},
  langid = {english},
}

@inproceedings{weinholdJVPFSAddingRobustness2011,
  title = {{{jVPFS}}: {{Adding Robustness}} to a {{Secure Stacked File System}} with {{Untrusted Local Storage Components}}},
  shorttitle = {\textbraceleft{{jVPFS}}\textbraceright},
  booktitle = {2011 {{USENIX Annual Technical Conference}} ({{USENIX ATC}} 11)},
  author = {Weinhold, Carsten},
  year = 2011,
  urldate = {2025-07-29},
  langid = {english},
}

@inproceedings{welchOptimizingHybridSSD2013,
  title = {Optimizing a Hybrid {{SSD}}/{{HDD HPC}} Storage System Based on File Size Distributions},
  booktitle = {2013 {{IEEE}} 29th {{Symposium}} on {{Mass Storage Systems}} and {{Technologies}} ({{MSST}})},
  author = {Welch, Brent and Noer, Geoffrey},
  year = 2013,
  month = may,
  pages = {1--12},
  publisher = {IEEE},
  address = {Long Beach, CA, USA},
  doi = {10.1109/MSST.2013.6558449},
  urldate = {2026-03-06},
  isbn = {978-1-4799-0218-7 978-1-4799-0217-0},
  langid = {english},
}

@inproceedings{wuAEGISFastAuthenticated2013,
  title = {{{AEGIS}}: {{A Fast Authenticated Encryption Algorithm}}},
  shorttitle = {{{AEGIS}}},
  booktitle = {Revised {{Selected Papers}} on {{Selected Areas}} in {{Cryptography}} -- {{SAC}} 2013 - {{Volume}} 8282},
  author = {Wu, Hongjun and Preneel, Bart},
  year = 2013,
  month = aug,
  pages = {185--201},
  publisher = {Springer-Verlag},
  address = {Berlin, Heidelberg},
  doi = {10.1007/978-3-662-43414-7_10},
  urldate = {2026-03-10},
  isbn = {978-3-662-43413-0}
}

@inproceedings{xuMlsDiskTrustedBlock2026,
  title = {{{MlsDisk}}: {{Trusted Block Storage}} for {{TEEs Based}} on {{Layered Secure Logging}}},
  shorttitle = {{{MlsDisk}}},
  booktitle = {24th {{USENIX Conference}} on {{File}} and {{Storage Technologies}} ({{FAST}} 26)},
  author = {Xu, Erci and Yu, Xinyi and Yin, Lujia and Luo, Xinyuan and Song, Shaowei and Chen, Qingsong and Yan, Shoumeng and Shu, Jiwu and Tian, Hongliang and Zhang, Yiming},
  year = 2026,
  pages = {443--459},
  urldate = {2026-02-27},
  isbn = {978-1-939133-53-3},
  langid = {english},
}

@misc{yangFastSecureDPDK2023,
  title = {Fast and {{Secure}}: {{DPDK Meets Confidential Computing}}},
  shorttitle = {Fast and {{Secure}}},
  author = {Yang, Zhifei},
  year = 2023,
  month = sep,
  address = {Dublin, Ireland},
  urldate = {2026-02-26}
}

@article{zadokCryptfsStackableVnode1998,
  title = {Cryptfs: {{A Stackable Vnode Level Encryption File System}}},
  author = {Zadok, Erez and Badulescu, Ion and Shender, Alex},
  year = 1998,
  doi = {10.7916/D82N5935},
  urldate = {2026-03-16},
  langid = {english},
  note = {technical report},
}

@inproceedings{zhangCitadelProtectingData2021,
  title = {Citadel: {{Protecting Data Privacy}} and {{Model Confidentiality}} for {{Collaborative Learning}}},
  shorttitle = {Citadel},
  booktitle = {Proceedings of the {{ACM Symposium}} on {{Cloud Computing}}},
  author = {Zhang, Chengliang and Xia, Junzhe and Yang, Baichen and Puyang, Huancheng and Wang, Wei and Chen, Ruichuan and Akkus, Istemi Ekin and Aditya, Paarijaat and Yan, Feng},
  year = 2021,
  month = nov,
  series = {{{SoCC}} '21},
  pages = {546--561},
  publisher = {Association for Computing Machinery},
  address = {New York, NY, USA},
  doi = {10.1145/3472883.3486998},
  urldate = {2026-05-06},
  isbn = {978-1-4503-8638-8},
}

@inproceedings{zhaoLargeScaleAnalysisDocker2019,
  title = {Large-{{Scale Analysis}} of the {{Docker Hub Dataset}}},
  booktitle = {2019 {{IEEE International Conference}} on {{Cluster Computing}} ({{CLUSTER}})},
  author = {Zhao, Nannan and Tarasov, Vasily and Albahar, Hadeel and Anwar, Ali and Rupprecht, Lukas and Skourtis, Dimitrios and Warke, Amit S. and Mohamed, Mohamed and Butt, Ali R.},
  year = 2019,
  month = sep,
  pages = {1--10},
  publisher = {IEEE},
  address = {Albuquerque, NM, USA},
  doi = {10.1109/CLUSTER.2019.8891000},
  urldate = {2026-03-06},
  copyright = {https://ieeexplore.ieee.org/Xplorehelp/downloads/license-information/IEEE.html},
  isbn = {978-1-7281-4734-5},
  langid = {english},
}

@inproceedings{arnautovSCONESecureLinux2016a,
  title = {{{SCONE}}: {{Secure Linux Containers}} with {{Intel SGX}}},
  shorttitle = {{{SCONE}}},
  booktitle = {12th {{USENIX Symposium}} on {{Operating Systems Design}} and {{Implementation}} ({{OSDI}} 16)},
  author = {Arnautov, Sergei and Trach, Bohdan and Gregor, Franz and Knauth, Thomas and Martin, Andre and Priebe, Christian and Lind, Joshua and Muthukumaran, Divya and O'Keeffe, Dan and Stillwell, Mark L. and Goltzsche, David and Eyers, Dave and Kapitza, R{\"u}diger and Pietzuch, Peter and Fetzer, Christof},
  year = 2016,
  pages = {689--703},
  urldate = {2026-08-18},
  isbn = {978-1-931971-33-1},
  langid = {english},
}

@article{balmauCharacterizingMachineLearning2022,
  title = {Characterizing {{I}}/{{O}} in {{Machine Learning}} with {{MLPerf Storage}}},
  author = {Balmau, Oana},
  year = 2022,
  month = nov,
  journal = {ACM SIGMOD Record},
  volume = {51},
  number = {3},
  pages = {47--48},
  issn = {0163-5808},
  doi = {10.1145/3572751.3572765},
  urldate = {2026-08-18},
}

@article{baumannShieldingApplicationsUntrusted2015,
  title = {Shielding {{Applications}} from an {{Untrusted Cloud}} with {{Haven}}},
  author = {Baumann, Andrew and Peinado, Marcus and Hunt, Galen},
  year = 2015,
  month = aug,
  journal = {ACM Transactions on Computer Systems (TOCS)},
  volume = {33},
  number = {3},
  pages = {8:1--8:26},
  issn = {0734-2071},
  doi = {10.1145/2799647},
  urldate = {2026-08-18},
}

@incollection{clarkeIncrementalMultisetHash2003,
  title = {Incremental {{Multiset Hash Functions}} and {{Their Application}} to {{Memory Integrity Checking}}},
  booktitle = {Advances in {{Cryptology}} - {{ASIACRYPT}} 2003},
  author = {Clarke, Dwaine and Devadas, Srinivas and Van Dijk, Marten and Gassend, Blaise and Suh, G. Edward},
  editor = {Goos, Gerhard and Hartmanis, Juris and Van Leeuwen, Jan and Laih, Chi-Sung},
  year = 2003,
  volume = {2894},
  pages = {188--207},
  publisher = {Springer Berlin Heidelberg},
  address = {Berlin, Heidelberg},
  doi = {10.1007/978-3-540-40061-5_12},
  urldate = {2026-08-21},
  isbn = {978-3-540-20592-0 978-3-540-40061-5},
  langid = {english},
}

@inproceedings{djokoNeXUSPracticalSecure2019,
  title = {{{NeXUS}}: {{Practical}} and {{Secure Access Control}} on {{Untrusted Storage Platforms}} Using {{Client-Side SGX}}},
  shorttitle = {{{NeXUS}}},
  booktitle = {2019 49th {{Annual IEEE}}/{{IFIP International Conference}} on {{Dependable Systems}} and {{Networks}} ({{DSN}})},
  author = {Djoko, Judicael B. and Lange, Jack and Lee, Adam J.},
  year = 2019,
  month = jun,
  pages = {401--413},
  issn = {1530-0889},
  doi = {10.1109/DSN.2019.00049},
  urldate = {2026-08-18},
}

@misc{IntelTrustDomain,
  title = {{{Intel}}\textregistered{} {{Trust Domain Extension Guest Kernel Hardening Documentation}} --- {{Intel}}\textregistered{} {{Trust Domain Extension}}},
  urldate = {2026-08-28},
  howpublished = {\url{https://intel.github.io/ccc-linux-guest-hardening-docs/index.html}},
  note = {Last accessed: August 2026}
}

@inproceedings{krahnPesosPolicyEnhanced2018,
  title = {Pesos: Policy Enhanced Secure Object Store},
  shorttitle = {Pesos},
  booktitle = {Proceedings of the {{Thirteenth EuroSys Conference}}},
  author = {Krahn, Robert and Trach, Bohdan and {Vahldiek-Oberwagner}, Anjo and Knauth, Thomas and Bhatotia, Pramod and Fetzer, Christof},
  year = 2018,
  month = apr,
  series = {{{EuroSys}} '18},
  pages = {1--17},
  publisher = {Association for Computing Machinery},
  address = {New York, NY, USA},
  doi = {10.1145/3190508.3190518},
  urldate = {2026-08-17},
  isbn = {978-1-4503-5584-1},
}

@misc{MoreInformationEncrypted,
  title = {More {{Information}} on {{Encrypted Memory Frameworks}}},
  year = 2025,
  journal = {Intel},
  urldate = {2026-08-28},
  howpublished = {\url{https://www.intel.com/content/www/us/en/developer/articles/news/more-information-encrypted-memory-frameworks.html}},
  langid = {english},
  note = {Last accessed: August 2026}
}

@misc{RedisPersistence,
  title = {Redis Persistence},
  year = 2026,
  journal = {Docs},
  urldate = {2026-08-28},
  chapter = {operate},
  howpublished = {\url{https://redis.io/docs/latest/operate/oss_and_stack/management/persistence/}},
  langid = {english},
  note = {Last accessed: August 2026}
}

@inproceedings{rodriguez-silvaVideoSurveillanceBased2012,
  title = {Video {{Surveillance Based}} on {{Cloud Storage}}},
  booktitle = {2012 {{IEEE Fifth International Conference}} on {{Cloud Computing}}},
  author = {{Rodr{\'\i}guez-Silva}, D.A. and {Adkinson-Orellana}, L. and {Gonz{\'a}lez-Casta{\~n}o}, F.J. and {Armi{\~n}o-Franco}, I. and {Gonz{\'a}lez-Mart{\'\i}nez}, D.},
  year = 2012,
  month = jun,
  pages = {991--992},
  issn = {2159-6190},
  doi = {10.1109/CLOUD.2012.44},
  urldate = {2026-08-28},
}

@article{silvaPACSArchiveArchitecture2012,
  title = {A {{PACS}} Archive Architecture Supported on Cloud Services},
  author = {Silva, Lu{\'i}s A. Basti{\~a}o and Costa, Carlos and Oliveira, Jos{\'e} Luis},
  year = 2012,
  month = may,
  journal = {International Journal of Computer Assisted Radiology and Surgery},
  volume = {7},
  number = {3},
  pages = {349--358},
  issn = {1861-6429},
  doi = {10.1007/s11548-011-0625-x},
  urldate = {2026-08-28},
  langid = {english}
}

@article{tarasovFilebenchFlexibleFramework2016,
  title = {Filebench: {{A Flexible Framework}} for {{File System Benchmarking}}},
  shorttitle = {Filebench},
  author = {Tarasov, Vasily and Zadok, E. and Shepler, S.},
  year = 2016,
  journal = {login Usenix Mag.},
  urldate = {2026-08-24},
}

@inproceedings{xuVStoreDataStore2019,
  title = {{{VStore}}: {{A Data Store}} for {{Analytics}} on {{Large Videos}}},
  shorttitle = {{{VStore}}},
  booktitle = {Proceedings of the {{Fourteenth EuroSys Conference}} 2019},
  author = {Xu, Tiantu and Botelho, Luis Materon and Lin, Felix Xiaozhu},
  year = 2019,
  month = mar,
  series = {{{EuroSys}} '19},
  pages = {1--17},
  publisher = {Association for Computing Machinery},
  address = {New York, NY, USA},
  doi = {10.1145/3302424.3303971},
  urldate = {2026-08-28},
  isbn = {978-1-4503-6281-8},
}

\appendices

\ifextended
\section{\texorpdfstring{\rev[rev1]{G.05, G.11}{Trusted Counter, Freshness \& Durability}}{Trusted Counter, Freshness \& Durability}}
\label{sec:appendix:counter}
\begin{revblock}[rev1]{G.05, G.11}%
\projectname{} treats the external trusted party as a pluggable service behind a small library (\autoref{subsec:trusted_counter}).

\myparagraph{Interface}
\projectname{} splits the counter into a \emph{local reservation} and an \emph{external commit} (\autoref{tab:appendix:counter-api}).
\texttt{next()} draws the next value of a monotonic sequence for every per-operation tag and per-epoch stamp, without I/O.
\texttt{commit()} publishes a reserved value and is the only operation that pays a round trip.
This keeps the cost of an external service affordable. Every object carries a distinct counter, yet only one commit per epoch reaches the service.
Coupling the two would add that round trip to every operation, the per-operation point of \autoref{fig:evaluation:trustedcounter}.

\projectname{}'s freshness guarantee then rests on three properties that state-continuity services already provide~\cite{rollback_parno,mateticROTERollbackProtection2017,martinADAMCSAdvancedAsynchronous2021,angelNimbleRollbackProtection2023}:
\emph{monotonicity}, a commit of $v$ is accepted only if it exceeds the stored value and is read back authenticated at start-up, which prevents replay of an older commit;
\emph{durability before acknowledgment}, since \projectname{} swaps its superblock on the strength of a commit;
and an \emph{authenticated, replay-protected channel} bound to the attested CVM through the sealed configuration of \autoref{fig:overview}, without which an attacker could forge or replay commits, and monotonicity would protect nothing.

\begin{table}[!tb]
    \centering
    \myfontsizeapi
    \setlength{\tabcolsep}{4pt}
    \revnoref[rev1]{%
    \begin{tabularx}{\linewidth}{@{}llX@{}}
    \toprule
    {Operation} & {Cost} & {Description} \\
    \midrule
    \multicolumn{3}{@{}l}{\emph{Local counter (in the CVM)}} \\
    \texttt{next()} & none & Reserve and return the next value \\
    \texttt{read()} & none & Last reserved value, without reserving \\
    \texttt{load/store} & none & Import/export state for provisioning and sealing \\
    \midrule
    \multicolumn{3}{@{}l}{\emph{External service (over the authenticated channel)}} \\
    \texttt{commit(v)} & 1 RTT & Publish reserved value $v$; asynchronous \\
    \bottomrule
    \end{tabularx}
    \vspace{-0.5em}
    \caption{\rev[rev1]{G.05, G.11}{\projectname{}'s trusted counter interface. Only \texttt{commit} leaves the CVM, and it is issued once per epoch rather than once per operation.}}
    \label{tab:appendix:counter-api}
    }
    \vspace{-1em}
\end{table}

\myparagraph{Epoch protocol}
An epoch is the unit of durability and freshness, and a \emph{trigger} controls length by choosing when a new epoch is opened.
The selection depends on the rollback window a workload can tolerate: \projectname{} supports operation-, time-, and event-based triggers (\autoref{tab:appendix:sync-modes}).
The engine then snapshots the metadata, reserves a value $c$, writes the encrypted snapshot to the cold blob (cost $D$), commits $c$ to the service (cost $\mathit{RTT}$), and finally swaps the superblock.
\emph{Coupling} controls how the metadata write and the commit are scheduled relative to each other.
\textsc{ordered} commits after the write ($D + \mathit{RTT}$, correct by construction), while \textsc{parallel} overlaps them ($\max(D,\mathit{RTT})$, best when $D \ll \mathit{RTT}$).
The service's value is at most \emph{one} metadata generation ahead of the durable metadata, since one epoch is in flight at a time.
The modifiers set what a completion includes.
The \emph{blocking} modifier holds subsequent mutations until the in-flight epoch is durable and its freshness is committed.
Without it, mutations continue while that epoch's persistence is still in flight, relaxing the crash-loss bound to a potential two generations, $\max(\text{window}, D + \mathit{RTT})$. 
With it, a crash loses at most one configured window, at a throughput of roughly $\text{window}/(\text{window} + D + \mathit{RTT})$.
\autoref{fig:evaluation:trustedcounter} measures this trade-off, with \emph{metadata-only} separating its durability curve from its freshness curve.
\end{revblock}

\begin{table}[!tb]
    \centering
    \myfontsizeapi
    \setlength{\tabcolsep}{4pt}
    \revnoref[rev1]{%
    \begin{tabularx}{\linewidth}{@{}llX@{}}
    \toprule
    {} & {Setting} & {Semantics and rollback window} \\
    \midrule
    \multirow{2}{*}{Coupling}
      & \emph{ordered} & Blob\,$\rightarrow$\,commit\,$\rightarrow$\,superblock: $D + \mathit{RTT}$ \\
      & \emph{parallel} & Commit with the blob write: $\max(D, \mathit{RTT})$ \\
    \midrule
    \multirow{5}{*}{Trigger}
      & \emph{shutdown} & Epoch at close only \\
      & \emph{barrier} & Epoch on an explicit barrier only \\
      & \emph{every-$N$} & Epoch every $N$ mutations\\
      & \emph{every-$T$} & Epoch every $T$\,\textmu s\\
      & \emph{per-op} & Epoch per mutation \\
    \midrule
    \multirow{2}{*}{Modifier}
      & \emph{blocking} & Admit further mutations only once the epoch is durable \\
      & \emph{metadata-only} & Persist without committing freshness \\
    \bottomrule
    \end{tabularx}
    \vspace{-0.5em}
    \caption{\rev[rev1]{G.05, G.11}{Durability and freshness configuration.}}
    \label{tab:appendix:sync-modes}
    }
    \vspace{-1em}
\end{table}

\fi

\ifextended
\section{\texorpdfstring{\rev[rev1]{G.12}{Instance Lifecycle}}{Instance Lifecycle}}
\label{app:lifecycle}

\begin{figure}[!tb]
    \centering
    \begin{revfloat}[rev1]{G.12}
    \includegraphics[width=\linewidth,trim=0 0 0 1.9,clip]{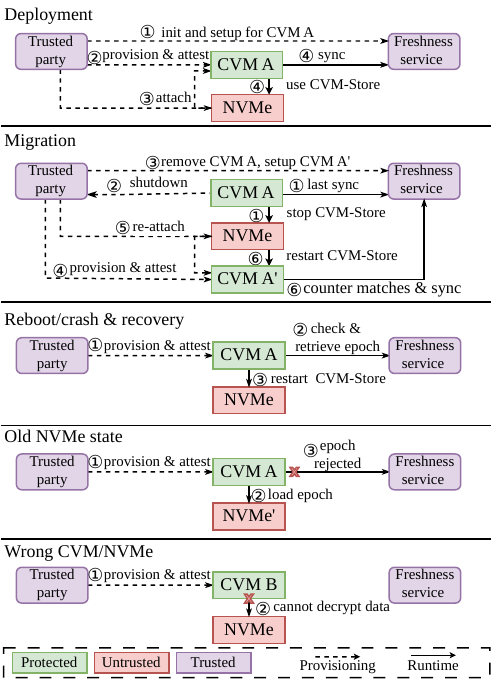}
    \vspace{-0.6em}
    \caption{\revnoref[rev1]{\projectname{} lifecycle across its common scenarios.}}
    \label{fig:appendix:lifecycle}
    \end{revfloat}
    \vspace{-1em}
\end{figure}

\begin{revblock}[rev1]{G.12}
\autoref{fig:appendix:lifecycle} shows the common lifecycle scenarios.
The trusted party uses the untrusted host's VMM (Virtual Machine Manager) to launch CVMs and to assign the NVMe devices to them.
For every instance, the sealing key pair is generated at boot and never leaves the CVM.
Thus, the identity the trusted party uses for provisioning is guaranteed to be unique and no two identical CVMs ever exist.

\myparagraph{Deployment}
The trusted party provides the freshness service with the new instance (\Circled{1}),
then uses the VMM to launch the CVM and attests it after boot.
After attestation, it provides the sealed configuration bound to the CVM identity with the symmetric disk-encryption key and freshness service parameters (\Circled{2}) 
and assigns the NVMe device to the CVM (\Circled{3}).
The CVM then creates the store and synchronizes the counter for each metadata epoch with the freshness service (\Circled{4}).

\myparagraph{Migration}
The migration must be coordinated with the trusted party, not done by the CVM instance itself.
The source instance synchronizes its last epoch and stops \projectname{} (\Circled{1}) and shuts down (\Circled{2}).
Afterwards, the trusted party removes the old instance from the freshness service and creates a new one with the same expected state (\Circled{3}).
Now, the target CVM is launched, attested, and provisioned (\Circled{4}).
Once it is ready for setup, the NVMe device is reassigned to it (\Circled{5}).
The new CVM then starts \projectname{} and resumes once its store's counter matches the freshness service (\Circled{6}).
Live migration is not supported, as \projectname{} expects only one user per store and CVM platforms offer only limited support for it at the moment.

\myparagraph{Reboot and crash recovery}
An intentionally or unintentionally restarted CVM is treated as a new CVM, because the original sealing key is gone.
Therefore, the CVM needs to be re-attested and provisioned by the trusted party (\Circled{1}).
After successful re-provisioning, \projectname{} retrieves the expected epoch from the freshness service and verifies the store against it (\Circled{2}), and the Service restarts \projectname{} (\Circled{3}).

\myparagraph{Old NVMe state}
A restored snapshot or copied image carries a stale epoch that lags behind the counter of the freshness service.
After CVM attestation and provisioning (\Circled{1}), the CVM loads the old epoch from the device (\Circled{2}), but detects the staleness via the freshness service (\Circled{3}).
Because our provisioning makes each CVM instance unique and CVMs cannot be directly cloned by design, no two instances can concurrently update the same entries in the freshness service.
Restoring an older state therefore requires coordination with the trusted party, which provisions the matching epoch into the freshness service.

\myparagraph{Wrong CVM or NVMe}
A store configuration is bound to a specific CVM instance by the trusted party.
A different CVM, even one provisioned by the trusted party with its own configuration (\Circled{1}), holds a different symmetric key and cannot decrypt the store on the attached device (\Circled{2}).
An unprovisioned CVM never holds any encryption keys or freshness service configuration.
\end{revblock}
\fi

\section{SPDK hardening \& DMA allocations}
\label{app:spdk}
We list the key SPDK structures that require DMA-accessible memory in \autoref{tab:dma-allocations}.
They cover NVMe queue management, blobstore metadata, and I/O buffer pools.
We identified these allocations by tracing SPDK memory allocations with GDB and manually cross-referencing the source code.

To harden the NVMe driver, we added bounds checks to the completion queues and split the \texttt{nvme\_tracker} struct so that only the required data is allocated in DMA memory.
\projectname{} uses SPDK's \texttt{iobuf} sized accordingly and allocated in DMA shared memory with its own module.
\begin{table}[!h]
\centering
\myfontsizeapi
\setlength{\tabcolsep}{3pt}
\begin{tabularx}{\linewidth}{@{}>{\hsize=0.853\hsize\raggedright\arraybackslash}X>{\hsize=1.147\hsize\raggedright\arraybackslash}X@{}}
\toprule
\textbf{Structure} & \textbf{Frequency and description} \\
\midrule
\texttt{pqpair->cmd} (\texttt{spdk\_nvme\_cmd[]}) & [Per queue pair] NVMe submission queue\\
\addlinespace[1.5pt]
\texttt{pqpair->cpl} (\texttt{spdk\_nvme\_cpl[]}) & [Per queue pair] NVMe completion queue\\
\addlinespace[1.5pt]
\texttt{pqpair->dma\_bufs} (\texttt{nvme\_\allowbreak{}dma\_\allowbreak{}prp\_\allowbreak{}sgl\_\allowbreak{}buffer[]}) & [Per queue pair] PRP/SGL scatter-gather descriptors (\qty{4}{\KB}, \qty{4}{\KB}-aligned), split from \texttt{nvme\_tracker}, which is now private\\
\addlinespace[1.5pt]
\texttt{spdk\_bs\_super\_block} & [On load/sync] Blobstore metadata (version, cluster size)\\
\addlinespace[1.5pt]
\texttt{spdk\_bs\_md\_mask} & [Periodic] Used-pages/blobids/clusters bitmasks (\qty{458}{\KB} ea.)\\
\addlinespace[1.5pt]
\texttt{spdk\_blob\_md\_page} & [Per opened blob] Blob extent maps and cluster metadata (\qty{4}{\KB})\\
\addlinespace[1.5pt]
\texttt{spdk\_nvme\_ctrlr\_data} / \texttt{spdk\_nvme\_ns\_data} & [Per identify command] Identify controller/namespace payload (\qty{4}{\KB})\\
\addlinespace[1.5pt]
\texttt{small\_pool\_base} / \texttt{large\_pool\_base} & [Once (init)] SPDK's modules with default names (unused), \projectname{}'s DMA buffers (resized based on configuration, e.g., \qty{180}{\mebi\byte} single-threaded w/ \qty{1}{\mebi\byte} chunks)\\
\addlinespace[1.5pt]
\texttt{g\_bdev\_mgr.\allowbreak{}zero\_buffer} & [Once (init)] Zero buffer for write-zero ops (\qty{1}{\mega\byte}) \\
\bottomrule
\end{tabularx}
\vspace{-0.5em}
\caption{Selection of SPDK main DMA memory allocations. All structures need to be allocated in DMA memory.}
\label{tab:dma-allocations}
\vspace{-1em}
\end{table}

\ifextended
\fi

\clearpage

\newcommand{\sys}{\projectname}
\newcommand{\file}[1]{{\def\UrlBreaks{\do\-}\def\UrlBigBreaks{}\path{#1}}}

\section{Artifact Appendix}
\label{app:ae}

This artifact contains the implementation and the scripts to reproduce the
experiments, figures, and tables of the paper. It consists of the \sys
library and its example clients, the patched guest kernel, the CVM-enabled
DPDK/SPDK forks, the applications ported to \sys (Redis, MLPerf Storage), and the benchmarking framework. 
The framework builds the guest images,
boots AMD SEV-SNP guests with NVMe passthrough, and automates the
measurements: the fio and crypto motivation experiments, the \sys evaluation
against ext4 and Rollbaccine (I/O characteristics, ablation, amplification,
epochs, seal/unseal recovery), and the end-to-end MLPerf Storage, Redis
and attestation benchmarks.

\subsection{Description \& Requirements}

\subsubsection{How to access}
The artifact is publicly available on GitHub at
\url{https://github.com/TUM-DSE/CVM-Store-NDSS27}%
\ifextended\else{} (DOI: \emph{TBD})\fi. This
repository is the evaluation harness and pins every other
component (\sys, DPDK/SPDK forks, guest kernel patches, MLPerf
Storage and Redis forks, and the orchestration
framework) as Nix flake inputs, so all components are built reproducibly. 

\subsubsection{Hardware dependencies}
\begin{itemize}
  \item An AMD SEV-SNP enabled server (EPYC Zen~4 and up, we mainly used EPYC 9334)
  \item At least 64 physical cores and 320\,GB of DDR5 RAM.
  \item A dedicated KIOXIA CM7-R PCIe Gen~5 NVMe SSD or similar with 3.8\,TB or more. It is erased by the experiments and must not hold any data. A slower or Gen~4 NVMe can demonstrate the functionality, not the absolute numbers. NVMe NUMA and guest configuration should match.
  \item Up to about 400\,GB of free disk space for traces and build artifacts.
\end{itemize}

\subsubsection{Software dependencies}
\begin{itemize}
  \item A Linux host with SEV-SNP support (7.2 Linux kernel upstream, or older 6.18 with custom patches), IOMMU and \texttt{vfio-pci} included. Hugepages should be allocatable on the host.
  \item Nix with flakes and ccache enabled. Everything else is built via Nix directly.
  \item Docker on the host, for the MLPerf Storage container in the guest
    image and Veraison.
  \item Root access, for booting VMs and binding the NVMe to VFIO.
\end{itemize}

\subsubsection{Benchmarks}
All workloads are included in the artifact repository or generated reproducibly.

\subsection{Artifact Installation \& Configuration}

\begin{enumerate}
\item Install Nix with flakes on the host and clone
  \texttt{CVM-Store-NDSS27}, then enter the shell with \texttt{nix develop}.
\item Add the host to \texttt{config.toml} with the correct NVMe.
\item Build OVMF, QEMU, and the guest image with MLPerf: \texttt{oma-arvio build.all} and
  \texttt{oma-arvio accelstore.build-guest --mlperf}
\item Set up Veraison for attestation:
  \texttt{oma-arvio veraison.setup}, then check with
  \texttt{oma-arvio veraison.status} that every service is up.
\item Smoke test that boots a SEV-SNP guest with NVMe:
  \texttt{oma-arvio-root vm.start --type snp --size medium -u}
\end{enumerate}

\subsection{Major Claims}
The artifact supports the following claims of the paper. C1 covers the
motivation, C2 to C7 correspond to the evaluation's Takeaways 1 to 6, and
each names the experiment and figure or table that it relates to.
\begin{itemize}
  \item (C1) CVM kernel I/O loses at least 10--23\% of VM throughput, authenticated
    block-level encryption (AES-XTS+sha256) a further 37--56\% of read
    bandwidth, while AEGIS exceeds NVMe bandwidth on one core. (E1),
    Figures~\ref{fig:dma-overhead}--\ref{fig:luks-overhead},
    Tables~\ref{tbl:luks-performance} and~\ref{tab:crypto-throughput}.
  \item (C2) For large objects, one core with full CIF is within
    $\approx$1\% (write) and 1--19\% (read) of raw NVMe bandwidth, while
    single-threaded ext4 and Rollbaccine reach 18--30\% on reads
    (Takeaway~1). (E2), Figure~\ref{fig:evaluation:io}.
  \item (C3) These characteristics carry through to applications: \sys
    matches or outperforms both baselines on large-object workloads and
    trails only on small-object or high-thread-count ones (Takeaway~2).
    (E3), Figure~\ref{fig:evalutation:e2e-performance}.
  \item (C4) Read/write amplification of the three abstractions stays
    below 1\textperthousand{} (Takeaway~3). (E4),
    Table~\ref{tab:evaluation:rw-amplification}.
  \item (C5) Full CIF costs $\approx$1\% runtime, while naively
    implementing any one part costs up to 76\% (Takeaway~4). (E4),
    Figure~\ref{fig:evaluation:ablation}.
  \item (C6) Attestation is a one-time overhead; sealing scales
    linearly with data (Takeaway~5).
    (E5), Tables~\ref{tab:evaluation:attestation}--\ref{tab:evaluation:api-recovery}.
  \item (C7) Widening the epoch takes filesystem and shared-log overhead
    from 8--10\% down to $\approx$2.5\% (Takeaway~6). (E6),
    Figure~\ref{fig:evaluation:trustedcounter}.
\end{itemize}

\subsection{Evaluation}
Every figure and table of the paper is tied to one \texttt{paper.<name>} task that
runs its experiment, exports results into \file{bench-result/parsed/},
and figures into \file{bench-result/plots/}.
All runs are automated, with minimal human intervention required.
\texttt{oma-arvio-root paper.all --quick} runs a representative subset of them in about a day. Without \texttt{--quick}, the full set may take several
weeks, mainly because of the fio runs, which require the NVMe to reach steady state.
The \file{README.md} in the artifact repository lists the exact differences between the quick subset and the paper, and a runtime breakdown of the tasks.
Quick runs use less data, so the absolute numbers may differ from those in the paper. However, the ratios between the arms and the shapes of the curves are expected to be similar on matching hardware.
The following table shows the experiments and tasks that are part of it.
\begin{center}
\footnotesize
\setlength{\tabcolsep}{3pt}
\begin{tabularx}{\columnwidth}{@{}l l >{\raggedright\arraybackslash}X l@{}}
\toprule
\multicolumn{2}{@{}l}{Task \texttt{paper.*}} & Output & Paper \\
\midrule
\multicolumn{4}{@{}l@{}}{\emph{Motivation}} \\
E1 & \texttt{crypto}$^*$    & \mbox{\file{crypto-summary.csv}}      & Tab.~\ref{tab:crypto-throughput} \\
   & \texttt{fio-iops}      & \mbox{\file{fio-iops_iops.pdf}}       & Fig.~\ref{fig:dma-overhead} \\
   & \texttt{fio-stacks}    & \mbox{\file{fio-luks_iops.pdf}}, \mbox{\file{fio-bandwidth.csv}} & Fig.~\ref{fig:luks-overhead}, Tab.~\ref{tbl:luks-performance} \\
\addlinespace
\multicolumn{4}{@{}l@{}}{\emph{I/O analysis of the three abstractions}} \\
E2 & \texttt{synthetic}$^*$ & \mbox{\file{synthetic.pdf}}           & Fig.~\ref{fig:evaluation:io} \\
\addlinespace
\multicolumn{4}{@{}l@{}}{\emph{End-to-end applications}} \\
E3 & \texttt{apps}$^*$      & \mbox{\file{apps.pdf}}                & Fig.~\ref{fig:evalutation:e2e-performance} \\
\addlinespace
\multicolumn{4}{@{}l@{}}{\emph{Ablation and amplification study}} \\
E4 & \texttt{ablation}$^*$  & \mbox{\file{ablation.pdf}}            & Fig.~\ref{fig:evaluation:ablation} \\
   & \texttt{amplification}$^*$ & \mbox{\file{amplification.csv}}   & Tab.~\ref{tab:evaluation:rw-amplification} \\
\addlinespace
\multicolumn{4}{@{}l@{}}{\emph{Attestation, sealing and recovery}} \\
E5 & \texttt{attestation}$^*$ & \mbox{\file{attestation-summary.csv}} & Tab.~\ref{tab:evaluation:attestation} \\
   & \texttt{seal-unseal}$^*$ & \mbox{\file{seal-unseal-summary.csv}} & Tab.~\ref{tab:evaluation:api-recovery} \\
\addlinespace
\multicolumn{4}{@{}l@{}}{\emph{Trusted counter epochs}} \\
E6 & \texttt{epoch}$^*$     & \mbox{\file{epoch-overhead.pdf}}      & Fig.~\ref{fig:evaluation:trustedcounter} \\
\bottomrule
\multicolumn{4}{@{}l@{}}{$^*$ included in \texttt{paper.all --quick}} \\
\end{tabularx}
\end{center}

\noindent Prerequisites:
\begin{itemize}
  \item All: make sure the build steps are done and no one else is using the NVMe.
  \item E1: the two fio tasks precondition the whole SSD (only run if time permits).
  \item E3: requires MLPerf in the guest image. If the runs fail, rebuild the guest image.
  \item E5: \texttt{paper.attestation} requires Veraison running on the host.
  Check with \texttt{oma-arvio veraison.status}.
\end{itemize}

\subsection{Customization}
The \file{README.md} and \file{docs/benchmarks.md} of the artifact
repository describe how to run every benchmark family manually, including how
to customize the VM size, the storage backends, crypto variants, durability
and freshness settings, and workloads.

\end{document}